\documentclass[journal]{IEEEtran}

\PassOptionsToPackage{table}{xcolor}
\usepackage{amsmath,amsfonts}
\usepackage{algorithmic}
\usepackage{array}
\usepackage[caption=false,font=normalsize,labelfont=sf,textfont=sf]{subfig}
\usepackage{textcomp}
\usepackage{stfloats}
\usepackage{url}
\usepackage{verbatim}
\usepackage{graphicx}
\usepackage{balance}
\usepackage{amsthm}
\theoremstyle{definition}
\newtheorem{definition}{Definition}[]
\usepackage{color}
\usepackage{xspace}
\usepackage{xparse}

\usepackage{hyperref}

\usepackage{multirow}
\usepackage{graphicx}
\usepackage{rotating}
\usepackage{array}
\usepackage{pifont}
\usepackage{float}
\usepackage[edges]{forest}
\usepackage{cite}

\definecolor{hidden-draw}{RGB}{20,68,106}
\definecolor{grey}{rgb}{0.5, 0.5, 0.5}

\usepackage[framemethod=tikz]{mdframed}
\usepackage{bbm}

\usepackage{caption}

\usepackage{xcolor}

\newcommand{\eat}[1]{}

\definecolor{shadecolor}{RGB}{220,220,220}

\definecolor{inputcolor}{RGB}{255,139,35}
\definecolor{outputcolor}{RGB}{120,212,252}
\definecolor{embedcolor}{RGB}{254,127,156}
\definecolor{maskcolor}{RGB}{122,128,255}
\definecolor{ecolor}{RGB}{58,149,54}

\definecolor{highcolor}{RGB}{255,153,153}
\definecolor{midcolor}{RGB}{255,204,204}
\definecolor{lowcolor}{RGB}{204,229,255}

\definecolor{green}{RGB}{0,128,0}

\definecolor{yellow}{RGB}{255,200,18}

\newcommand{\addd}[1]{\textcolor{black}{#1}}
\renewcommand{\marginpar}[1]{}

\newcommand{\stab}{\vspace{1.2ex}\noindent}

\newcommand{\bi}{\begin{itemize}}
\newcommand{\ei}{\end{itemize}}

\newcommand{\be}{\begin{enumerate}}
\newcommand{\ee}{\end{enumerate}}
\newcommand{\beqn}{\begin{eqnarray*}}
\newcommand{\eeqn}{\end{eqnarray*}}

\newcommand{\stitle}[1]{\stab\noindent{\bf #1}}
\newcommand{\etitle}[1]{\vspace{1mm}\noindent{\underline{\em #1}}}

\newcommand{\ie}{\textit{i.e.,}\xspace}
\newcommand{\eg}{\textit{e.g.,}\xspace}

\newcommand{\aka}{\emph{a.k.a.}\xspace}

\newcommand{\db}{{\sc db}\xspace}

\newcommand{\nlq}{{\sc nl}\xspace}
\newcommand{\sql}{{\sc sql}\xspace}

\newcommand{\nlsql}{{Text-to-SQL}\xspace}

\newcommand{\knowledge}{{\sc k}\xspace}

\NewDocumentCommand{\nan}{ mO{} }{\textcolor{blue}{\textsuperscript{\textit{Nan}}\textsf{\textbf{\small[#1]}}}}

\NewDocumentCommand{\yuyu}{ mO{} }{\textcolor{green}{\textsuperscript{\textit{Yuyu}}\textsf{\textbf{\small[#1]}}}}

\NewDocumentCommand{\xinyu}{ mO{} }{\textcolor{red}{\textsuperscript{\textit{Xinyu}}\textsf{\textbf{\small[#1]}}}}

\NewDocumentCommand{\boyan}{ mO{} }{\textcolor{yellow}{\textsuperscript{\textit{Boyan}}\textsf{\textbf{\small[#1]}}}}

\makeatletter
\apptocmd{\@maketitle}{\centering\insertfig}{}{}
\makeatother

\newcommand{\greencheck}{{\color{green}\ding{51}}}

\usepackage[skins]{tcolorbox}
\definecolor{colorcommentfg}{HTML}{336633}
\definecolor{colorcommentbg}{HTML}{ededed}
\definecolor{colorcommentframe}{HTML}{336633}

\begin{document}


\title{A Survey of Text-to-SQL in the Era of LLMs: Where are we, and where are we going?}

\author{Xinyu~Liu, Shuyu~Shen, Boyan~Li, Peixian~Ma, Runzhi~Jiang, Yuxin Zhang,  Ju~Fan, \\ Guoliang~Li,~\IEEEmembership{Fellow,~IEEE},~Nan~Tang, and Yuyu~Luo* \\
\textit{\small \sf \textcolor{blue}{{Text-to-SQL Handbook}:  \url{https://github.com/HKUSTDial/NL2SQL_Handbook}}}
\thanks{Xinyu~Liu, Shuyu~Shen, Boyan~Li, Peixian~Ma, Runzhi~Jiang, Nan~Tang and~Yuyu~Luo are with The Hong Kong University of Science and Technology (Guangzhou), China. E-mail: \{xliu371, sshen190, bli303, rjiang073, pma929\}@connect.hkust-gz.edu.cn, \{yuyuluo, nantang\}@hkust-gz.edu.cn.}
\thanks{Yuxin~Zhang and Ju~Fan are with Renmin University of China, Beijing, China. E-mail: \{zhangyuxin159, fanj\}@ruc.edu.cn. Guoliang~Li is with Tsinghua University, Beijing, China. E-mail: liguoliang@tsinghua.edu.cn.}
\thanks{$^*$Corresponding Author: Yuyu~Luo (yuyuluo@hkust-gz.edu.cn).}

}

\markboth{Journal of \LaTeX\ Class Files,~Vol.~18, No.~9, September~2020}%
{How to Use the IEEEtran \LaTeX \ Templates}

\newcommand{\insertfig}{
	\begin{center}
		\setcounter{figure}{0}
		\captionsetup{type=figure}
		\includegraphics[width=\textwidth]{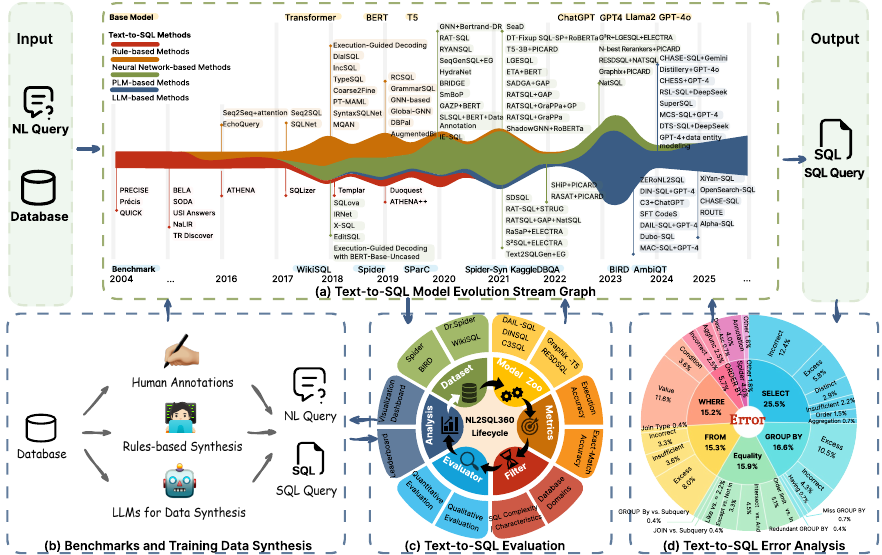}
		\captionof{figure}{\addd{An Overview of the Survey: The Lifecycle of the Text-to-SQL Task.}}
		\vspace{-3em}
		\label{fig:overview}
	\end{center}
}

\maketitle

\begin{abstract}
Translating users' natural language queries (NL) into SQL queries (\ie Text-to-SQL, \aka NL2SQL) can significantly reduce barriers to accessing relational databases and support various commercial applications. 
The performance of Text-to-SQL has been greatly enhanced with the emergence of Large Language Models (LLMs). In this survey, we provide a comprehensive review of Text-to-SQL techniques powered by LLMs, covering its entire lifecycle from the following four aspects:
(1) \textbf{\textit{Model:}} Text-to-SQL translation techniques that tackle not only NL ambiguity and under-specification, but also properly map NL with database schema and instances;
(2) \textbf{\textit{Data:}} From the collection of training data, data synthesis due to training data scarcity, to Text-to-SQL benchmarks;
(3) \textbf{\textit{Evaluation:}} Evaluating Text-to-SQL methods from multiple angles using different metrics and granularities; and
(4) \textbf{\textit{Error Analysis:}} analyzing Text-to-SQL errors to find the root cause and guiding Text-to-SQL models to evolve.
Moreover, we offer a rule of thumb for developing Text-to-SQL solutions.
Finally, we discuss the research challenges and open problems of Text-to-SQL in the LLMs era. 
\end{abstract}



\section{Introduction}
\label{sec:intro}

\IEEEPARstart{N}{atural} Language to SQL (\ie \nlsql), which converts a natural language query (\nlq) into an \sql query, is a key technique toward lowering the barrier to accessing relational databases
~\cite{gu2023few, DBLP:conf/acl/ChenCWMPSS023, DBLP:conf/kdd/WangQHLYWLSHSL22,DBLP:conf/kdd/LiuH0W22,pourreza2024din,gao2023text,li2023resdsql}.
This technique supports various applications such as business intelligence and natural language interfaces for databases, making it a key step toward democratizing data science
~\cite{DBLP:conf/cidr/0001YF0LH24,  DBLP:journals/corr/abs-2406-07815, DBLP:journals/corr/abs-2406-11033, DBLP:journals/tvcg/ShenSLYHZTW23, DBLP:journals/tkde/LuoQCTLL22, DBLP:journals/tvcg/LuoTLTCQ22, DBLP:conf/sigmod/TangLOLC22, DBLP:conf/sigmod/Luo00CLQ21, DBLP:journals/vldb/QinLTL20, DBLP:conf/icde/LuoQ0018, DBLP:conf/sigmod/LuoQ00W18}.
Recent advancements in language models have significantly extended the frontiers of research and application in \nlsql. Concurrently, the trend among database vendors to offer \nlsql solutions has evolved from a mere notion to a necessary strategy~\cite{DBLP:journals/dase/ZhouSL24, DBLP:journals/sigmod/AmerYahiaBCLSXY23}.
Therefore, we need to understand the fundamentals, techniques, and challenges regarding \nlsql. 

In this survey, we systematically review recent \nlsql techniques through a new framework, as shown in Figure~\ref{fig:overview}.

\bi
    \item {\bf Text-to-SQL with Language Models.}
        We will first review existing \nlsql solutions from the perspective of language models, categorizing them into four major categories ({see Figure~\ref{fig:overview}(a)}). 
        We will then focus on the recent advances in  Pre-trained Language Models (PLMs) and  Large Language Models (LLMs) for \nlsql.
    \item {\bf Benchmarks and Training Data Synthesis}.
        Undoubtedly, the performance of PLM- and LLM-based \nlsql models is highly dependent on the amount and quality of the training data. Therefore, we will first summarize the characteristics of existing benchmarks and analyze their statistical information (\eg database complexity) in detail. 
        We will then discuss methods for collecting and synthesizing high-quality training data, emphasizing this as a research opportunity (see Figure~\ref{fig:overview}(b)). 
    \item {\bf Evaluation.}
        Comprehensively evaluating \nlsql models is crucial for optimizing and selecting models for different usage scenarios. We will discuss the multi-angle evaluation and scenario-based evaluation for the \nlsql task ({see Figure~\ref{fig:overview}(c)}). For example, we can assess the \nlsql model in specific contexts by filtering benchmarks based on SQL characteristics, \nlq variants, database domains, and so on.
    \item {\bf Text-to-SQL Error Analysis.}
    Error analysis is essential in \nlsql research for identifying limitations and improving the model robustness. We review existing error taxonomies, analyze their limitations, and propose principles for designing comprehensive taxonomies for \nlsql output errors. Using these principles, we create a two-level error taxonomy and utilize it to summarize and analyze \nlsql output errors
     (see Figure~\ref{fig:overview}(d)).
    
\ei

In addition to the above, we will provide practical guidance for developing \nlsql solutions, including a roadmap for optimizing LLMs for \nlsql tasks and a decision flow for selecting \nlsql modules tailored to various \nlsql scenarios. Finally, we will discuss key open problems in the field, such as open \nlsql tasks, cost-effective \nlsql with LLMs, and trustworthy \nlsql solutions.

\stitle{Differences from Existing Surveys.}
Our survey distinguishes itself from existing \nlsql surveys~\cite{zhang2024natural,katsogiannis2023survey,deng-etal-2022-recent,kim2020natural, shi2024surveyemployinglargelanguage, mohammadjafari2024naturallanguagesqlreview,zhu2024largelanguagemodelenhanced, hong2024next} and  tutorials~\cite{Hozcan2020state, DBLP:conf/sigmod/LiR17, DBLP:journals/pvldb/Katsogiannis-Meimarakis23} in five aspects. 
\bi
\item We systematically review \textit{{the entire lifecycle of Text-to-SQL problem}}, as shown in Figure~\ref{fig:overview}.
This lifecycle includes various \nlsql translation methodologies powered by language models (Figure~\ref{fig:overview}(a)), 
training data collection and synthesis methods (Figure~\ref{fig:overview}(b)),
multi-angle and scenarios-based evaluations (Figure~\ref{fig:overview}(c)), and \nlsql error analysis techniques (Figure~\ref{fig:overview}(d)). 
\item We provide a more detailed and comprehensive summary of the inherent challenges in \nlsql. Additionally, we analyze the technical challenges when developing a robust \nlsql solution for real-world scenarios, which are often overlooked in other surveys.
\item We particularly focus on recent advances in \textit{{LLM-based}} \nlsql methods,
 summarizing key modules and comparing different strategies within this scope. {We are the first survey to provide a modular summary of methods and provide detailed analyses for each key module (\eg database content retrieval).}
\item We highlight the importance of \textit{evaluating \nlsql methods in a multi-angle way}, analyze the \nlsql error patterns, and provide a two-level error taxonomy. 
\item We provide practitioners with a roadmap for optimizing LLMs to \nlsql and a decision flow for selecting the suitable \nlsql modules for various scenarios.
\ei

\stitle{Contributions.}
We make the following contributions.

\bi

\item {\em Text-to-SQL with Language Models.} 
We comprehensively review existing \nlsql techniques from a lifecycle perspective (Figure~\ref{fig:overview}). 
We introduce the \nlsql task definition, discuss challenges (Figure~\ref{fig:Task_challenge}), provide a taxonomy of  \nlsql solutions based on language models (Figure~\ref{fig:evoluation_process}), and summarize the key modules of language model-powered \nlsql solutions (Figure~\ref{fig:overview_Model} and Table~\ref{tab:Methods}).
Next, we elaborate on each module of language model-powered \nlsql methods, including the pre-processing strategies (Section~\ref{sec:preprocessing}), \nlsql translation methods (Section~\ref{sec: translation}), and post-processing techniques (Section~\ref{sec:postprocessing}).

\item {\em Benchmarks and Training Data Synthesis.} We summarize existing \nlsql benchmarks based on their characteristics (Figure~\ref{fig:dataset_timeline}). We analyze each benchmark in depth and discuss its pros and cons  
 (Table~\ref{tab:datasets}). (Section~\ref{sec:benchmark})

\item {\em Evaluation and Errors Analysis.} We highlight the importance of evaluation in developing practical \nlsql solutions. We review widely used evaluation metrics and toolkits for assessing \nlsql solutions. 
We provide a taxonomy to summarize typical errors produced by \nlsql methods. (Section~\ref{sec:evaluation})

\item {\em Practical Guidance for Developing Text-to-SQL Solutions.} 
We provide a roadmap for optimizing existing LLMs to \nlsql tasks (Figure~\ref{fig:NL2SQL_Guidance}(a)). In addition, we design a decision flow to guide the selection of appropriate modules for different scenarios (Figure~\ref{fig:NL2SQL_Guidance}(b)).

\item \addd{{\em Open Problems in Text-to-SQL.} We analyze the limitations of LLM-based methods and discuss new research opportunities, including the open-world \nlsql problem and cost-effective solutions (Section~\ref{sec:openproblem}).}

\item {\em Text-to-SQL Handbook.} 
We maintain an online handbook (\url{https://github.com/HKUSTDial/NL2SQL_Handbook})  to help readers stay current with \nlsql advancements.
\ei

\section{Text-to-SQL Problem and Background}
\label{sec: taskoverview}

In this section, we first formalize the definition of the \nlsql task (Section~\ref{sub:nlsql_def}).
We then introduce the workflow of how humans perform the \nlsql task (Section~\ref{sub:workflow}) and discuss the key challenges (Section~\ref{sub:task_challenge}). 
Finally, we describe the evolution of \nlsql solutions based on the development of language models (Section~\ref{sub:history}).

\subsection{Problem Formulation}
\label{sub:nlsql_def}

\begin{definition} [\textbf{Natural Language to SQL (Text-to-SQL)}]
Natural Language to SQL (\nlsql), also known as NL2SQL, is the task of converting natural language queries (\nlq) into corresponding SQL queries (\sql) that can be executed on a relational database (\db). 
Specifically, given an \nlq and a \db, the goal of \nlsql is to generate an \sql that accurately reflects the user's intent and returns the appropriate results when executed on the database. 
\end{definition}

\etitle{Discussion.}
In some cases, the corresponding \sql query to an \nlq may be multiple due to the ambiguity or underspecification of the \nlq, or ambiguity in the database schema. In addition, even when the \nlq, database schema and database content are clear and specific, there may still be multiple equivalent \sql queries that can satisfy the given \nlq question.

\begin{figure}[t]
	\centering
	\includegraphics[width=\linewidth]{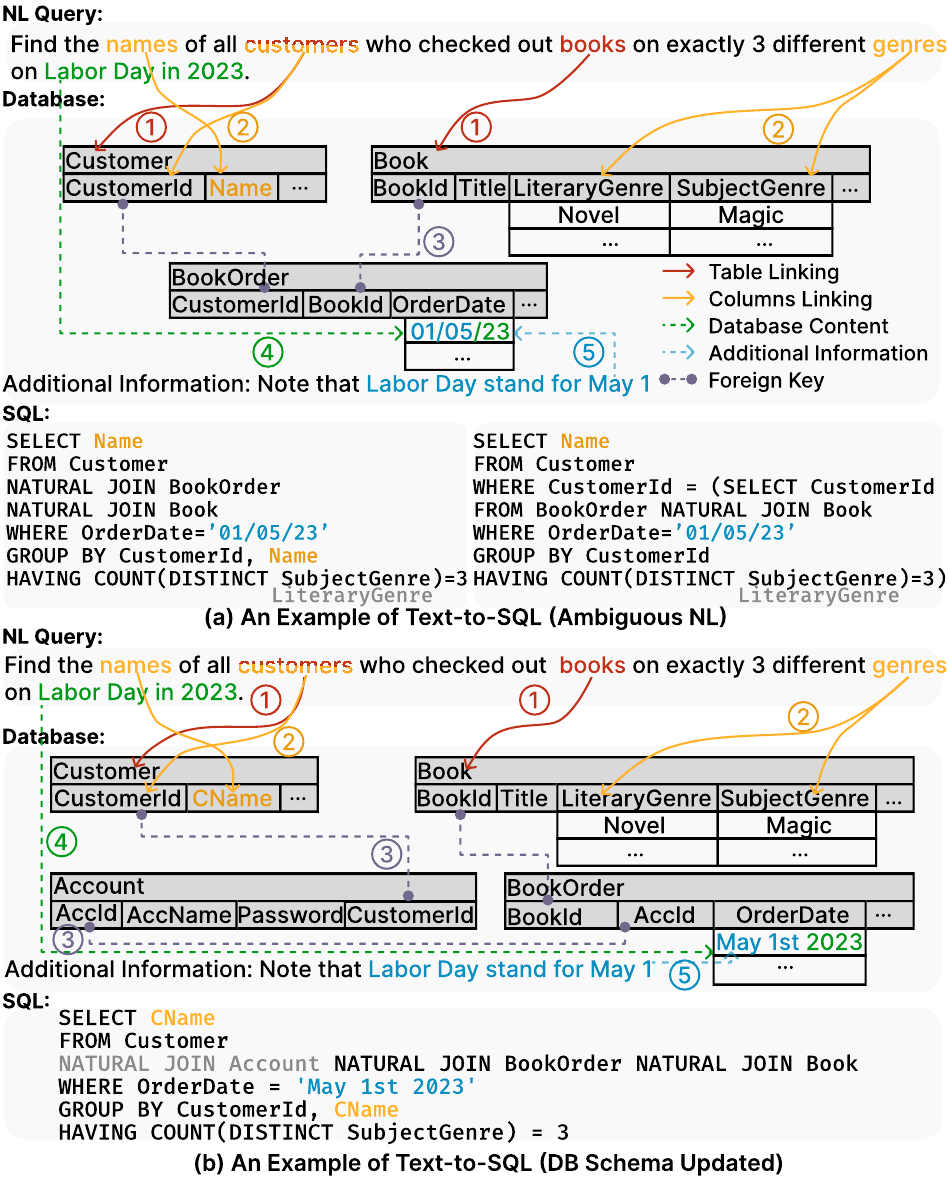}
	\caption{Examples of the \nlsql Task and Its Challenges. }
	\label{fig:Task_challenge}
\end{figure}

\subsection{Text-to-SQL Human Workflow}
\label{sub:workflow}

When professional users (\eg DBAs) perform the \nlsql task, they first interpret the \nlq question, examine the database schema and contents, and then construct the corresponding \sql based on their SQL expertise.
Below, we outline this process in detail, as illustrated in Figure~\ref{fig:Task_challenge}(a).

\etitle{Step-1: Understanding Natural Language Query:} 
Given the \nlq query ``{\em Find the names of all customers who checked out books on exactly 3 different genres on Labor Day in 2023}'', the DBA's first task is to grasp the user's intent and identify key components. Key elements include:
1) \textit{Entities or Attributes}: ``names'',  ``customers'', ``books'',  and ``genres''; 
2) \textit{Temporal Context}: ``Labor Day in 2023''; and 
3) \textit{Specific Conditions}: ``exactly 3 different genres''.
Then, the DBA may further understand the overall purpose of the \nlq query. In this case, the DBA should retrieve a list of customer names based on specific borrowing behavior on a particular date.

\etitle{Step-2: Finding Relevant Tables, Columns, and Cell Values:} 
Next, the DBA examines the database schema and contents to identify the relevant tables, columns, and cell values for constructing the \sql.
For example, the DBA may determine that the ``Customer'' and ``Book'' tables are relevant based on their understanding of the \nlq (see Figure~\ref{fig:Task_challenge}(a)-\ding{172}). The DBA then decides which columns should be mentioned. For example, the keyword ``genres'' can refer to either ``LiteraryGenre'' or ``SubjectGenre'' (see Figure~\ref{fig:Task_challenge}(a)-\ding{173}).
Furthermore, the DBA should interpret ``Labor Day in 2023'' based on the context. In the US, ``Labor Day in 2023'' refers to ``September 4th, 2023'', while in China, it refers to ``May 1st, 2023''. This judgment relies on domain knowledge or available additional information (see Figure~\ref{fig:Task_challenge}(a)-\ding{176}).

Note that Step-2 aligns with the concepts of {\em schema linking}, {\em database content retrieval}, and {\em additional information acquisition} in recent \nlsql solutions powered by language models (please refer to Figure~\ref{fig:overview_Model} for more details).

\etitle{Step-3:  Writing SQL based on NL and DB Understanding:} Finally, the DBA writes the corresponding \sql based on the insights gained in Steps-1 and -2. This process, known as ``\nlsql Translation'', relies heavily on the DBA's SQL expertise. However, this process can be very challenging due to the ambiguity of the \nlq or the complexity of the database. For example, as shown in Figure~\ref{fig:Task_challenge}(a), despite understanding the need to link the \textit{Customer} and \textit{Book} tables, one must be familiar with the usage and norms of employing either a natural join or a subquery. In addition, there may be multiple possible \sql queries because ``genres'' can refer to either ``LiteraryGenre'' or ``SubjectGenre''.

\etitle{Takeways.} 
From the above steps, we intuitively identify three \textit{inherent} challenges in the \nlsql task: the uncertainty of the natural language, the complexity of the database, and the translation from the ``free-form'' natural language queries to the ``constrained and formal'' \sql queries.

\subsection{Text-to-SQL Task Challenges}
\label{sub:task_challenge}

In this section, we will first discuss the fundamental challenges of the \nlsql task. We will then analyze the \textit{technical} challenges, \ie the challenges we face when developing a strong \nlsql solution in real-world scenarios.

\stitle{C1: Uncertain Natural Language Query.}
Natural language queries often contain uncertainties due to ambiguity and underspecification~\cite{DBLP:conf/sigmod/Katsogiannis-Meimarakis21}.
In \nlsql tasks, the challenges related to \nlq can be summarized as follows:

\bi
    \item \textit{\underline{Lexical Ambiguity}}: This occurs when a single word has multiple meanings. For example, the word ``bat'' can refer to an {\em animal}, or a {\em baseball bat}, or the action of {\em swinging}.

    \item 
    \textit{\underline{Syntactic Ambiguity}}: This occurs when a sentence can be parsed in multiple ways. For example, in the sentence ``Mary saw the man with the telescope'', the phrase ``with the telescope'' can mean either that Mary used a telescope to see the man or that the man had a telescope.

    \item \textit{\underline{Under-specification}}: This occurs when linguistic expressions lack sufficient detail to convey specific intentions or meanings clearly. For example, ``Labor Day in 2023'' refers to September 4th in the US but May 1st in China.
\ei

\begin{figure*}[!t]  
	\centering  
	\includegraphics[width=\textwidth]{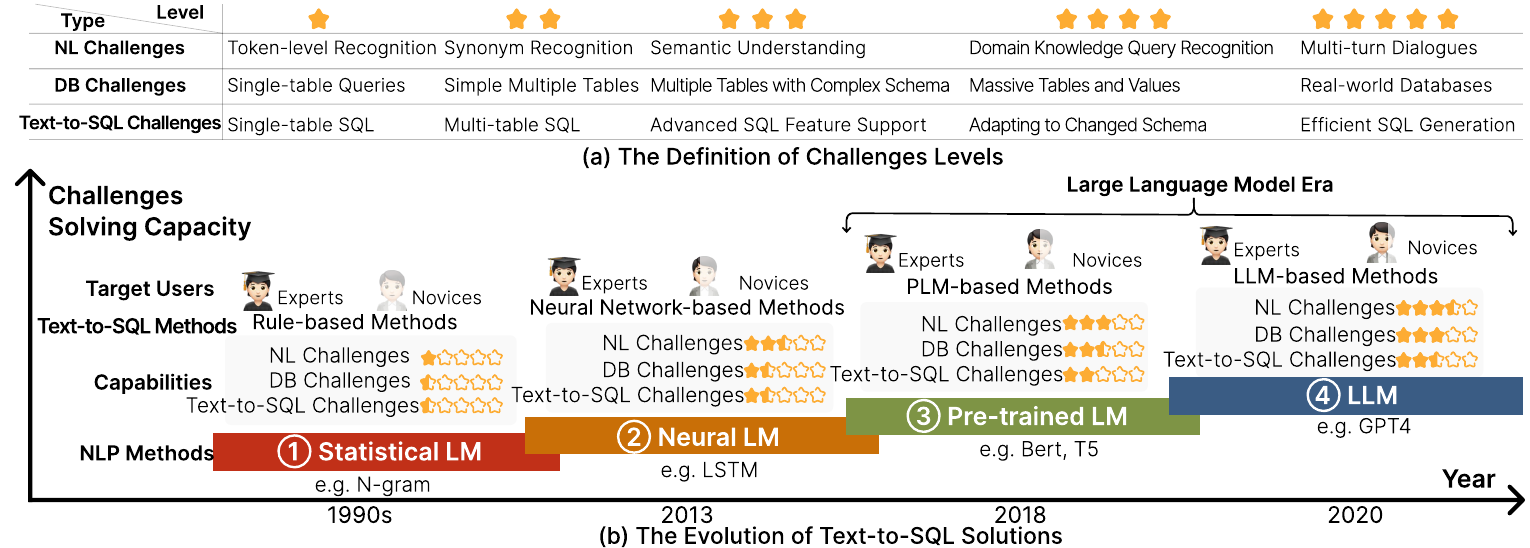}  
	\caption{The Evolution of Text-to-SQL Solutions from the Perspective of Language Models.}  
	\label{fig:evoluation_process}  
\end{figure*}

\stitle{C2: Complex Database and Dirty Content.}
The \nlsql task requires a deep understanding of the database schema, including table names, columns, relationships, and data attributes. The complexity of modern schemas and large data volumes make this task especially challenging.

\bi
\item \textit{\underline{Complex Relationships Among Tables}}: 
Databases often contain hundreds of tables with complex interrelationships. \nlsql solutions must accurately comprehend and leverage these relationships when generating \sql.
 
\item \textit{\underline{Ambiguity in Attributes and Values}}:
Ambiguous values and attributes  in a database can complicate \nlsql systems' ability to identify the correct context.

\item \textit{\underline{Domain-Specific Schema Designs}}: 
Different domains often have unique database designs and schema patterns. 
The variations in schema design across domains make it difficult to develop a {\em one-size-fits-all} \nlsql model.

\item \textit{\underline{Large and Dirty Database Values}}: Efficiently handling vast data volumes in large databases is critical, as processing all data as input is impractical. Additionally, dirty data, such as missing values, duplicates, or inconsistencies, can lead to erroneous query results (e.g., affecting {\tt WHERE} clauses) if not properly managed.
\ei

\stitle{C3: Text-to-SQL Translation.}
The \nlsql task differs from the compilation of a high-level programming language to a low-level machine language, as it usually has a \textit{one-to-many} mapping between the input \nlq, \db and output \sql. Specifically, the \nlsql task faces several unique challenges:

\begin{itemize}
    \item \textit{\underline{Free-form \nlq \textit{vs.} Constrained and Formal \sql}}: 
    Natural language is flexible, while \sql queries must adhere to strict syntax. Translating \nlq into \sql requires precision to ensure the generated queries are executable.

    \item \textit{\underline{Multiple Possible \sql Queries}}: 
    A single \nlq query can correspond to multiple \sql queries that fulfill the query intent, leading to ambiguity in determining appropriate \sql translation (see the example in Figure~\ref{fig:Task_challenge}(a)).

    \item \textit{\underline{Database Schema Dependency}}: 
    The \nlsql translation is highly dependent on the underlying database schema. As shown in Figure~\ref{fig:Task_challenge} (a) and (b), the same \nlq may produce different \sql queries based on schema variations. This requires \nlsql models to bridge gaps between training data and real-world schema differences.
    
\end{itemize}

Beyond the intrinsic challenges, developers must also overcome several technical obstacles to build reliable and efficient \nlsql systems, as discussed below.

\stitle{C4: Technical Challenges in Developing Text-to-SQL Solutions.}  
Developing robust \nlsql solutions requires addressing several key technical challenges, including:

\bi

\item \textit{\underline{Cost-effective Solution}}:
Deploying \nlsql models, particularly those using large language models, demands significant resources, such as hardware and/or API costs. Achieving an optimal balance between model performance and cost efficiency remains a crucial challenge.

\item \textit{\underline{Model Efficiency}}:
A trade-off often exists between model size and performance, with larger models generally yielding better results. Optimizing efficiency without compromising accuracy is essential, especially in interactive querying scenarios requiring low latency.

\item \textit{\underline{SQL Efficiency}}: 
The \sql generated by \nlsql models must be both correct and optimized for performance. This includes optimizing join operations, index usage, and query structures. Efficient queries reduce database load, improving system responsiveness and throughput.

\item \textit{\underline{Insufficient and Noisy Training Data}}: 
High-quality \nlsql training data is challenging to obtain. Public datasets are often limited and may include noisy annotations, affecting model performance~\cite{liu2025nl2sql, yang2025automated}. Annotation requires database expertise, increasing costs, and the complexity of \nlsql tasks often leads to errors.

\item \textit{\underline{Trustworthiness and Reliability}}:
\nlsql models must be trustworthy and reliable, consistently producing accurate results across diverse datasets and scenarios. Trustworthiness requires transparency, allowing users to understand and verify the generated \sql. 

\ei

\subsection{Challenges Solving with Large Language Models}
\label{sub:history}

\etitle{Difficulty Levels.} We categorize the difficulty of \nlsql into five levels, each addressing specific hurdles, as shown in Figure~\ref{fig:evoluation_process}(a). The first three levels cover challenges that have been or are currently being addressed, highlighting the gradual progress in \nlsql capabilities. The fourth level includes challenges that are the focus of current LLM-based solutions, while the fifth level represents future challenges, showing our vision for \nlsql advancements over the next five years. 

\etitle{The Evolution of \nlsql Solutions.}
The development of \nlsql solutions, illustrated in Figure~\ref{fig:evoluation_process}(b), progresses through four distinct stages:
the rule-based stage, the neural network-based stage, the PLM-based stage, and the LLM-based stage. 
At each stage, we analyze shifts in target users, \ie from experts to broader user groups, and the extent to which various \nlsql challenges are addressed. 

\begin{figure}[t]
	\centering        \includegraphics[width=.9\linewidth]{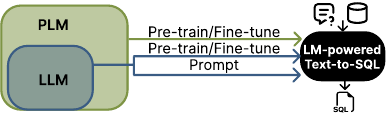}
	\caption{The Categorization of PLM and LLM in Text-to-SQL.}
	\label{fig: PLM_LLM}
\end{figure}

\subsubsection{Rule-based Stage}
In the early stages, statistical language models (\eg semantic parsers) were used to interpret \nlq queries and convert them into \sql queries using predefined rules~\cite{DBLP:journals/corr/abs-2204-00498, 10.1145/2588555.2594519, DBLP:conf/sigmod/Katsogiannis-Meimarakis21, yu2021grappa}. 
However, rule-based 
\nlsql methods face challenges in adaptability, scalability, and generalization. At this stage, natural language understanding was limited to the token level, with research primarily focused on single-table \sql queries (see Figure~\ref{fig:evoluation_process}(b)-\ding{172}).

\subsubsection{Neural Network-based Stage}
To alleviate the limitations of rule-based methods, researchers explored neural networks for the \nlsql task. This led to the development of models based on sequence-to-sequence architectures and graph neural networks~\cite{xiao2016sequence, lin2019grammar, bogin2019representing}, which enhanced the handling of synonyms and intent understanding. Thus, research advanced from single-table scenarios to more complex multi-table scenarios (see Figure~\ref{fig:evoluation_process}(b)-\ding{173}).
However, the generalization ability of these methods is still limited by model size and the availability of sufficient training data.

\subsubsection{PLM-based Stage}
The introduction of PLMs like BERT~\cite{devlin-etal-2019-bert} and T5~\cite{raffel2020exploring} in 2018 led to significant advancements in \nlsql methods based on PLMs~\cite{li2023graphix, li2023resdsql, gu2023interleaving}, achieving competitive performance on various benchmarks (see Figure~\ref{fig:evoluation_process}(b)-\ding{174}).
At this stage, PLM-based \nlsql models trained on large corpora have greatly enhanced natural language understanding, resolving approximately 80\% of cases in the Spider dataset~\cite{dataset-spider}. However, accuracy drops to about 50\% on the extra hard cases of Spider~\cite{nlsql360}.
In addition, these models still face challenges in handling complex schemas.

\etitle{Remark: PLMs vs. LLMs} {\em
Figure~\ref{fig: PLM_LLM} shows the key differences between LLMs and PLMs. 
LLMs are a subset of PLMs, distinguished by their advanced language understanding and emergent capabilities~\cite{zhao2023survey, minaee2024largelanguagemodelssurvey}. The emergent abilities allow LLMs to perform \nlsql tasks directly using prompts. In contrast, PLMs generally require additional pre-training or fine-tuning for acceptable \nlsql performance.
}

\subsubsection{LLM-based Stage}
\label{subsub:LLM-based Stage}
LLMs demonstrate unique emergent capabilities that surpass traditional PLMs in NLP tasks, marking a new paradigm for \nlsql solutions. These LLM-based \nlsql methods have become the most representative solutions in the current \nlsql landscape~\cite{pourreza2024din, li2024codes, gao2023text, dataset-bull}. Current research focuses on optimizing prompt design~\cite{gao2023text} and fine-tuning LLMs~\cite{li2024codes}.
For example, DAIL-SQL~\cite{gao2023text} utilizes the GPT-4 with effective prompt engineering techniques, achieving strong results on the Spider dataset~\cite{dataset-spider}. 
Meanwhile, CodeS~\cite{li2024codes} builds an LLM specifically for \nlsql tasks by pretraining StarCoder~\cite{li2023starcoder} on a large \nlsql-related corpus, showing solid performance on benchmarks like BIRD~\cite{dataset-bird}.
At this stage, LLMs' emergent capabilities have significantly improved natural language understanding, shifting the task's focus toward database-specific challenges. New benchmarks like BIRD~\cite{dataset-bird} and BULL~\cite{dataset-bull} emphasize handling massive tables and domain-specific solutions  (see Figure~\ref{fig:evoluation_process}(b)-\ding{175}).

\stitle{Text-to-SQL Solutions in LLMs Era.}  
Broadly speaking, there are two major approaches to leverage the capabilities of LLMs for \nlsql: 1) in-context learning, and 2) pre-train/fine-tune LLMs specialized for \nlsql.

\etitle{In-Context Learning for Text-to-SQL.}
For in-context learning methods, the goal is to optimize the prompt function $P$ to guide the LLMs, which can be formulated as follows:
\begin{equation*}
	\mathcal{F}_{\text{LLM}}(P \mid \text\nlq, \text\db, \text\knowledge )\rightarrow \text\sql,
\end{equation*}

where \knowledge denotes additional information or domain-specific knowledge related to \nlq or \db. $P$ is a \textit{prompt function} that transforms the input (\nlq, \db, \knowledge) into a suitable \textit{textual prompt} for the LLMs. A well-designed $P$ can effectively guide the LLMs to perform the \nlsql task more accurately.

Employing in-context learning strategies for \nlsql treats LLMs as \textit{off-the-shelf} tools, without modifying their parameters. However, if users have sufficient training data or hardware resources, calibrating the LLMs’ parameters can enhance performance and accuracy, allowing the model to be more closely tailored to the specific \nlsql task.

\etitle{Pre-train and Fine-tune LLMs for Text-to-SQL.} Fully optimizing the parameters of LLMs for \nlsql involves two key stages, pre-train and fine-tune, formulated as follows:
\begin{equation*}
	\text{LLM}^* = \mathcal{F}_{\text{fine-tune}} \left( \mathcal{F}_{\text{pre-train}}(\text{LLM}, \mathcal{D}_{p}), \mathcal{D}_{f} \right)
\end{equation*}

During pre-train, the LLM is trained on a large-scale and diverse dataset $\mathcal{D}_{p}$ that includes a broad range of linguistic patterns and domain-general knowledge, enabling the model to develop robust understanding capabilities.

In the subsequent fine-tune stage, the pre-trained model is further adjusted on a more specialized dataset $\mathcal{D}_{f}$, which is closely aligned with the \nlsql task. 
This targeted training refines the model's capabilities, enabling it to more effectively interpret and generate \sql based on \nlq queries.

\section{Language Model-powered \nlsql Overview}
\label{sec:lm4nlsql}

\begin{figure}[t!]
	\centering
	\includegraphics[width=\linewidth]{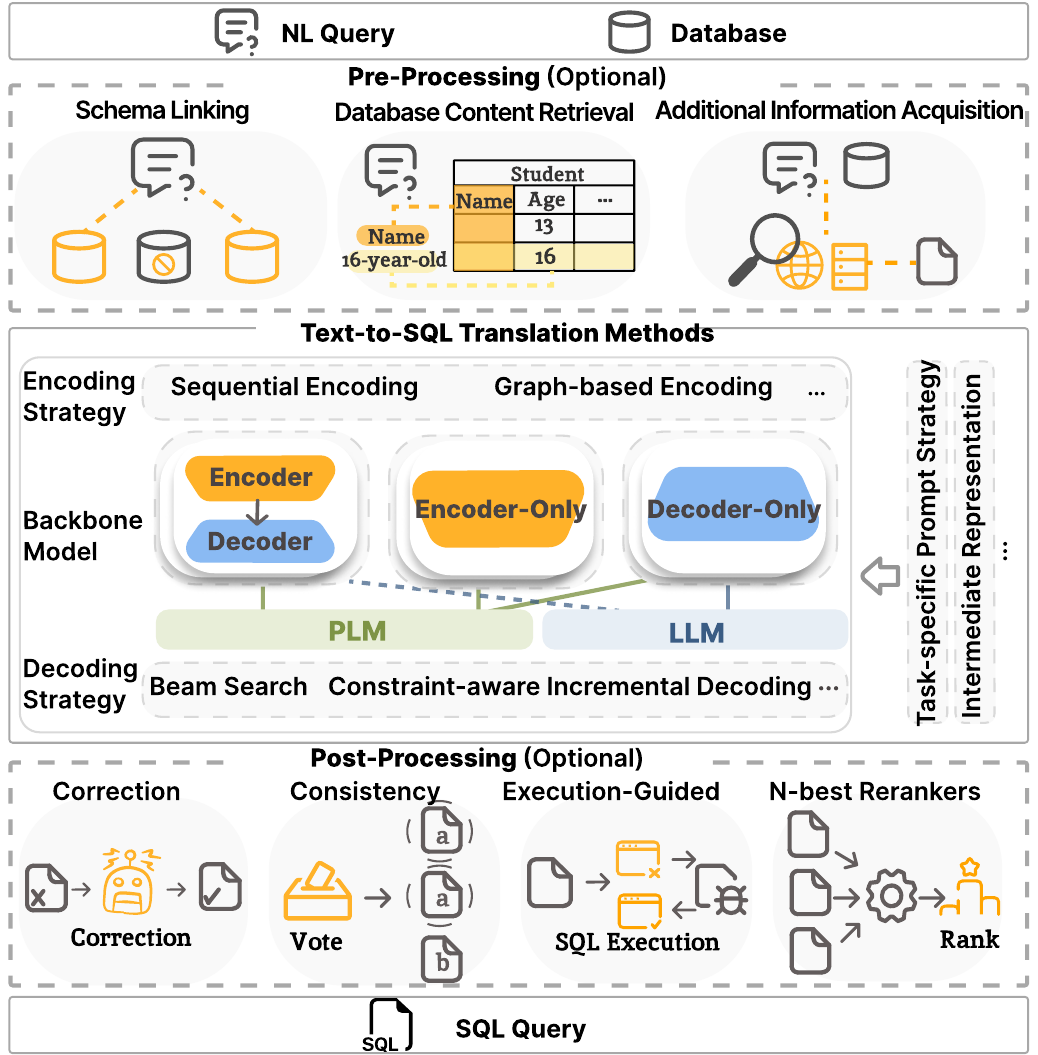}
	\caption{An Overview of Text-to-SQL {Modules} in LLM Era.}
	\label{fig:overview_Model}
\end{figure}

While \nlsql was initially designed as an end-to-end task, recent advances, particularly in the LLM era, have shifted towards a modular design. As shown in Figure~\ref{fig:overview_Model}, modern PLM- and LLM-based solutions typically decompose the task into three stages: Pre-Processing, Translation, and Post-Processing. Each stage contains specialized modules such as schema linking, intermediate representation, and execution-guided correction.
This design reflects the increasing complexity of \nlsql and aligns with the rising trend of multi-agent or multi-module collaboration. Table~\ref{tab:Methods} further compares the key design choices across recent solutions.

\stitle{Pre-Processing  Methods.}
Pre-processing enhances inputs and plays a significant role in improving Text-to-SQL parsing~\cite{lei2020re}. 

\bi
    \item Schema Linking: This module identifies the most relevant tables and columns for \nlsql (Section~\ref{subsec:schema-linking}).

    \item Database Content Retrieval: This key module accesses the appropriate database contents or cell values needed for formulating \sql (Section~\ref{subsec: DB Content}).

    \item Additional Information Acquisition: This key module enriches the contextual backdrop by integrating domain-specific knowledge  (Section~\ref{subsec:Additional Information}). 

\ei

\stitle{Translation Methods.} 
This is the core of \nlsql solution, responsible for converting input \nlq queries into \sql.  

\bi
    \item Encoding Strategy: This crucial module transforms the input \nlq and database schema into an internal representation, capturing both the semantic and structural information of the input data 
    (Section~\ref{subsec:encoding}).

    \item Decoding Strategy: This key module transforms the internal representation into \sql queries
    (Section~\ref{subsec:decoding}).

    \item Task-specific Prompt Strategy: This module provides tailored guidance for the \nlsql model, optimizing the \nlsql translation workflow
    (Section~\ref{subsec:Prompt_Strategy}).

    \item Intermediate Representation: This module serves as a bridge between \nlq and \sql translation, providing a structured approach to abstract, align, and optimize \nlq understanding, simplify complex reasoning, and guide the generation of accurate \sql queries
    (Section~\ref{subsec:IR}).
\ei

\stitle{Post-Processing Methods.}
Post-processing is a crucial step to refine the generated \sql queries for better accuracy. 
\bi
    \item SQL Correction Strategy: This aims to identify and correct syntax errors in generated \sql  (Section~\ref{subsec:sql_correction}).
     
    \item Output Consistency: This module ensures the uniformity of \sql by sampling multiple reasoning results and selecting the most consistent result (Section~\ref{subsec:output_consistency}).
    
    \item Execution-Guided Strategy: It uses the execution results of \sql to guide subsequent refinements (Section~\ref{subsec:execution_guided}).
    
    \item N-best Rankers Strategy: It aims to rerank the top-$k$ results generated by the \nlsql model to enhance query accuracy (Section~\ref{subsec:nbest_ranker}).
\ei

\etitle{\addd{Multi-agent Collaboration for Text-to-SQL in LLM Era.}} \addd{Building upon modular design principles, recent research has further introduced multi-agent architectures to tackle the Text-to-SQL task. In contrast to traditional monolithic systems, multi-agent frameworks assign specialized responsibilities to distinct agents, each dedicated to handling a specific subtask. This design facilitates enhanced division of labor and more effective coordination among components. For example, MAC-SQL~\cite{wang2023mac} adopts a three-agent architecture, with separate agents for schema linking, query decomposition and generation, and execution-guided refinement. Similarly, CHASE-SQL~\cite{pourreza2024chase} employs a divide-and-conquer approach, selecting relevant database content during preprocessing, generating SQL queries through multiple chain-of-thought pathways, and iteratively refining outputs through self-correction and ranking.
Pushing the boundary further, Alpha-SQL~\cite{li2025alpha} proposes a planning-centric autonomous agent framework that leverages LLMs in combination with Monte Carlo Tree Search (MCTS). This agent dynamically selects and activates the appropriate modules, such as schema linking and SQL generation, based on contextual reasoning and execution-based feedback. Alpha-SQL's strategy-driven exploration and adaptive control offer robust generalization, avoiding the rigidity of pipeline-based approaches.
}

\begin{sidewaystable*}
\begin{center}
\caption{Comparisons of Existing Text-to-SQL Solutions.}
\label{tab:Methods}
\renewcommand{\arraystretch}{1.5} 
\resizebox{\textwidth}{!}{
\begin{tabular}{|c|c|c|ccc|ccccc|cccc|}
\hline
\multirow{2}{*}{\textbf{Methods}} &
  \multirow{2}{*}{\textbf{Years}} &
  \multirow{2}{*}{\textbf{Finetuning}} &
  \multicolumn{3}{c|}{\textbf{Pre-Processing}} &
  \multicolumn{5}{c|}{\textbf{Text-to-SQL Translation Methods}} &
  \multicolumn{4}{c|}{\textbf{Post-Processing}} \\ \cline{4-15}
 &
   &
   &
  \multicolumn{1}{c|}{\textbf{\begin{tabular}[c]{@{}c@{}}Schema\\ Linking\end{tabular}}} &
  \multicolumn{1}{c|}{\textbf{\begin{tabular}[c]{@{}c@{}}DB Content\\ Retrieval\end{tabular}}} &
  \textbf{\begin{tabular}[c]{@{}c@{}}Additional Information\\ Acquisition\end{tabular}} &
  \multicolumn{1}{c|}{\textbf{\begin{tabular}[c]{@{}c@{}}Backbone\\ Model\end{tabular}}} &
  \multicolumn{1}{c|}{\textbf{\begin{tabular}[c]{@{}c@{}}Encoding \\ Strategy\end{tabular}}} &
  \multicolumn{1}{c|}{\textbf{\begin{tabular}[c]{@{}c@{}}Intermediate\\ Representation\end{tabular}}} &
  \multicolumn{1}{c|}{\textbf{\begin{tabular}[c]{@{}c@{}}Task-specific\\ Prompt Strategy\end{tabular}}} &
  \textbf{\begin{tabular}[c]{@{}c@{}}Decoding\\ Strategy\end{tabular}} &
  \multicolumn{1}{c|}{\textbf{Correction}} &
  \multicolumn{1}{c|}{\textbf{Consistency}} &
  \multicolumn{1}{c|}{\textbf{\begin{tabular}[c]{@{}c@{}}Execution\\ -Guided\end{tabular}}} &
  \textbf{\begin{tabular}[c]{@{}c@{}}N-best\\ Rerankers\end{tabular}} \\ \hline
OpenSearch-SQL~\cite{xie2025opensearch} &
     2025 &
     - &
     \multicolumn{1}{c|}{\greencheck} &
     \multicolumn{1}{c|}{\greencheck} &
     \greencheck &
     \multicolumn{1}{c|}{Decoder-Only} &
     \multicolumn{1}{c|}{Sequential Encoding} &
     \multicolumn{1}{c|}{Syntax Language} &
     \multicolumn{1}{c|}{COT} &
     Greedy Search &
     \multicolumn{1}{c|}{\greencheck} &
     \multicolumn{1}{c|}{\greencheck} &
     \multicolumn{1}{c|}{\greencheck} &
     - \\ \hline
CHASE-SQL~\cite{pourreza2024chase} &
     2025 &
     - &
     \multicolumn{1}{c|}{\greencheck} &
     \multicolumn{1}{c|}{\greencheck} &
     \greencheck &
     \multicolumn{1}{c|}{Decoder-Only} &
     \multicolumn{1}{c|}{Sequential Encoding} &
     \multicolumn{1}{c|}{-} &
     \multicolumn{1}{c|}{Multi-COT} &
     Greedy Search &
     \multicolumn{1}{c|}{\greencheck} &
     \multicolumn{1}{c|}{\greencheck} &
     \multicolumn{1}{c|}{\greencheck} &
     - \\ \hline
ROUTE~\cite{qin2024route} &
     2025 &
     \greencheck &
     \multicolumn{1}{c|}{\greencheck} &
     \multicolumn{1}{c|}{\greencheck} &
     \greencheck &
     \multicolumn{1}{c|}{Decoder-Only} &
     \multicolumn{1}{c|}{Sequential Encoding} &
     \multicolumn{1}{c|}{-} &
     \multicolumn{1}{c|}{-} &
     Greedy Search &
     \multicolumn{1}{c|}{\greencheck} &
     \multicolumn{1}{c|}{\greencheck} &
     \multicolumn{1}{c|}{\greencheck} &
     - \\ \hline
Alpha-SQL~\cite{li2025alpha} &
     2025 &
     - &
     \multicolumn{1}{c|}{\greencheck} &
     \multicolumn{1}{c|}{\greencheck} &
     \greencheck &
     \multicolumn{1}{c|}{Decoder-Only} &
     \multicolumn{1}{c|}{Sequential Encoding} &
     \multicolumn{1}{c|}{-} &
     \multicolumn{1}{c|}{COT} &
     Greedy Search &
     \multicolumn{1}{c|}{\greencheck} &
     \multicolumn{1}{c|}{\greencheck} &
     \multicolumn{1}{c|}{\greencheck} &
     - \\ \hline
CHESS~\cite{talaei2024chess} &
  2024 &
  - &
  \multicolumn{1}{c|}{\greencheck} &
  \multicolumn{1}{c|}{\greencheck} &
  \greencheck &
  \multicolumn{1}{c|}{Decoder-Only} &
  \multicolumn{1}{c|}{Sequential Encoding} &
  \multicolumn{1}{c|}{-} &
  \multicolumn{1}{c|}{COT} &
  Greedy Search &
  \multicolumn{1}{c|}
  {\greencheck} &
  \multicolumn{1}{c|}{\greencheck} &
  \multicolumn{1}{c|}{\greencheck} &
  - \\ \hline
CodeS~\cite{li2024codes} &
  2024 &
  - &
  \multicolumn{1}{c|}{\greencheck} &
  \multicolumn{1}{c|}{\greencheck} &
  - &
  \multicolumn{1}{c|}{Decoder-Only} &
  \multicolumn{1}{c|}{Sequential Encoding} &
  \multicolumn{1}{c|}{-} &
  \multicolumn{1}{c|}{-} &
  Greedy Search &
  \multicolumn{1}{c|}{-} &
  \multicolumn{1}{c|}{-} &
  \multicolumn{1}{c|}{\greencheck} &
  - \\ \hline
SFT CodeS~\cite{li2024codes} &
  2024 &
  \greencheck &
  \multicolumn{1}{c|}{\greencheck} &
  \multicolumn{1}{c|}{\greencheck} &
  \greencheck &
  \multicolumn{1}{c|}{Decoder-Only} &
  \multicolumn{1}{c|}{Sequential Encoding} &
  \multicolumn{1}{c|}{-} &
  \multicolumn{1}{c|}{-} &
  Greedy Search &
  \multicolumn{1}{c|}{-} &
  \multicolumn{1}{c|}{-} &
  \multicolumn{1}{c|}{\greencheck} &
  - \\ \hline
FinSQL~\cite{dataset-bull} &
     2024 &
     \multicolumn{1}{c|}{\greencheck} &
     \multicolumn{1}{c|}{\greencheck} &
     \multicolumn{1}{c|}{-} &
     \multicolumn{1}{c|}{\greencheck} &
     \multicolumn{1}{c|}{Decoder-Only} &
     \multicolumn{1}{c|}{Sequential Encoding} &
     \multicolumn{1}{c|}{-} &
     \multicolumn{1}{c|}{-} &
     Greedy Search &
     \multicolumn{1}{c|}{\greencheck} &
     \multicolumn{1}{c|}{\greencheck} &
     \multicolumn{1}{c|}{-} &
     \multicolumn{1}{c|}{-} \\ \hline
DTS-SQL~\cite{pourreza2024dts} &
  2024 &
  \greencheck &
  \multicolumn{1}{c|}{\greencheck} &
  \multicolumn{1}{c|}{-} &
  - &
  \multicolumn{1}{c|}{Decoder-Only} &
  \multicolumn{1}{c|}{Sequential Encoding} &
  \multicolumn{1}{c|}{-} &
  \multicolumn{1}{c|}{-} &
  Greedy Search &
  \multicolumn{1}{c|}{-} &
  \multicolumn{1}{c|}{-} &
  \multicolumn{1}{c|}{-} &
  - \\ \hline
TA-SQL~\cite{qu2024before} &
  2024 &
  - &
  \multicolumn{1}{c|}{\greencheck} &
  \multicolumn{1}{c|}{-} &
  - &
  \multicolumn{1}{c|}{Decoder-Only} &
  \multicolumn{1}{c|}{Sequential Encoding} &
  \multicolumn{1}{c|}{Sketch Structure} &
  \multicolumn{1}{c|}{COT} &
  Greedy Search &
  \multicolumn{1}{c|}{-} &
  \multicolumn{1}{c|}{-} &
  \multicolumn{1}{c|}{-} &
  - \\ \hline
SuperSQL~\cite{nlsql360} &
  2024 &
  - &
  \multicolumn{1}{c|}{\greencheck} &
  \multicolumn{1}{c|}{\greencheck} &
  \greencheck &
  \multicolumn{1}{c|}{Decoder-Only} &
  \multicolumn{1}{c|}{Sequential Encoding} &
  \multicolumn{1}{c|}{-} &
  \multicolumn{1}{c|}{-} &
  Greedy Search &
  \multicolumn{1}{c|}{-} &
  \multicolumn{1}{c|}{\greencheck} &
  \multicolumn{1}{c|}{-} &
  - \\ \hline
ZeroNL2SQL~\cite{gu2023interleaving} &
  2024 &
  \greencheck &
  \multicolumn{1}{c|}{-} &
  \multicolumn{1}{c|}{-} &
  - &
  \multicolumn{1}{c|}{Encoder-Decoder} &
  \multicolumn{1}{c|}{Sequential Encoding} &
  \multicolumn{1}{c|}{Sketch Structure} &
  \multicolumn{1}{c|}{Decomposition} &
  Beam Search &
  \multicolumn{1}{c|}{\greencheck} &
  \multicolumn{1}{c|}{-} &
  \multicolumn{1}{c|}{\greencheck} &
  - \\ \hline
PET-SQL~\cite{li2024pet} &
  2024 &
  \greencheck &
  \multicolumn{1}{c|}{\greencheck} &
  \multicolumn{1}{c|}{-} &
  \greencheck &
  \multicolumn{1}{c|}{Decoder-Only} &
  \multicolumn{1}{c|}{Sequential Encoding} &
  \multicolumn{1}{c|}{-} &
  \multicolumn{1}{c|}{-} &
  Greedy Search &
  \multicolumn{1}{c|}{-} &
  \multicolumn{1}{c|}{\greencheck} &
  \multicolumn{1}{c|}{-} &
  - \\ \hline
CoE-SQL~\cite{zhang2024coe} &
     2024 &
     \multicolumn{1}{c|}{-} &
     \multicolumn{1}{c|}{-} &
     \multicolumn{1}{c|}{-} &
     \multicolumn{1}{c|}{\greencheck} &
     \multicolumn{1}{c|}{Decoder-Only} &
     \multicolumn{1}{c|}{Sequential Encoding} &
     \multicolumn{1}{c|}{-} &
     \multicolumn{1}{c|}{CoT} &
     Greedy Search &
     \multicolumn{1}{c|}{\greencheck} &
     \multicolumn{1}{c|}{-} &
     \multicolumn{1}{c|}{-} &
     - \\ \hline
PURPLE~\cite{ren2024purple} &
  2024 &
  - &
  \multicolumn{1}{c|}{\greencheck} &
  \multicolumn{1}{c|}{-} &
  \multicolumn{1}{c|}{\greencheck} &
  \multicolumn{1}{c|}{Decoder-Only} &
  \multicolumn{1}{c|}{Sequential Encoding} &
  \multicolumn{1}{c|}{-} &
  \multicolumn{1}{c|}{-} &
  Greedy Search &
  \multicolumn{1}{c|}{\greencheck} &
  \multicolumn{1}{c|}{\greencheck} &
  \multicolumn{1}{c|}{\greencheck} &
  - \\ \hline
MetaSQL~\cite{fan2024metasql} &
  2024 &
  - &
  \multicolumn{1}{c|}{\greencheck} &
  \multicolumn{1}{c|}{-} &
  \multicolumn{1}{c|}{\greencheck} &
  \multicolumn{1}{c|}{Decoder-Only} &
  \multicolumn{1}{c|}{Sequential Encoding} &
  \multicolumn{1}{c|}{-} &
  \multicolumn{1}{c|}{Decomposition} &
  Greedy Search &
  \multicolumn{1}{c|}{-} &
  \multicolumn{1}{c|}{-} &
  \multicolumn{1}{c|}{-} &
  \multicolumn{1}{c|}{\greencheck} \\ \hline
DEA-SQL~\cite{xie2024decomposition} &
  2024 &
  - &
  \multicolumn{1}{c|}{\greencheck} &
  \multicolumn{1}{c|}{-} &
  \multicolumn{1}{c|}{\greencheck} &
  \multicolumn{1}{c|}{Decoder-Only} &
  \multicolumn{1}{c|}{Sequential Encoding} &
  \multicolumn{1}{c|}{-} &
  \multicolumn{1}{c|}{Decomposition} &
  Greedy Search &
  \multicolumn{1}{c|}{\greencheck} &
  \multicolumn{1}{c|}{-} &
  \multicolumn{1}{c|}{-} &
  - \\ \hline
DIN-SQL~\cite{pourreza2024din} &
  2023 &
  - &
  \multicolumn{1}{c|}{\greencheck} &
  \multicolumn{1}{c|}{-} &
  \greencheck &
  \multicolumn{1}{c|}{Decoder-Only} &
  \multicolumn{1}{c|}{Sequential Encoding} &
  \multicolumn{1}{c|}{Syntax Language} &
  \multicolumn{1}{c|}{Decomposition} &
  Greedy Search &
  \multicolumn{1}{c|}{\greencheck} &
  \multicolumn{1}{c|}{-} &
  \multicolumn{1}{c|}{-} &
  - \\ \hline
DAIL-SQL~\cite{gao2023text} &
  2023 &
  - &
  \multicolumn{1}{c|}{-} &
  \multicolumn{1}{c|}{-} &
  {\greencheck} &
  \multicolumn{1}{c|}{Decoder-Only} &
  \multicolumn{1}{c|}{Sequential Encoding} &
  \multicolumn{1}{c|}{-} &
  \multicolumn{1}{c|}{-} &
  Greedy Search &
  \multicolumn{1}{c|}{-} &
  \multicolumn{1}{c|}{\greencheck} &
  \multicolumn{1}{c|}{-} &
  - \\ \hline
C3-SQL~\cite{dong2023c3} &
  2023 &
  - &
  \multicolumn{1}{c|}{\greencheck} &
  \multicolumn{1}{c|}{-} &
  - &
  \multicolumn{1}{c|}{Decoder-Only} &
  \multicolumn{1}{c|}{Sequential Encoding} &
  \multicolumn{1}{c|}{-} &
  \multicolumn{1}{c|}{COT} &
  Greedy Search &
  \multicolumn{1}{c|}{-} &
  \multicolumn{1}{c|}{\greencheck} &
  \multicolumn{1}{c|}{-} &
  - \\ \hline
RESDSQL~\cite{li2023resdsql} &
  2023 &
  \greencheck &
  \multicolumn{1}{c|}{\greencheck} &
  \multicolumn{1}{c|}{\greencheck} &
  - &
  \multicolumn{1}{c|}{Encoder-Decoder} &
  \multicolumn{1}{c|}{Sequential Encoding} &
  \multicolumn{1}{c|}{Syntax Language} &
  \multicolumn{1}{c|}{Decomposition} &
  Beam Search &
  \multicolumn{1}{c|}{-} &
  \multicolumn{1}{c|}{-} &
  \multicolumn{1}{c|}{-} &
  - \\ \hline
T5-3B+NatSQL+Token Preprocessing~\cite{rai-etal-2023-improving} &
  2023 &
  \greencheck &
  \multicolumn{1}{c|}{\greencheck} &
  \multicolumn{1}{c|}{\greencheck} &
  - &
  \multicolumn{1}{c|}{Encoder-Decoder} &
  \multicolumn{1}{c|}{Sequential Encoding} &
  \multicolumn{1}{c|}{Syntax Language} &
  \multicolumn{1}{c|}{-} &
  Greedy Search &
  \multicolumn{1}{c|}{-} &
  \multicolumn{1}{c|}{-} &
  \multicolumn{1}{c|}{-} &
  - \\ \hline
ACT-SQL~\cite{zhang2023act} &
  2023 &
  - &
  \multicolumn{1}{c|}{\greencheck}&
  \multicolumn{1}{c|}{-} &
  \multicolumn{1}{c|}{\greencheck} &
  \multicolumn{1}{c|}{Decoder-Only} &
  \multicolumn{1}{c|}{Sequential Encoding} &
  \multicolumn{1}{c|}{-} &
  \multicolumn{1}{c|}{CoT} &
  Greedy Search &
  \multicolumn{1}{c|}{-} &
  \multicolumn{1}{c|}{-} &
  \multicolumn{1}{c|}{-} &
  - \\ \hline
ODIS~\cite{chang2023selective} &
  2023 &
  - &
  \multicolumn{1}{c|}{-} &
  \multicolumn{1}{c|}{-} &
  \multicolumn{1}{c|}{\greencheck} &
  \multicolumn{1}{c|}{Decoder-Only} &
  \multicolumn{1}{c|}{Sequential Encoding} &
  \multicolumn{1}{c|}{-} &
  \multicolumn{1}{c|}{-} &
  Greedy Search &
  \multicolumn{1}{c|}{-} &
  \multicolumn{1}{c|}{-} &
  \multicolumn{1}{c|}{-} &
  - \\ \hline
MAC-SQL~\cite{wang2023mac} &
  2023 &
  - &
  \multicolumn{1}{c|}{\greencheck} &
  \multicolumn{1}{c|}{-} &
  - &
  \multicolumn{1}{c|}{Decoder-Only} &
  \multicolumn{1}{c|}{Sequential Encoding} &
  \multicolumn{1}{c|}{-} &
  \multicolumn{1}{c|}{Decomposition} &
  Greedy Search &
  \multicolumn{1}{c|}{\greencheck} &
  \multicolumn{1}{c|}{-} &
  \multicolumn{1}{c|}{\greencheck} &
  - \\ \hline
SC-Prompt~\cite{gu2023few} &
  2023 &
  \greencheck &
  \multicolumn{1}{c|}{-} &
  \multicolumn{1}{c|}{-} &
  - &
  \multicolumn{1}{c|}{Encoder-Decoder} &
  \multicolumn{1}{c|}{Separate Encoding} &
  \multicolumn{1}{c|}{Sketch Structure} &
  \multicolumn{1}{c|}{-} &
  Beam Search &
  \multicolumn{1}{c|}{\greencheck} &
  \multicolumn{1}{c|}{-} &
  \multicolumn{1}{c|}{-} &
  - \\ \hline
CatSQL~\cite{fu2023catsql} &
  2023 &
  \greencheck &
  \multicolumn{1}{c|}{-} &
  \multicolumn{1}{c|}{-} &
  - &
  \multicolumn{1}{c|}{Encoder-Only} &
  \multicolumn{1}{c|}{Sequential Encoding} &
  \multicolumn{1}{c|}{Sketch Structure} &
  \multicolumn{1}{c|}{-} &
  Beam Search &
  \multicolumn{1}{c|}{\greencheck} &
  \multicolumn{1}{c|}{-} &
  \multicolumn{1}{c|}{-} &
  - \\ \hline
SQLFormer~\cite{bazaga2023sqlformer} &
  2023 &
  \greencheck &
  \multicolumn{1}{c|}{\greencheck} &
  \multicolumn{1}{c|}{-} &
  - &
  \multicolumn{1}{c|}{Encoder-Decoder} &
  \multicolumn{1}{c|}{Graph-based Encoding} &
  \multicolumn{1}{c|}{-} &
  \multicolumn{1}{c|}{-} &
  Beam Search &
  \multicolumn{1}{c|}{-} &
  \multicolumn{1}{c|}{-} &
  \multicolumn{1}{c|}{-} &
  - \\ \hline
G³R~\cite{xiang2023g3r} &
  2023 &
  \greencheck &
  \multicolumn{1}{c|}{\greencheck} &
  \multicolumn{1}{c|}{\greencheck} &
  - &
  \multicolumn{1}{c|}{Encoder-Only} &
  \multicolumn{1}{c|}{Graph-based Encoding} &
  \multicolumn{1}{c|}{-} &
  \multicolumn{1}{c|}{COT} &
  Beam Search &
  \multicolumn{1}{c|}{-} &
  \multicolumn{1}{c|}{-} &
  \multicolumn{1}{c|}{-} &
  \greencheck \\ \hline
Graphix-T5~\cite{li2023graphix} &
  2022 &
  \greencheck &
  \multicolumn{1}{c|}{\greencheck} &
  \multicolumn{1}{c|}{\greencheck} &
  - &
  \multicolumn{1}{c|}{Encoder-Decoder} &
  \multicolumn{1}{c|}{Graph-based Encoding} &
  \multicolumn{1}{c|}{-} &
  \multicolumn{1}{c|}{-} &
  Constraint-aware Incremental &
  \multicolumn{1}{c|}{-} &
  \multicolumn{1}{c|}{-} &
  \multicolumn{1}{c|}{-} &
  - \\ \hline
SHiP~\cite{zhao2022importance} &
  2022 &
  \greencheck &
  \multicolumn{1}{c|}{-} &
  \multicolumn{1}{c|}{\greencheck} &
  - &
  \multicolumn{1}{c|}{Encoder-Decoder} &
  \multicolumn{1}{c|}{Graph-based Encoding} &
  \multicolumn{1}{c|}{-} &
  \multicolumn{1}{c|}{-} &
  Constraint-aware Incremental &
  \multicolumn{1}{c|}{-} &
  \multicolumn{1}{c|}{-} &
  \multicolumn{1}{c|}{-} &
  - \\ \hline
N-best List Rerankers~\cite{zeng2023n} &
  2022 & 
  \greencheck &
  \multicolumn{1}{c|}{\greencheck} &
  \multicolumn{1}{c|}{\greencheck} &
  - &
  \multicolumn{1}{c|}{Encoder-Decoder} &
  \multicolumn{1}{c|}{Sequential Encoding} &
  \multicolumn{1}{c|}{-} &
  \multicolumn{1}{c|}{-} &
  Constraint-aware Incremental &
  \multicolumn{1}{c|}{-} &
  \multicolumn{1}{c|}{-} &
  \multicolumn{1}{c|}{-} &
  \greencheck \\ \hline
RASAT~\cite{qi2022rasat} &
  2022 &
  \greencheck &
  \multicolumn{1}{c|}{-} &
  \multicolumn{1}{c|}{\greencheck} &
  - &
  \multicolumn{1}{c|}{Encoder-Decoder} &
  \multicolumn{1}{c|}{Graph-based Encoding} &
  \multicolumn{1}{c|}{-} &
  \multicolumn{1}{c|}{-} &
  Constraint-aware Incremental &
  \multicolumn{1}{c|}{-} &
  \multicolumn{1}{c|}{-} &
  \multicolumn{1}{c|}{-} &
  - \\ \hline
PICARD~\cite{scholak2021picard} &
  2022 &
  \greencheck &
  \multicolumn{1}{c|}{-} &
  \multicolumn{1}{c|}{\greencheck} &
  - &
  \multicolumn{1}{c|}{Encoder-Decoder} &
  \multicolumn{1}{c|}{Sequential Encoding} &
  \multicolumn{1}{c|}{-} &
  \multicolumn{1}{c|}{-} &
  Constraint-aware Incremental &
  \multicolumn{1}{c|}{-} &
  \multicolumn{1}{c|}{-} &
  \multicolumn{1}{c|}{-} &
  - \\ \hline
TKK~\cite{gao2022towards} &
  2022 &
  \greencheck &
  \multicolumn{1}{c|}{-} &
  \multicolumn{1}{c|}{\greencheck} &
  - &
  \multicolumn{1}{c|}{Encoder-Decoder} &
  \multicolumn{1}{c|}{Separate Encoding} &
  \multicolumn{1}{c|}{Sketch Structure} &
  \multicolumn{1}{c|}{Decomposition} &
  Constraint-aware Incremental &
  \multicolumn{1}{c|}{-} &
  \multicolumn{1}{c|}{-} &
  \multicolumn{1}{c|}{-} &
  - \\ \hline
S²SQL~\cite{hui2022s} &
  2022 &
  \greencheck &
  \multicolumn{1}{c|}{\greencheck} &
  \multicolumn{1}{c|}{\greencheck} &
  - &
  \multicolumn{1}{c|}{Encoder-Only} &
  \multicolumn{1}{c|}{Graph-based Encoding} &
  \multicolumn{1}{c|}{-} &
  \multicolumn{1}{c|}{-} &
  \multicolumn{1}{c|}{Greedy Search} &
  \multicolumn{1}{c|}{-} &
  \multicolumn{1}{c|}{-} &
  \multicolumn{1}{c|}{-} &
  - \\ \hline
RAT-SQL~\cite{DBLP:conf/acl/WangSLPR20} &
  2021 &
  \greencheck &
  \multicolumn{1}{c|}{\greencheck} &
  \multicolumn{1}{c|}{\greencheck} &
  - &
  \multicolumn{1}{c|}{Encoder-Only} &
  \multicolumn{1}{c|}{Graph-based Encoding} &
  \multicolumn{1}{c|}{Syntax Language} &
  \multicolumn{1}{c|}{-} &
  Beam Search &
  \multicolumn{1}{c|}{-} &
  \multicolumn{1}{c|}{-} &
  \multicolumn{1}{c|}{-} &
  - \\ \hline
SmBoP~\cite{rubin2020smbop} &
  2021 &
  \greencheck &
  \multicolumn{1}{c|}{-} &
  \multicolumn{1}{c|}{\greencheck} &
  - &
  \multicolumn{1}{c|}{Encoder-Only} &
  \multicolumn{1}{c|}{Graph-based Encoding} &
  \multicolumn{1}{c|}{-} &
  \multicolumn{1}{c|}{-} &
  Beam Search &
  \multicolumn{1}{c|}{-} &
  \multicolumn{1}{c|}{-} &
  \multicolumn{1}{c|}{-} &
  - \\ \hline
RaSaP~\cite{huang2021relation} &
  2021 &
  \greencheck &
  \multicolumn{1}{c|}{\greencheck} &
  \multicolumn{1}{c|}{\greencheck} &
  - &
  \multicolumn{1}{c|}{Encoder-Only} &
  \multicolumn{1}{c|}{Graph-based Encoding} &
  \multicolumn{1}{c|}{-} &
  \multicolumn{1}{c|}{-} &
  Beam Search &
  \multicolumn{1}{c|}{-} &
  \multicolumn{1}{c|}{-} &
  \multicolumn{1}{c|}{-} &
  - \\ \hline
BRIDGE~\cite{lin2020bridging} &
  2020 &
  \greencheck &
  \multicolumn{1}{c|}{-} &
  \multicolumn{1}{c|}{\greencheck} &
  - &
  \multicolumn{1}{c|}{Encoder-Only} &
  \multicolumn{1}{c|}{Sequential Encoding} &
  \multicolumn{1}{c|}{-} &
  \multicolumn{1}{c|}{-} &
  Constraint-aware Incremental  &
  \multicolumn{1}{c|}{-} &
  \multicolumn{1}{c|}{-} &
  \multicolumn{1}{c|}{-} &
  - \\ \hline

\end{tabular}
}
\end{center}
\end{sidewaystable*}

\section{Pre-Processing Strategies for Text-to-SQL}
\label{sec:preprocessing}

The pre-processing step is crucial in the \nlsql translation process, as it identifies relevant tables and columns (\ie Schema Linking) and retrieves necessary database contents or cell values (\ie DB Content Retrieval) to support \sql query generation.
What's more, it enriches context by incorporating domain-specific knowledge (\ie Additional Information Acquisition), which can improve the understanding of the query context and correct errors to prevent their propagation.

\subsection{Schema Linking}
\label{subsec:schema-linking}
Schema linking aims to identify the tables and columns relevant to the given \nlq query, ensuring accurate mapping and processing of key information within the limited input. This step is essential for improving the performance of the \nlsql task. In the LLM era, schema linking has become even more critical due to the input length limitations of LLMs.

We categorize existing schema linking strategies into three groups based on their characteristics: 
\textit{1) string matching-based schema linking}, 
\textit{2) neural network-based schema linking}, and 
\textit{3) in-context learning for schema linking}.

\subsubsection{String Matching-based Schema Linking}
Early research~\cite{guo2019towards, yu2018typesql, lin2019grammar} primarily focused on string matching techniques for schema linking. These methods use similarity measures between the \nlq queries and \db schemas to identify relevant mappings. 
IRNet~\cite{guo2019towards} adopts \textit{exact matching}, identifying links when candidates are identical or one is a substring of the other. While effective for simple cases, it may yield false positives due to shared words. To handle spelling variations, ValueNet~\cite{brunner2021valuenet} applies \textit{approximate matching} via the Damerau–Levenshtein distance~\cite{damerau1964technique}.

However, these methods struggle with handling synonyms and are not robust enough to manage vocabulary variations, limiting their effectiveness in complex \nlsql tasks.

\subsubsection{Neural Network-based Schema Linking} 
To alleviate the above limitations, researchers have employed deep neural networks to align database schemas with natural language queries~\cite{ lei2020re, li2023resdsql}. These methods can better parse complex semantic relationships between \nlq queries and database schema. 

DAE~\cite{dong2019data} frames schema linking as a sequential tagging problem, using a two-stage anonymization model to capture semantic relationships between schema and \nlq. 
SLSQL~\cite{lei2020re} annotates schema linking information in the Spider dataset~\cite{dataset-spider}, enabling a systematic, data-driven study.
RESDSQL~\cite{li2023resdsql} introduces a ranking-enhanced encoding framework for schema linking, using a cross-encoder to prioritize tables and columns based on classification probabilities. FinSQL~\cite{dataset-bull} uses a parallel cross-encoder to retrieve relevant schema elements, significantly reducing linking time.

However, neural network-based methods often struggle to generalize across databases with diverse schemas or domains, especially when training data is scarce.

\subsubsection{In-Context Learning for Schema Linking}
With the advancement of LLMs like GPT-4, research is exploring how to leverage their strong reasoning capabilities for schema linking, \ie directly identifying and linking relevant database schema components from the \nlq query.
A key technique is In-Context Learning  (ICL) technique~\cite{brown2020language}, which utilizes LLMs' ability to understand complex language patterns and relationships within data schemas, enabling a more dynamic and flexible schema linking process~\cite{pourreza2024din, dong2023c3, lee2024mcs, wang2023mac, talaei2024chess}.

C3-SQL~\cite{dong2023c3} employs zero-shot prompts with GPT-3.5 using self-consistency for table and column linking. For table linking, tables are ranked by relevance and listed; for column linking, columns are ranked within relevant tables and outputted as a dictionary, prioritizing matches with question terms or foreign keys.
MAC-SQL~\cite{wang2023mac} proposes a multi-agent collaborative framework for \nlsql, where the \textit{Selector} agent handles schema linking, activated only when the database schema prompt exceeds a specified length. CHESS~\cite{talaei2024chess} utilizes GPT-4 to extract keywords from both \nlq and evidence (additional information from BIRD~\cite{dataset-bird}), implementing a three-stage schema pruning protocol with different prompts.

Employing ICL for schema linking has shown promising performance. However, LLMs have inherent limitations in the amount of context they can process, meaning complex schemas with many tables and columns may exceed this limit.

\subsection{Database Content Retrieval}
\label{subsec: DB Content}
Database content retrieval focuses on efficiently extracting cell values for specific \sql clauses such as \texttt{WHERE}.
We categorize existing database content retrieval strategies into three groups based on their characteristics: 
\textit{1) String Matching-based Methods}, 
\textit{2) Neural Network-based Methods}, and 
\textit{3) Index Strategy for Database Content Retrieval}.

\subsubsection{String Matching-based Methods}
String matching-based methods identify and compare cell values related to the \nlq query through string matching~\cite{guo2019towards, brunner2021valuenet, li2023resdsql, li2023graphix, scholak2021picard, lin2020bridging}.

IRNet~\cite{guo2019towards} uses n-grams, treating text between quotes as cell values. 
BRIDGE~\cite{lin2020bridging} advances this with an anchor text matching technique that automatically extracts cell values from the \nlq. Using heuristics, it calculates the maximum sequence match to define matching boundaries, excluding irrelevant substrings, and adjusts thresholds for accuracy.

However, while string matching methods are effective, they struggle with synonyms and can be computationally expensive when handling large databases.

\subsubsection{Neural Network-based Methods}
These methods aim to capture complex data and semantic features through layers of nonlinear transformations, helping to resolve synonym issues.

TABERT~\cite{yin2020tabert} uses a method called \textit{database content snapshots} to encode relevant database content for the \nlq query, employing attention mechanisms to manage information across cell value representations in different rows. Another approach leverages graph relationships to represent database content. IRNet~\cite{guo2019towards} uses the knowledge graph ConceptNet~\cite{speer2012representing} to find and link relevant cell values, assigning types based on exact or partial matches. RAT-SQL~\cite{DBLP:conf/acl/WangSLPR20} further enhances structural reasoning by modeling the relationship between cell values and the \nlq query, identifying column-value relationships where the query value is part of the column’s candidate cell value. 

While these methods capture semantic features, they may struggle with ambiguous or context-dependent \nlq, leading to inaccurate cell value retrieval. Moreover, the training of neural networks demands substantial computational resources.

\subsubsection{Index Strategy for Database Content Retrieval}

Efficiently retrieving relevant cell values is crucial for the performance of \nlsql systems, especially with large datasets. Indexing is a key method for improving retrieval efficiency by enabling faster access to relevant cell values~\cite{talaei2024chess,li2024codes}.

CHESS~\cite{talaei2024chess} uses Locality-sensitive Hashing~\cite{indyk1998approximate} for approximate nearest neighbor searches, indexing unique cell values to quickly find the top matches related to the \nlq query. This approach speeds up the process of comparing edit distances and semantic embeddings. CodeS~\cite{li2024codes} employs a coarse-to-fine matching strategy. 
It uses BM25~\cite{10.1561/1500000019} to build an index for coarse-grained searches, identifying candidate values, which are then refined by applying the Longest Common Substring algorithm~\cite{aho1975efficient} to assess similarity with the \nlq, thereby pinpointing the most relevant cell values.

While indexing significantly improves retrieval efficiency, building indexes is time-consuming, and frequent changes to database content require continuous updates, adding overhead.

\subsection{Additional Information Acquisition}
\label{subsec:Additional Information}
Additional information, such as domain knowledge, plays a crucial role in enhancing \nlsql models' understanding of \nlq queries, schema linking, and overall \nlsql translation.
This information can provide demonstration examples, domain knowledge, formulaic evidence, and format information for the \nlsql backbone model or specific modules, thereby enhancing the quality of generated results. 
We categorize existing strategies into the following two groups: 
\textit{1) Sample-based Methods}, and
\textit{2) Retrieval-based Methods}.

\subsubsection{Sample-based Methods}
With advancements in LLMs and in-context learning techniques, researchers often incorporate additional information into the textual inputs (\ie prompts) alongside demonstration examples~\cite{chang2023selective, li2024codes, talaei2024chess}.
DIN-SQL~\cite{pourreza2024din} integrates additional information through few-shot learning across multiple stages. 
This helps DIN-SQL to handle challenges like complex schema links, multiple table joins, and nested queries.
In practice, real-world databases often contain rich cross-domain knowledge that can serve as external evidence to support query generation. For example, BIRD~\cite{dataset-bird} contains domain knowledge which is crucial for various \nlsql works~\cite{gao2023text, talaei2024chess}.
Recent works~\cite{chang2023selective, li2024codes} also encode schema metadata (e.g., data types) into natural language to enhance context understanding.

\tikzstyle{my-box}=[
    rectangle,
    draw=hidden-draw,
    rounded corners,
    align=left,
    text opacity=1,
    minimum height=1.3em,
    minimum width=5em,
    inner sep=2pt,
    fill opacity=.8,
    line width=0.6pt,
]

\tikzstyle{leaf-head}=[my-box, minimum height=1.3em,
    draw=grey!80, 
    fill=grey!15, 
    text=black, font=\normalsize,
    inner xsep=1pt,
    inner ysep=2pt,
    line width=0.6pt,
    align=center,
]

\tikzstyle{leaf-encoding_strategy}=[my-box, minimum height=1.3em,
    draw=red!80,
    fill=red!15, 
    text=black, font=\normalsize,
    inner xsep=1pt,
    inner ysep=2pt,
    line width=0.6pt,
    align=center,
]
\tikzstyle{leaf-decoding_strategy}=[my-box, minimum height=1.3em,
    draw=orange!70,
    fill=orange!15, 
    text=black, font=\normalsize,
    inner xsep=1pt,
    inner ysep=2pt,
    line width=0.6pt,
    align=center,
]

\tikzstyle{leaf-generation_strategy}=[my-box, minimum height=1.3em,
    draw=green!80,
    fill=green!15, 
    text=black, font=\normalsize,
    inner xsep=1pt,
    inner ysep=2pt,
    line width=0.6pt,
    align=center,
]
\tikzstyle{leaf-IR}=[my-box, minimum height=1.3em,
    draw=cyan!80,
    fill=cyan!15, 
    text=black, font=\normalsize,
    inner xsep=1pt,
    inner ysep=2pt,
    line width=0.6pt,
    align=center,
]

\tikzstyle{modelnode-encoding_strategy}=[my-box, minimum height=1.3em,
    draw=red!100,
    fill=white, 
    text=black, font=\normalsize,
    inner xsep=1pt,
    inner ysep=2pt,
    line width=0.6pt,
]

\tikzstyle{modelnode-decoding_strategy}=[my-box, minimum height=1.3em,
    draw=orange!100, 
    fill=white, 
    text=black, font=\normalsize,
    inner xsep=1pt,
    inner ysep=2pt,
    line width=0.6pt,
]

\tikzstyle{modelnode-generation_strategy}=[my-box, minimum height=1.3em,
    draw=green!100,
    fill=white, 
    text=black, font=\normalsize,
    inner xsep=1pt,
    inner ysep=2pt,
    line width=0.6pt,
]
\tikzstyle{modelnode-IR}=[my-box, minimum height=1.3em,
    draw=cyan!100,
    fill=white, 
    text=black, font=\normalsize,
    inner xsep=1pt,
    inner ysep=2pt,
    line width=0.6pt,
]

\begin{figure*}
    \centering
    \vspace{-1em}
    \resizebox{1\textwidth}{!}{
        \begin{forest}
            forked edges,
            for tree={
                grow=east,
                reversed=true,
                anchor=base west,
                parent anchor=east,
                child anchor=west,
                base=left,
                font=\normalsize,
                rectangle,
                draw=hidden-draw,
                rounded corners,
                align=left,
                minimum width=1em,
                edge+={darkgray, line width=1pt},
                s sep=3pt,
                inner xsep=0pt,
                inner ysep=3pt,
                line width=0.6pt,
                ver/.style={rotate=90, child anchor=north, parent anchor=south, anchor=center},
            },
            [Text-to-SQL\\Translation Methods, leaf-head, text width=9em
                [Encoding Strategy \\ (\S\ref{subsec:encoding}), leaf-encoding_strategy, text width=9em
                    [Sequential Encoding, leaf-encoding_strategy, text width=9.5em
                        [C3-SQL~\cite{dong2023c3}{, }PET-SQL~\cite{li2024pet}{, }CHESS~\cite{talaei2024chess}{, }DAIL-SQL~\cite{gao2023text}{, }DIN-SQL~\cite{pourreza2024din}{, }CatSQL~\cite{fu2023catsql} {, } BRIDGE~\cite{lin2020bridging}{, }MAC-SQL~\cite{wang2023mac}{, }\\TA-SQL~\cite{qu2024before}{, }RESDSQL~\cite{li2023resdsql}{, }CodeS~\cite{li2024codes}{, }N-best List Rerankers~\cite{zeng2023n} {, } T5+NatSQL+Token Preprocessing~\cite{rai-etal-2023-improving}, modelnode-encoding_strategy, text width=53em]
                    ]
                    [Graph-based Encoding , leaf-encoding_strategy, text width=9.5em
                        [SQLformer~\cite{bazaga2023sqlformer}{, }RASAT~\cite{qi2022rasat}{, }SHiP~\cite{zhao2022importance}{, }SmBoP~\cite{rubin2020smbop}{, }RaSaP~\cite{huang2021relation} {, } S²SQL~\cite{hui2022s}{, }Graphix-T5~\cite{li2023graphix}{, }G³R~\cite{xiang2023g3r}{, }RATSQL~\cite{DBLP:conf/acl/WangSLPR20}, modelnode-encoding_strategy, text width=53em]
                    ]
                    [Separate Encoding , leaf-encoding_strategy, text width=9.5em
                        [SC-Prompt~\cite{gu2023few}{, }TKK~\cite{gao2022towards}, modelnode-encoding_strategy, text width=53em]
                    ]
                ]
                [Decoding Strategy \\ (\S\ref{subsec:decoding}), leaf-decoding_strategy, text width=9em
                    [Greedy Search-based \\Decoding, leaf-decoding_strategy, text width=9.5em
                        [DAIL-SQL~\cite{gao2023text}{, }DIN-SQL~\cite{pourreza2024din}{, }C3-SQL~\cite{dong2023c3}{, }PET-SQL~\cite{li2024pet}{, }CHESS~\cite{talaei2024chess} {, }CodeS~\cite{li2024codes}{, }SuperSQL~\cite{nlsql360}{, } TA-SQL~\cite{qu2024before} {, }\\MAC-SQL~\cite{wang2023mac}{, }T5+NatSQL+Token Preprocessing~\cite{rai-etal-2023-improving}, modelnode-decoding_strategy, text width=53em]
                    ]
                    [Beam Search-based \\Decoding, leaf-decoding_strategy, text width=9.5em
                        [CatSQL~\cite{fu2023catsql}{, }ZeroNL2SQL~\cite{gu2023interleaving}{, }SQLformer~\cite{bazaga2023sqlformer}{, }SC-Prompt~\cite{gu2023few}{, }SmBoP~\cite{rubin2020smbop} {, } RaSaP~\cite{huang2021relation}{, }G³R~\cite{xiang2023g3r}{, }RESDSQL~\cite{li2023resdsql}{, }\\RATSQL~\cite{DBLP:conf/acl/WangSLPR20}, modelnode-decoding_strategy, text width=53em]
                    ]
                    [Constraint-aware \\Incremental Decoding , leaf-decoding_strategy, text width=9.5em
                        [TKK~\cite{gao2022towards}{, }PICARD~\cite{scholak2021picard}{, }RASAT~\cite{qi2022rasat}{, }SHiP~\cite{zhao2022importance}{, }Graphix-T5~\cite{li2023graphix} {, } BRIDGE~\cite{lin2020bridging}{, }N-best List Rerankers~\cite{zeng2023n}, modelnode-decoding_strategy, text width=53em]
                    ]
                ]
                [Task-specific Prompt \\Strategy (\S\ref{subsec:Prompt_Strategy}), leaf-generation_strategy, text width=9em
                    [Chain-of-Thought, leaf-generation_strategy, text width=9.5em
                        [CHESS~\cite{talaei2024chess}{, }
                        ACT-SQL~\cite{zhang2023act}{, }
                        COE-SQL~\cite{zhang2024coe}{, }
                        TA-SQL~\cite{qu2024before}{, }
                        C3-SQL~\cite{dong2023c3}{, }
                        G³R~\cite{xiang2023g3r}{, }
                        MAC-SQL~\cite{wang2023mac}
                        ,modelnode-generation_strategy, text width=53em]
                    ]
                    [Decomposition, leaf-generation_strategy, text width=9.5em
                        [
                        TKK~\cite{gao2022towards}{, }
                        G³R~\cite{xiang2023g3r}{, }
                        DEA-SQL~\cite{xie2024decomposition}{, }
                        MAC-SQL~\cite{wang2023mac}{, }
                        DIN-SQL~\cite{pourreza2024din}
                        , modelnode-generation_strategy, text width=53em]
                    ]
                ]
                [Intermediate \\ Representation \\ (\S\ref{subsec:IR}), leaf-IR, text width=9em
                    [SQL-like \\ Syntax Language, leaf-IR, text width=9.5em
                        [Schema-free SQL~\cite{li2014schema}{, }
                        SyntaxSQLNet~\cite{yu2018syntaxsqlnet}{, }
                        SemQL~\cite{lee1999semql}{, }
                        EditSQL~\cite{zhang2019editing}{, }
                        RAT-SQL~\cite{DBLP:conf/acl/WangSLPR20}{, }
                        NatSQL~\cite{NatSQL}{, } 
                        SHiP~\cite{zhao2022importance}{, } \\
                        QPL~\cite{zhang2019editing}{, } 
                        RESDSQL~\cite{li2023resdsql}{, }
                        QDMR~\cite{wolfson2020break}{, } 
                        OpenSearch-SQL~\cite{xie2025opensearch}
                        ,modelnode-IR, text width=53em]
                    ]
                    [SQL-like\\ Sketch Structure, leaf-IR, text width=9.5em
                        [SyntaxSQLNet~\cite{yu2018syntaxsqlnet}{, }
                        SC-prompt~\cite{gu2023few}{, }
                        CatSQL~\cite{fu2023catsql}{, }
                        ZeroNL2SQL~\cite{gu2023interleaving}{, }
                        TA-SQL~\cite{qu2024before}{, }
                        RESDSQL~\cite{li2023resdsql}
                        ,modelnode-IR, text width=53em]
                    ]
                ]
            ]
            \end{forest}}
     \vspace{-1.5em}
    \caption{\addd{A Taxonomy of Text-to-SQL Translation Methods based on their Design Choices.}}
    \vspace{-1 em}
    \label{fig:taxonomy}
\end{figure*}

\subsubsection{Retrieval-based Methods}
Extracting relevant knowledge and few-shot examples from extensive domain knowledge bases can increase token usage, impacting efficiency and computational cost~\cite{pourreza2024din, gao2023text}. To enhance accuracy and efficiency, some researchers employ similarity-based retrieval methods. For example, PET-SQL~\cite{li2024pet} builds a pool of question frames and question-SQL pairs, selecting the $k$ most similar examples to the target query, which are then used in prompts.

When databases lack text-based additional information, researchers devise methods to retrieve and convert external knowledge into natural language. 
For example, REGROUP~\cite{dou2023towards} creates a formulaic knowledge base across domains (e.g., finance, transportation) and uses Dense Passage Retriever~\cite{karpukhin2020dense} to compute similarity scores, integrating related entities with \nlq and schema through an Erasing-Then-Awakening model~\cite{DBLP:conf/acl/LiuYZGZL21}.
ReBoost~\cite{sui2023reboost} uses a two-phase Explain-Squeeze Schema Linking strategy, first presenting a generalized schema to LLMs, then applying targeted prompts to improve query-to-entity mapping accuracy.

Retrieval-based methods improve the effectiveness of acquiring additional information but increase computing costs. Moreover, current research mostly relies on domain-specific text, with limited use of structured knowledge. Thus, integrating diverse information sources could further enhance \nlsql performance, especially for domain-specific databases.

\section{Text-to-SQL Translation Methods}
\label{sec: translation}

In this section, we elaborate on \nlsql translation methods using language models. As shown in Figure~\ref{fig:taxonomy}, we will detail their encoding (Section~\ref{subsec:encoding}), decoding (Section~\ref{subsec:decoding}), and task-specific prompt strategies (Section~\ref{subsec:Prompt_Strategy}). 
Moreover, we will discuss how the intermediate representation can benefit the \nlsql translation process (Section~\ref{subsec:IR}).

\subsection{Encoding Strategy}
\label{subsec:encoding}

\begin{figure}[t!]
	\centering
	\includegraphics[width=\linewidth]{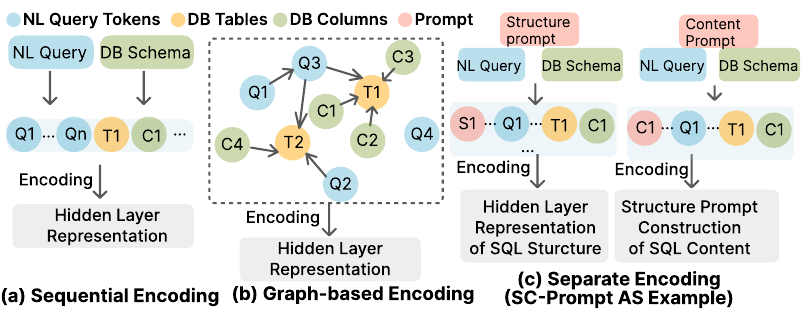}
	\caption{\addd{An Overview of the Encoding Strategies.}}
	\label{fig:encoding_overview}	
\end{figure}

\marginpar{AE C1}
\marginpar{R3 C1}
\addd{
In the \nlsql task, encoding refers to transforming \nlq and database schema into a structured format suitable for language model processing. This step is essential for converting unstructured data into a form usable for \sql generation, capturing the \nlq's semantics and the schema's structure to help the model map user intent to appropriate \sql.
As shown in Figure~\ref{fig:encoding_overview},  primary encoding strategies include
\textit{1) Sequential Encoding}, 
\textit{2) Graph-based Encoding}, and 
\textit{3) Separate Encoding}.
}

\subsubsection{Sequential Encoding Strategy}
\addd{
Sequential encoding is a strategy in \nlsql models where both the \nlq and the database schema are treated as token sequences.  As shown in Figure~\ref{fig:encoding_overview}(a), the model processes the entire input as a linear token sequence using standard Transformer-based architectures.}

\addd{
Models like T5~\cite{raffel2020exploring} are used to encode \nlq and database schema sequentially in works~\cite{rai-etal-2023-improving, scholak2021picard}.
BRIDGE~\cite{lin2020bridging} improves the alignment between the \nlq and database schema by representing both as a tagged sequence and inserting matched database cell values (called anchor texts) next to corresponding fields. 
RESDSQL~\cite{li2023resdsql} uses a ranking-enhanced encoder to sort and filter schema items, prioritizing the most relevant ones and reducing schema linking complexity. Although LLM-based \nlsql systems often do not explicitly define an input encoding strategy, they typically implicitly adopt a sequential form by concatenating queries and schema components. These models leverage self-attention to model dependencies across the entire sequence.
}

\addd{While this allows for flexible contextualization, such approaches may struggle to capture complex relational structures, limiting their performance on deeply nested \sql queries.}

\subsubsection{Graph-based Encoding Strategy}
\addd{Graph-based encoding in \nlsql models represents both \nlq and database schema as interconnected graphs, leveraging the relational structure of databases and inter-dependencies in the input data, as shown in Figure~\ref{fig:encoding_overview}(b).  
Unlike sequential encoding, this approach preserves the schema's topology, offering richer context for each element and enhancing the model's ability to produce accurate \sql queries ~\cite{DBLP:conf/acl/WangSLPR20, rubin2020smbop,  hui2022s, qi2022rasat, zhao2022importance, bazaga2023sqlformer, li2023graphix}.}

\addd{RAT-SQL~\cite{DBLP:conf/acl/WangSLPR20} introduces a relation-aware self-attention mechanism, explicitly using relational information in a graph structure to jointly encode the question and schema, enhancing the model's understanding of structural information. 
S$^2$SQL~\cite{hui2022s} injects syntactic structure information at the encoding stage using the ELECTRA~\cite{clark2020electra} model, enhancing semantic understanding. G$^3$R~\cite{xiang2023g3r} uses the LGESQL\cite{cao2021lgesql} encoder and Graph Attention Network (GAT)~\cite{velivckovic2017graph} to capture multi-source heterogeneous information. 
Graphix-T5~\cite{li2023graphix} adds graph-aware layers, incorporating structural information directly into the encoding process, significantly improving \sql query generation across multiple benchmarks.}

\addd{However, this strategy typically involves more intricate graph construction and processing algorithms. It also tends to rely on large-scale training data to achieve optimal performance, which limits its applicability in low-resource scenarios.}

\subsubsection{Separate Encoding Strategy}

\addd{The separate encoding strategy in \nlsql refers to independently encoding different parts of the input (typically \nlq and the \db schema) rather than combining them into a single sequence. This strategy has evolved significantly over time and can be broadly categorized into traditional and modern forms.}

\addd{In traditional separate encoding, early models like SQLNet~\cite{xu2017sqlnet} and Seq2SQL~\cite{dataset-wikisql}, processed the \nlq and database schema separately due to format mismatches. However, this lack of interaction between the two components hindered effective schema linking, leading to limited performance and making this approach less common in recent research.
Modern separate encoding strategies, as illustrated in Figure~\ref{fig:encoding_overview}(c), focus on modular representation learning by decomposing the \nlsql task into subtasks and encoding different aspects separately. TKK~\cite{gao2022towards} employs task decomposition and multi-task learning strategies by breaking down the complex \nlsql task into subtasks and progressively integrating knowledge. Similarly, SC-Prompt~\cite{gu2023few} divides text encoding into two stages: structure and content, each encoded separately.}

\addd{While separate encoding may increase computational overhead due to multiple processing steps, it allows for more refined understanding of different aspects of queries. This modularity provides the model with greater flexibility to handle various query tasks, thereby enhancing overall performance.}

\subsection{Decoding Strategy}
\label{subsec:decoding}

\begin{figure}[t!]
	\centering
	\includegraphics[width=\linewidth]{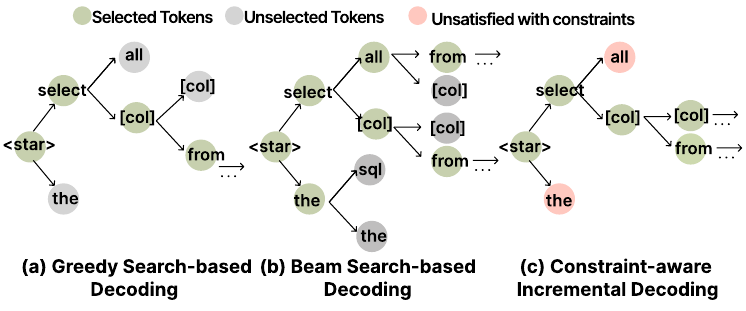}
	\caption{\addd{An Overview of the Decoding Strategies.}}
	\label{fig:decoding_overview}
\end{figure}

\addd{Decoding is a crucial step in \nlsql translation, transforming encoder-generated representations into \sql. An effective decoding strategy ensures that the generated \sql queries are not only syntactically correct but also semantically aligned with the \nlq queries, while optimizing \sql execution efficiency.
Figure~\ref{fig:decoding_overview} introduces several key decoding strategies.}

\subsubsection{Greedy Search-based Decoding Strategy}
\addd{The greedy search-based decoding strategy is a simple and efficient method that selects the token with the highest probability at each decoding step. As illustrated in Figure~\ref{fig:decoding_overview}(a), it constructs the output sequence by making a series of locally optimal choices, without considering future possibilities.}

Since GPT models (\eg GPT-4) default to greedy search-based decoding, many \nlsql solutions using GPT fall into this category. 
DTS-SQL~\cite{pourreza2024dts}, based on DeepSeek~\cite{bi2024deepseek}, uses the same approach. Early models like SQLNet~\cite{xu2017sqlnet} and Seq2SQL~\cite{dataset-wikisql} also rely on greedy search for \sql generation.

\marginpar{AE C1}
\marginpar{R3 C1}
\addd{Despite its efficiency, greedy search has notable limitations. By focusing only on immediate token probabilities, it fails to account for long-term dependencies or global sequence coherence. As a result, errors made early in the decoding process may propagate and lead to suboptimal or incorrect SQL queries, particularly for complex or multi-step questions.}

\subsubsection{Beam Search-based Decoding Strategy}
\addd{Beam search is a widely used decoding strategy that explores a broader search space compared to greedy decoding, often leading to higher-quality results. Instead of selecting only the top token at each step, it retains a fixed number of top-ranked partial sequences (known as beams) and expands each by considering the top-$k$ most probable next tokens, as illustrated in Figure~\ref{fig:decoding_overview}(b).}

Given its advantages, several \nlsql models employ beam search~\cite{huang2021relation, fu2023catsql, li2023resdsql}. 
RAT-SQL~\cite{DBLP:conf/acl/WangSLPR20} combines relation-aware graph structure encoding with beam search to generate multiple \sql candidates, reranking them based on graph structure information. 
Unlike RAT-SQL, EditSQL~\cite{zhang2019editing} uses beam search alongside dialogue history to generate and refine candidate \sql queries.
SmBoP~\cite{rubin2020smbop} employs a semi-autoregressive bottom-up decoding approach, improving efficiency by parallelizing sub-tree construction and scoring, with logarithmic time complexity. 
ZeroNL2SQL~\cite{gu2023interleaving} retains the top-$k$ hypotheses during the \sql sketch generation stage, which are then refined for query and predicate calibration.

\addd{Compared to greedy decoding, beam search improves the ability to generate syntactically and semantically valid \sql queries, especially in complex scenarios, by considering multiple hypotheses at each step. However, this benefit comes at the cost of increased computational complexity and memory usage, potentially slowing down the decoding process.}

\subsubsection{Constraint-aware Incremental Decoding Strategy}

\addd{Constraint-aware incremental decoding strategies aim to ensure the structural and syntactic validity of \sql queries by applying explicit constraints during the decoding process. As shown in Figure~\ref{fig:decoding_overview}(c), these strategies incrementally generate \sql while enforcing \sql grammar constraints at each step.}

\addd{A representative implementation is PICARD~\cite{scholak2021picard} (Parsing Incrementally for Constrained Auto-Regressive Decoding), which integrates \sql grammar constraints into the decoding loop. It verifies the syntactic validity of the partially generated query at every step, ensuring that each token adheres to the \sql grammar. This significantly reduces the generation of invalid or incomplete queries. Many models~\cite{gao2022towards, qi2022rasat, scholak2021picard, zhao2022importance, zeng2023n, li2023graphix} have adopted this paradigm to improve performance.}

\addd{In addition to grammar-level constraints, some models incorporate schema-level constraints during decoding. BRIDGE~\cite{lin2020bridging} introduces Schema-Consistency Guided Decoding, which enforces alignment between the generated \sql query and the underlying database schema by verifying their consistency and adjusting the decoding path accordingly.}

\addd{While this strategy introduces additional computational overhead due to per-token constraint evaluation, it offers strong guarantees of syntactic correctness and is particularly effective for generating structurally complex \sql queries.}

\subsection{Task-specific Prompt Strategy}
\label{subsec:Prompt_Strategy}

In the era of LLMs, prompt engineering has become a powerful method for harnessing LLM capabilities across diverse tasks. In \nlsql, task-specific prompts are crafted to guide LLMs in optimizing \nlsql translation, enhancing the accuracy of translating complex \nlq queries into precise \sql queries.
Broadly speaking, there are two main types of task-specific prompt strategies:
\textit{1)  Chain-of-Thought prompting}, and
\textit{2) Decomposition Strategy}.

\subsubsection{Chain-of-Thought  (CoT) Prompting} 
The CoT prompting~\cite{wei2022chain}, known for its effectiveness, showcases the LLM’s reasoning process, improving both the accuracy and interpretability of the generated results. In \nlsql, CoT enhances model performance and ensures that the generated \sql statements are more aligned with human expectations~\cite{tai2023exploring}.

CHESS~\cite{talaei2024chess} transforms \nlq into \sql statements through a streamlined pipeline that utilizes LLMs and CoT. This process includes entity and context retrieval, schema selection, \sql generation, and revision.	
In addition, the integration of CoT with other techniques can enhance the performance of \nlsql models. These techniques include in-context learning~\cite{zhang2023act, zhang2024coe}, logical synthesis~\cite{qu2024before}, calibration with hints~\cite{dong2023c3, xiang2023g3r} and multi-agent system~\cite{wang2023mac}. 
Specifically, in-context learning and logical synthesis enrich CoT by embedding a deeper linguistic understanding, enabling precise semantic mapping to \sql constructs~\cite{zhang2023act, zhang2024coe}.   
Calibration with hints fine-tunes model responses, aligning them closely with \nlq nuances for accurate intent translation~\cite{dong2023c3, xiang2023g3r}.
Furthermore, integrating the multi-agent framework with CoT fosters a collaborative approach, with specialized agents handling tasks like schema linking and \sql generation, which speeds up reasoning and enhances adaptability~\cite{wang2023mac}.

Overall, these techniques create a more robust \nlsql framework, offering better precision and reliability in translating complex \nlq queries into accurate \sql statements. However, CoT prompting may introduce longer reasoning chains and latency, and its effectiveness can be sensitive to prompt design and task complexity.

\subsubsection{Decomposition Strategy} 

The decomposition strategy divides the \nlsql task into sequential subtasks, allowing each sub-module to concentrate on a specific generation step, thereby enhancing accuracy, quality, and interpretability.

Different approaches vary in subtask decomposition granularity~\cite{pourreza2024dts, gao2022towards, xiang2023g3r, xie2024decomposition, wang2023mac}. 
TKK~\cite{gao2022towards} applies finer-grained decomposition by breaking down \nlsql parsing into subtasks like mapping \nlq to {\tt SELECT}, {\tt FROM}, and {\tt WHERE} clauses. This approach helps the model concentrate on each clause, enhancing understanding of the problem, schema, and \sql alignment. Similar strategies are used in G$^3$R~\cite{xiang2023g3r} and DEA-SQL~\cite{xie2024decomposition}. 
Moreover, decomposition also reduces model complexity.
For example, MAC-SQL~\cite{wang2023mac} introduces a Decomposer agent to split the user's query into subproblems, making \sql generation for each part more manageable.

In general, the decomposition strategy divides the \nlsql translation task into multiple subtasks, enabling each sub-module to focus on enhancing its specific output. However, this approach also raises computational costs, making model training and deployment more complex and resource-intensive.

\begin{figure}[t!]
	\centering
\includegraphics[width=\linewidth]{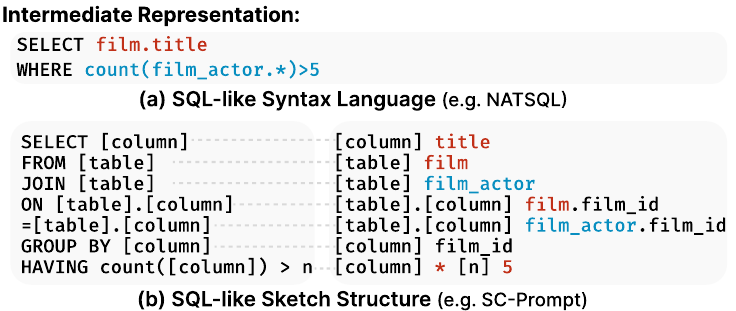}
	\caption{An Example of the Intermediate Representation.}
\label{fig:Intermediate_Representation}
\end{figure}

\subsection{Intermediate Representation for Text-to-SQL Translation}
\label{subsec:IR}

The \nlsql is challenging due to the complexity and ambiguity of \nlq queries, coupled with the syntax-constrained nature of \sql. To simplify this process, researchers have developed a \textit{grammar-free} intermediate representation (IR) to bridge the ``free-form'' \nlq question and the ``constrained and formal'' \sql. This IR provides a structured yet flexible format, capturing the essential components and relationships within an \nlq query without the strict syntax requirements of \sql.
Figure~\ref{fig:Intermediate_Representation} shows two types of IR strategies, discussed below.

\subsubsection{SQL-like syntax language} 
As shown in Figure~\ref{fig:Intermediate_Representation}(a), SQL-like syntax language is a simplified \sql-like structure.
Early approaches used information retrieval techniques to map the original question and schema data into this syntax~\cite{li2014schema, dataset-academic}. 
Subsequent research efforts have focused on consolidating or eliminating partial clauses or operations in \sql queries to simplify SQL-like syntax language~\cite{lee1999semql, yu2018syntaxsqlnet}. 
SyntaxSQLNet~\cite{yu2018syntaxsqlnet} simplifies the syntax language by removing parts of the {\tt FROM} and {\tt JOIN} clauses. SemQL~\cite{lee1999semql} removes the entire {\tt FROM}, {\tt JOIN}, {\tt ON}, and {\tt GROUP BY} clauses, and further merges the {\tt WHERE} and {\tt HAVING} conditions into a unified filtering representation.
Recent research has focused on simplifying syntax languages to improve parsing efficiency~\cite{eyal2023semantic}. NatSQL~\cite{NatSQL}, a widely used SQL-like syntax language, eliminates uncommon \sql operators and keywords, streamlining schema linking by minimizing necessary schema items. Combined with PLMs, NatSQL has achieved strong results on various benchmarks~\cite{li2023resdsql, rai-etal-2023-improving}. 

SQL-like syntax languages have demonstrated potential in bridging user queries and databases. However, previous studies face challenges due to high complexity and limited coverage of database structures~\cite{NatSQL}. As databases grow in size and domain specificity, maintaining the simplicity of SQL-like syntax languages becomes increasingly difficult. Moreover, some of these languages require manual construction and adjustments, raising deployment costs and complexity.

\subsubsection{SQL-like sketch structure} Leveraging the structural characteristics of \sql, researchers have developed SQL-like sketches that mirror \sql structure for parsing, enabling diverse \nlq queries to be mapped into a specific sketch space, as shown in Figure~\ref{fig:Intermediate_Representation}(b). 
This approach reduces parsing complexity.

Early works applied fixed sketch rules and neural networks to map the \nlq into SQL-like sketch structure~\cite{lee2019clause, yu2018syntaxsqlnet}. SyntaxSQLNet~\cite{yu2018syntaxsqlnet} uses a syntax tree and a corresponding decoder, dividing the decoding into nine sub-modules that separately predict operators, keywords, and entities before combining them to generate the final \sql.
In recent years, the development of language models has allowed researchers to design more elaborate SQL-like sketch structures for parsing~\cite{gu2023few, fu2023catsql, gu2023interleaving, qu2024before}. 
CatSQL~\cite{fu2023catsql} constructs a more general template sketch with slots serving as initial placeholders.
Its base model focuses on the parsing of \nlq to fill these placeholders, consequently decreasing the computational cost. 
Moreover, several recent works cover both SQL-like syntax language and SQL-like sketch transition methods. For instance, RESDSQL~\cite{li2023resdsql} introduces a rank-enhanced encoding and skeleton-aware decoding framework.  During the decoding phase, its decoder initially generates the \sql skeleton and then transforms it into the actual \sql query. When combined with NatSQL, RESDSQL demonstrates the ability further to enhance the quality of \sql query generation.

In general, SQL-like sketch structure can be more easily combined with other strategies, such as decomposition strategy or SQL-like syntax language strategy. In addition, it can more fully utilize the comprehension and cloze capabilities of existing LLMs and reduce the dependence on professionals. 

\section{Post-Processing Strategies for Text-to-SQL}
\label{sec:postprocessing}

After the \nlsql model generates the \sql, post-processing can refine it to better meet user expectations. This step involves leveraging additional information or models to enhance the \sql, with a focus on \sql correction, ensuring output consistency, and execution-guided checking.

\begin{figure*}[!t]
 \centering
 \includegraphics[width=.8\textwidth]{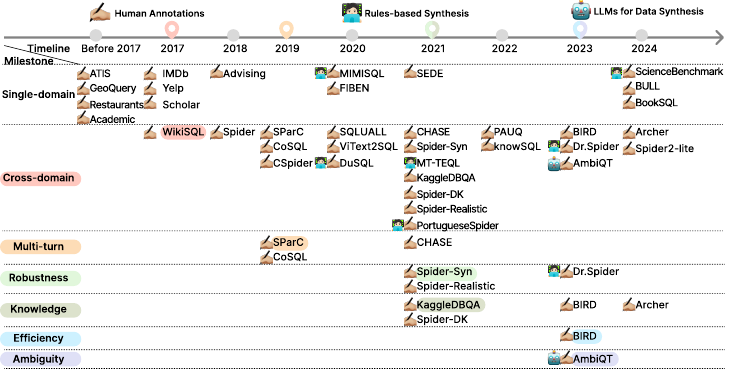}
 \caption{Timeline for Text-to-SQL Benchmarks.}
 \label{fig:dataset_timeline}
\end{figure*}

\subsection{SQL Correction Strategies}
\label{subsec:sql_correction}

The \sql generated by \nlsql models may contain syntax errors. 
DIN-SQL~\cite{pourreza2024din} introduces a self-correction module that operates in a zero-shot setting, where the model receives only the faulty \sql and attempts to correct it. Two prompts are used: a general prompt for CodeX, which directly asks for error identification and correction, and a mild prompt for GPT-4, which seeks potential issues without presuming errors.
To handle errors in predicate predictions, such as incorrect columns or values, ZeroNL2SQL~\cite{gu2023interleaving} employs a multi-level matching approach. This method incrementally expands matching across columns, tables, and databases, allowing matched values to be returned to the LLMs to generate \sql queries consistent with the database content.

While these methods focus on fixing syntax errors, they often overlook semantic errors~\cite{liu2025nl2sql}, such as incorrect table joins, misaligned conditions, or inaccurate aggregations, which are essential for improving accuracy.

\subsection{Output Consistency}
\label{subsec:output_consistency}
To enhance output consistency,  self-consistency~\cite{wang2022self} has been introduced, based on the idea that complex reasoning tasks may have multiple valid paths to a single correct answer. This approach samples various reasoning paths and selects the most consistent answer to improve output quality. 

DAIL-SQL~\cite{gao2023text} integrates self-consistency, achieving a $0.4\%$ performance improvement over configurations without it. 
To reduce LLM randomness, FinSQL~\cite{dataset-bull} generates 
$n$ candidate \sql queries in parallel, clusters them based on keyword consistency, and selects a query from the largest cluster. 
The self-consistency strategy enhances LLM output diversity by increasing the temperature and selecting the final result through majority voting. However, recent studies~\cite{renze2024effect} indicate that relying on a single model may still yield limited diversity. To overcome this limitation, PET-SQL~\cite{li2024pet} introduces a cross-consistency strategy, 
in which multiple LLMs generate \sql at lower temperatures and vote based on execution results.

While these methods improve accuracy by enforcing consistency across multiple executions, they significantly increase inference cost and time.

\subsection{Execution-Guided Strategies}
\label{subsec:execution_guided}
In \nlsql tasks, the execution result of an \sql query provides critical feedback on \nlsql translation accuracy. For example, errors or {\tt NULL} values in execution results can signal potential issues with the \sql query.

To reflect human behavior in writing complex \sql queries, CHESS~\cite{talaei2024chess} provides LLMs with the database schema, question, candidate \sql queries, and their execution results. CHESS starts with a draft query and refines it based on execution feedback, adjusting for syntax errors as needed. CodeS~\cite{li2024codes}, on the other hand, generates a complete \sql statements through beam search, producing four \sql candidates and selecting the first executable one as the final result.

Execution-Guided Strategies refine \sql based on execution results, ensuring the query retrieves data correctly. However, this approach can significantly increase \sql generation time, especially with large databases.

\subsection{N-best Rerankers Strategies}
\label{subsec:nbest_ranker}
In \nlsql tasks, especially in cross-domain scenarios, generated \sql queries can vary subtly in structure and semantics. N-best reranking reorders the top-$N$ model outputs, often leveraging a larger model or additional knowledge sources. For example, fine-tuning a BERT-based reranker, as demonstrated by Bertrand-dr on the Spider dataset~\cite{kelkar2020bertrand}, has effectively improved the performance of several \nlsql models.

However, the effectiveness of Bertrand-dr’s reranking can be unstable and sensitive to threshold settings, sometimes even yielding negative effects. To address these limitations, G$^3$R~\cite{xiang2023g3r} introduces a feature-enhanced reranker using PLM-based hybrid prompt tuning, which bridges domain gaps without extra parameters. Contrastive learning then sharpens distinctions between candidate queries~\cite{DBLP:journals/pacmmod/LuoZ00CS23}. Similarly, ReFSQL~\cite{zhang2023refsql} retrieves the most relevant results from the retriever and generator modules to improve final answer quality.

While N-best reranking is widely used in PLM-based methods to refine \sql candidates, it is less common in LLM-based methods, which typically have stronger inference capabilities.

\section{Text-to-SQL Benchmarks}
\label{sec:benchmark}

In this section, we will first elaborate on the different types of \nlsql datasets, highlighting their characteristics, as shown in Figure~\ref{fig:dataset_timeline} (Section~\ref{subsec:dataset_overview}).
We will then perform an in-depth analysis of existing datasets (Section~\ref{subsec:dataset_stats}).

\subsection{An Overview of Text-to-SQL Benchmarks}
\label{subsec:dataset_overview}
With advancements in \nlsql, various datasets have emerged to address the evolving challenges, as shown in Figure~\ref{fig:dataset_timeline}.
These range from single-domain databases with simple queries to cross-domain, multi-turn, multilingual, and domain-specific scenarios, reflecting the progress and the emergence of new challenges for \nlsql solutions.

\begin{table*}[!t]
\centering
\caption{Statistics of Text-to-SQL Benchmarks.}
\label{tab:datasets}
\resizebox{0.95\textwidth}{!}{%
\begin{tabular}{lccccccccccccc}
\hline
\rowcolor[HTML]{FFFFFF} 
\cellcolor[HTML]{FFFFFF} &
  \multicolumn{3}{c}{\cellcolor[HTML]{FFFFFF}\textbf{Redundancy Measure}} &
  \multicolumn{5}{c}{\cellcolor[HTML]{FFFFFF}\textbf{DB Complexity}} &
  \multicolumn{5}{c}{\cellcolor[HTML]{FFFFFF}\textbf{Query Complexity}} \\ \cline{2-14} 
\rowcolor[HTML]{FFFFFF} 
\multirow{-2}{*}{\cellcolor[HTML]{FFFFFF}\textbf{Dataset}} &
  \textbf{\#-Questions} &
  \textbf{\begin{tabular}[c]{@{}c@{}}\#-Unique\\ SQLs\end{tabular}} &
  \textbf{\begin{tabular}[c]{@{}c@{}}\#-Questions\\ / \#-SQLs\end{tabular}} &
  \textbf{\#-DBs} &
  \textbf{\#-Tables} &
  \textbf{\begin{tabular}[c]{@{}c@{}}\#-Tables\\ / DB\end{tabular}} &
  \textbf{\begin{tabular}[c]{@{}c@{}}\#-Cols\\ / Table\end{tabular}} &
  \textbf{\begin{tabular}[c]{@{}c@{}}\#-Records\\ / DB\end{tabular}} &
  \textbf{Tables} &
  \textbf{Selects} &
  \textbf{Agg} &
  \textbf{\begin{tabular}[c]{@{}c@{}}Scalar\\ Func\end{tabular}} &
  \textbf{\begin{tabular}[c]{@{}c@{}}Math\\ Comp\end{tabular}} \\ \hline
ATIS~\cite{dataset-atis} &
  \cellcolor[HTML]{FBFCFE}5280 &
  \cellcolor[HTML]{FBFCFE}947 &
  \cellcolor[HTML]{F6FAFA}5.6 &
  \cellcolor[HTML]{FCFCFF}1 &
  \cellcolor[HTML]{FCFCFF}25 &
  \cellcolor[HTML]{E4F3EB}25 &
  \cellcolor[HTML]{CEEAD7}5.24 &
  \cellcolor[HTML]{FCFCFF}162243 &
  \cellcolor[HTML]{63BE7B}8.39 &
  \cellcolor[HTML]{97D3A8}1.79 &
  \cellcolor[HTML]{F5F9F9}0.22 &
  \cellcolor[HTML]{FCFCFF}0 &
  \cellcolor[HTML]{FCFCFF}0 \\
GeoQuery~\cite{dataset-geoquery} &
  \cellcolor[HTML]{FCFCFF}877 &
  \cellcolor[HTML]{FCFCFF}246 &
  \cellcolor[HTML]{F9FBFC}3.6 &
  \cellcolor[HTML]{FCFCFF}1 &
  \cellcolor[HTML]{FCFCFF}7 &
  \cellcolor[HTML]{F6FAFA}7 &
  \cellcolor[HTML]{DAEFE2}4.14 &
  \cellcolor[HTML]{FCFCFF}937 &
  \cellcolor[HTML]{E3F2EA}2.22 &
  \cellcolor[HTML]{97D3A8}2.19 &
  \cellcolor[HTML]{6DC284}0.92 &
  \cellcolor[HTML]{FCFCFF}0 &
  \cellcolor[HTML]{F9FBFD}0.01 \\
Restaurants~\cite{dataset-restaurant} &
  \cellcolor[HTML]{FCFCFF}378 &
  \cellcolor[HTML]{FCFCFF}23 &
  \cellcolor[HTML]{E6F4EC}16.4 &
  \cellcolor[HTML]{FCFCFF}1 &
  \cellcolor[HTML]{FCFCFF}3 &
  \cellcolor[HTML]{FAFCFE}3 &
  \cellcolor[HTML]{DCEFE3}4.00 &
  \cellcolor[HTML]{FCFCFF}19295 &
  \cellcolor[HTML]{DFF1E6}2.43 &
  \cellcolor[HTML]{E7F4ED}1.17 &
  \cellcolor[HTML]{DCEFE3}0.35 &
  \cellcolor[HTML]{FCFCFF}0 &
  \cellcolor[HTML]{FCFCFF}0 \\
Academic~\cite{dataset-academic} &
  \cellcolor[HTML]{FCFCFF}196 &
  \cellcolor[HTML]{FCFCFF}185 &
  \cellcolor[HTML]{FCFCFF}1.1 &
  \cellcolor[HTML]{FCFCFF}1 &
  \cellcolor[HTML]{FCFCFF}17 &
  \cellcolor[HTML]{ECF6F2}17 &
  \cellcolor[HTML]{E5F3EB}3.12 &
  \cellcolor[HTML]{C0E4CB}58249674 &
  \cellcolor[HTML]{C9E8D3}3.48 &
  \cellcolor[HTML]{F7FAFB}1.04 &
  \cellcolor[HTML]{B7E0C3}0.54 &
  \cellcolor[HTML]{FCFCFF}0 &
  \cellcolor[HTML]{FCFCFF}0 \\
IMDb~\cite{dataset-yelp-imdb} &
  \cellcolor[HTML]{FCFCFF}131 &
  \cellcolor[HTML]{FCFCFF}89 &
  \cellcolor[HTML]{FCFCFF}1.5 &
  \cellcolor[HTML]{FCFCFF}1 &
  \cellcolor[HTML]{FCFCFF}17 &
  \cellcolor[HTML]{ECF6F2}17 &
  \cellcolor[HTML]{DCEFE4}3.94 &
  \cellcolor[HTML]{D3ECDC}40147386 &
  \cellcolor[HTML]{D5ECDD}2.91 &
  \cellcolor[HTML]{FBFCFE}1.01 &
  \cellcolor[HTML]{E5F3EB}0.30 &
  \cellcolor[HTML]{FCFCFF}0 &
  \cellcolor[HTML]{FCFCFF}0 \\
Yelp~\cite{dataset-yelp-imdb} &
  \cellcolor[HTML]{FCFCFF}128 &
  \cellcolor[HTML]{FCFCFF}110 &
  \cellcolor[HTML]{FCFCFF}1.2 &
  \cellcolor[HTML]{FCFCFF}1 &
  \cellcolor[HTML]{FCFCFF}8 &
  \cellcolor[HTML]{F5FAF9}8 &
  \cellcolor[HTML]{D1EBDA}5 &
  \cellcolor[HTML]{F7FAFB}4823945 &
  \cellcolor[HTML]{DFF1E6}2.41 &
  \cellcolor[HTML]{FCFCFF}1 &
  \cellcolor[HTML]{C8E7D2}0.45 &
  \cellcolor[HTML]{FCFCFF}0 &
  \cellcolor[HTML]{FCFCFF}0 \\
Scholar~\cite{dataset-scholar} &
  \cellcolor[HTML]{FCFCFF}817 &
  \cellcolor[HTML]{FCFCFF}193 &
  \cellcolor[HTML]{F8FBFB}4.2 &
  \cellcolor[HTML]{FCFCFF}1 &
  \cellcolor[HTML]{FCFCFF}10 &
  \cellcolor[HTML]{F3F9F8}10 &
  \cellcolor[HTML]{ECF6F1}2.50 &
  \cellcolor[HTML]{63BE7B}147416275 &
  \cellcolor[HTML]{CBE9D5}3.38 &
  \cellcolor[HTML]{FAFBFD}1.02 &
  \cellcolor[HTML]{9CD5AC}0.68 &
  \cellcolor[HTML]{FCFCFF}0 &
  \cellcolor[HTML]{F6FAFA}0.02 \\
WikiSQL~\cite{dataset-wikisql} &
  \cellcolor[HTML]{E3F2EA}80654 &
  \cellcolor[HTML]{63BE7B}80257 &
  \cellcolor[HTML]{FCFCFF}1 &
  \cellcolor[HTML]{F4F9F8}26531 &
  \cellcolor[HTML]{FBFCFE}26531 &
  \cellcolor[HTML]{FCFCFF}1 &
  \cellcolor[HTML]{C2E5CD}6.34 &
  \cellcolor[HTML]{FCFCFF}17 &
  \cellcolor[HTML]{FCFCFF}1 &
  \cellcolor[HTML]{FCFCFF}1 &
  \cellcolor[HTML]{E9F5EF}0.28 &
  \cellcolor[HTML]{FCFCFF}0 &
  \cellcolor[HTML]{FCFCFF}0 \\
Advising~\cite{dataset-advising} &
  \cellcolor[HTML]{FBFCFE}4387 &
  \cellcolor[HTML]{FCFCFF}205 &
  \cellcolor[HTML]{DFF1E6}21.4 &
  \cellcolor[HTML]{FCFCFF}1 &
  \cellcolor[HTML]{FCFCFF}15 &
  \cellcolor[HTML]{EEF7F3}15 &
  \cellcolor[HTML]{B6E0C3}7.40 &
  \cellcolor[HTML]{FCFCFF}332596 &
  \cellcolor[HTML]{CBE8D4}3.41 &
  \cellcolor[HTML]{E2F2E8}1.21 &
  \cellcolor[HTML]{D2EBDB}0.40 &
  \cellcolor[HTML]{FCFCFF}0 &
  \cellcolor[HTML]{DAEFE2}0.11 \\
Spider~\cite{dataset-spider} &
  \cellcolor[HTML]{F9FBFC}11840 &
  \cellcolor[HTML]{F0F8F5}6448 &
  \cellcolor[HTML]{FBFCFE}1.8 &
  \cellcolor[HTML]{FCFCFF}206 &
  \cellcolor[HTML]{FCFCFF}1056 &
  \cellcolor[HTML]{F8FBFC}5.13 &
  \cellcolor[HTML]{D1EBDA}5.01 &
  \cellcolor[HTML]{FCFCFF}8980 &
  \cellcolor[HTML]{EBF6F1}1.83 &
  \cellcolor[HTML]{E7F4ED}1.17 &
  \cellcolor[HTML]{B7E0C3}0.54 &
  \cellcolor[HTML]{FCFCFF}0 &
  \cellcolor[HTML]{FCFCFF}0 \\
SParC~\cite{dataset-sparc} &
  \cellcolor[HTML]{F9FBFD}10228 &
  \cellcolor[HTML]{EBF6F1}8981 &
  \cellcolor[HTML]{FCFCFF}1.1 &
  \cellcolor[HTML]{FCFCFF}166 &
  \cellcolor[HTML]{FCFCFF}876 &
  \cellcolor[HTML]{F8FBFC}5.28 &
  \cellcolor[HTML]{CFEAD8}5.14 &
  \cellcolor[HTML]{FCFCFF}9665 &
  \cellcolor[HTML]{F0F8F5}1.58 &
  \cellcolor[HTML]{F0F7F4}1.10 &
  \cellcolor[HTML]{CAE8D4}0.44 &
  \cellcolor[HTML]{FCFCFF}0 &
  \cellcolor[HTML]{FCFCFF}0 \\
CoSQL~\cite{dataset-cosql} &
  \cellcolor[HTML]{FAFBFD}8350 &
  \cellcolor[HTML]{EDF6F2}8007 &
  \cellcolor[HTML]{FCFCFF}1 &
  \cellcolor[HTML]{FCFCFF}166 &
  \cellcolor[HTML]{FCFCFF}876 &
  \cellcolor[HTML]{F8FBFC}5.28 &
  \cellcolor[HTML]{CFEAD8}5.14 &
  \cellcolor[HTML]{FCFCFF}9665 &
  \cellcolor[HTML]{F1F8F6}1.54 &
  \cellcolor[HTML]{EEF7F3}1.11 &
  \cellcolor[HTML]{CEEAD7}0.42 &
  \cellcolor[HTML]{FCFCFF}0 &
  \cellcolor[HTML]{FCFCFF}0 \\
CSpider~\cite{dataset-cspider} &
  \cellcolor[HTML]{F9FBFC}11840 &
  \cellcolor[HTML]{F0F8F5}6408 &
  \cellcolor[HTML]{FBFCFE}1.8 &
  \cellcolor[HTML]{FCFCFF}206 &
  \cellcolor[HTML]{FCFCFF}1056 &
  \cellcolor[HTML]{F8FBFC}5.13 &
  \cellcolor[HTML]{D1EBDA}5.01 &
  \cellcolor[HTML]{FCFCFF}8980 &
  \cellcolor[HTML]{EBF6F1}1.83 &
  \cellcolor[HTML]{E7F4ED}1.17 &
  \cellcolor[HTML]{B7E0C3}0.54 &
  \cellcolor[HTML]{FCFCFF}0 &
  \cellcolor[HTML]{FCFCFF}0 \\
MIMICSQL~\cite{dataset-mimicsql} &
  \cellcolor[HTML]{F6FAFA}20000 &
  \cellcolor[HTML]{E9F5EF}10000 &
  \cellcolor[HTML]{FBFCFE}2 &
  - &
  - &
  - &
  - &
  - &
  \cellcolor[HTML]{EDF6F2}1.74 &
  \cellcolor[HTML]{FCFCFF}1 &
  \cellcolor[HTML]{7DC991}0.84 &
  \cellcolor[HTML]{FCFCFF}0 &
  \cellcolor[HTML]{FCFCFF}0 \\
SQUALL~\cite{dataset-squall} &
  \cellcolor[HTML]{F9FBFC}11276 &
  \cellcolor[HTML]{EDF6F2}8296 &
  \cellcolor[HTML]{FCFCFF}1.4 &
  \cellcolor[HTML]{FCFCFF}2108 &
  \cellcolor[HTML]{FCFCFF}4028 &
  \cellcolor[HTML]{FCFCFF}1.91 &
  \cellcolor[HTML]{A3D8B2}9.18 &
  \cellcolor[HTML]{FCFCFF}71 &
  \cellcolor[HTML]{F8FBFC}1.22 &
  \cellcolor[HTML]{D7EDDF}1.29 &
  \cellcolor[HTML]{D2EBDB}0.40 &
  \cellcolor[HTML]{F3F9F7}0.03 &
  \cellcolor[HTML]{CBE8D4}0.16 \\
FIBEN~\cite{sen2020athena++} &
  \cellcolor[HTML]{FCFCFF}300 &
  \cellcolor[HTML]{FCFCFF}233 &
  \cellcolor[HTML]{FCFCFF}1.3 &
  \cellcolor[HTML]{FCFCFF}1 &
  \cellcolor[HTML]{FCFCFF}152 &
  \cellcolor[HTML]{63BE7B}152 &
  \cellcolor[HTML]{FCFCFF}2.46 &
  \cellcolor[HTML]{F0F8F5}11668125 &
  \cellcolor[HTML]{9DD6AE}5.59 &
  \cellcolor[HTML]{B4DFC1}1.56 &
  \cellcolor[HTML]{63BE7B}0.97 &
  \cellcolor[HTML]{FCFCFF}0 &
  \cellcolor[HTML]{F0F7F5}0.04 \\
ViText2SQL~\cite{dataset-vitext2sql} &
  \cellcolor[HTML]{FAFBFD}9693 &
  \cellcolor[HTML]{F3F8F7}5223 &
  \cellcolor[HTML]{FBFCFE}1.9 &
  \cellcolor[HTML]{FCFCFF}166 &
  \cellcolor[HTML]{FCFCFF}876 &
  \cellcolor[HTML]{F8FBFC}5.28 &
  \cellcolor[HTML]{CFEAD8}5.14 &
  \cellcolor[HTML]{FCFCFF}9665 &
  \cellcolor[HTML]{F9FBFC}1.17 &
  \cellcolor[HTML]{EDF6F2}1.12 &
  \cellcolor[HTML]{B7E0C3}0.54 &
  \cellcolor[HTML]{FCFCFF}0 &
  \cellcolor[HTML]{FCFCFF}0 \\
DuSQL~\cite{dataset-dusql} &
  \cellcolor[HTML]{F5F9F9}25003 &
  \cellcolor[HTML]{D6EDDE}20308 &
  \cellcolor[HTML]{FCFCFF}1.2 &
  \cellcolor[HTML]{FCFCFF}208 &
  \cellcolor[HTML]{FCFCFF}840 &
  \cellcolor[HTML]{F9FBFD}4.04 &
  \cellcolor[HTML]{CDE9D7}5.29 &
  \cellcolor[HTML]{FCFCFF}20 &
  \cellcolor[HTML]{F2F8F7}1.49 &
  \cellcolor[HTML]{DCEFE4}1.25 &
  \cellcolor[HTML]{92D1A4}0.73 &
  \cellcolor[HTML]{FCFCFF}0 &
  \cellcolor[HTML]{9FD7AF}0.30 \\
PortugueseSpider~\cite{dataset-portuguese-spider} &
  \cellcolor[HTML]{FAFBFD}9693 &
  \cellcolor[HTML]{F2F8F7}5275 &
  \cellcolor[HTML]{FBFCFE}1.8 &
  \cellcolor[HTML]{FCFCFF}166 &
  \cellcolor[HTML]{FCFCFF}876 &
  \cellcolor[HTML]{F8FBFC}5.28 &
  \cellcolor[HTML]{CFEAD8}5.14 &
  \cellcolor[HTML]{FCFCFF}9665 &
  \cellcolor[HTML]{EBF5F0}1.85 &
  \cellcolor[HTML]{E7F4ED}1.17 &
  \cellcolor[HTML]{B7E0C3}0.54 &
  \cellcolor[HTML]{FCFCFF}0 &
  \cellcolor[HTML]{FCFCFF}0 \\
CHASE~\cite{dataset-chase} &
  \cellcolor[HTML]{F8FBFB}15408 &
  \cellcolor[HTML]{E2F2E9}13900 &
  \cellcolor[HTML]{FCFCFF}1.1 &
  \cellcolor[HTML]{FCFCFF}350 &
  \cellcolor[HTML]{FCFCFF}1609 &
  \cellcolor[HTML]{F9FBFC}4.60 &
  \cellcolor[HTML]{CFEAD8}5.19 &
  \cellcolor[HTML]{FCFCFF}4594 &
  \cellcolor[HTML]{ECF6F1}1.81 &
  \cellcolor[HTML]{E8F4EE}1.16 &
  \cellcolor[HTML]{E3F2EA}0.31 &
  \cellcolor[HTML]{FCFCFF}0 &
  \cellcolor[HTML]{FCFCFF}0 \\
Spider-Syn~\cite{spider-syn} &
  \cellcolor[HTML]{FCFCFF}1034 &
  \cellcolor[HTML]{FBFCFF}550 &
  \cellcolor[HTML]{FBFCFE}1.9 &
  \cellcolor[HTML]{FCFCFF}166 &
  \cellcolor[HTML]{FCFCFF}876 &
  \cellcolor[HTML]{F8FBFC}5.28 &
  \cellcolor[HTML]{CFEAD8}5.14 &
  \cellcolor[HTML]{FCFCFF}9665 &
  \cellcolor[HTML]{EEF7F3}1.68 &
  \cellcolor[HTML]{E7F4ED}1.17 &
  \cellcolor[HTML]{ADDCBB}0.59 &
  \cellcolor[HTML]{FCFCFF}0 &
  \cellcolor[HTML]{FCFCFF}0 \\
Spider-DK~\cite{dataset-spider-dk} &
  \cellcolor[HTML]{FCFCFF}535 &
  \cellcolor[HTML]{FCFCFF}283 &
  \cellcolor[HTML]{FBFCFE}1.9 &
  \cellcolor[HTML]{FCFCFF}169 &
  \cellcolor[HTML]{FCFCFF}887 &
  \cellcolor[HTML]{F8FBFC}5.25 &
  \cellcolor[HTML]{CFEAD8}5.14 &
  \cellcolor[HTML]{FCFCFF}9494 &
  \cellcolor[HTML]{EEF7F3}1.71 &
  \cellcolor[HTML]{E8F4EE}1.16 &
  \cellcolor[HTML]{B7E0C3}0.54 &
  \cellcolor[HTML]{FCFCFF}0 &
  \cellcolor[HTML]{FCFCFF}0 \\
Spider-Realistic~\cite{dataset-spider-realistic} &
  \cellcolor[HTML]{FCFCFF}508 &
  \cellcolor[HTML]{FCFCFF}290 &
  \cellcolor[HTML]{FBFCFE}1.8 &
  \cellcolor[HTML]{FCFCFF}166 &
  \cellcolor[HTML]{FCFCFF}876 &
  \cellcolor[HTML]{F8FBFC}5.28 &
  \cellcolor[HTML]{CFEAD8}5.14 &
  \cellcolor[HTML]{FCFCFF}9665 &
  \cellcolor[HTML]{ECF6F1}1.79 &
  \cellcolor[HTML]{E2F2E8}1.21 &
  \cellcolor[HTML]{BFE3CA}0.50 &
  \cellcolor[HTML]{FCFCFF}0 &
  \cellcolor[HTML]{FCFCFF}0 \\
KaggleDBQA~\cite{dataset-kaggledbqa} &
  \cellcolor[HTML]{FCFCFF}272 &
  \cellcolor[HTML]{FCFCFF}249 &
  \cellcolor[HTML]{FCFCFF}1.1 &
  \cellcolor[HTML]{FCFCFF}8 &
  \cellcolor[HTML]{FCFCFF}17 &
  \cellcolor[HTML]{FBFCFF}2.12 &
  \cellcolor[HTML]{94D2A5}10.53 &
  \cellcolor[HTML]{FCFCFF}595075 &
  \cellcolor[HTML]{F7FAFB}1.25 &
  \cellcolor[HTML]{F6FAFA}1.05 &
  \cellcolor[HTML]{9AD4AA}0.69 &
  \cellcolor[HTML]{FCFCFF}0 &
  \cellcolor[HTML]{F0F7F5}0.04 \\
SEDE~\cite{dataset-sede} &
  \cellcolor[HTML]{F9FBFC}12023 &
  \cellcolor[HTML]{E7F4ED}11421 &
  \cellcolor[HTML]{FCFCFF}1.1 &
  \cellcolor[HTML]{FCFCFF}1 &
  \cellcolor[HTML]{FCFCFF}29 &
  \cellcolor[HTML]{E0F1E7}29 &
  \cellcolor[HTML]{B8E1C4}7.28 &
  - &
  \cellcolor[HTML]{EAF5EF}1.90 &
  \cellcolor[HTML]{D7EDDF}1.29 &
  \cellcolor[HTML]{69C181}0.94 &
  \cellcolor[HTML]{B8E1C4}0.49 &
  \cellcolor[HTML]{9FD7AF}0.49 \\
MT-TEQL~\cite{dataset-mt-teql} &
  \cellcolor[HTML]{63BE7B}489076 &
  \cellcolor[HTML]{F4F9F8}4525 &
  \cellcolor[HTML]{63BE7B}108.1 &
  \cellcolor[HTML]{63BE7B}489076 &
  \cellcolor[HTML]{63BE7B}3279004 &
  \cellcolor[HTML]{F7FAFB}6.70 &
  \cellcolor[HTML]{CBE8D5}5.51 &
  - &
  \cellcolor[HTML]{EEF7F3}1.69 &
  \cellcolor[HTML]{E9F5EF}1.15 &
  \cellcolor[HTML]{B9E1C5}0.53 &
  \cellcolor[HTML]{FCFCFF}0 &
  \cellcolor[HTML]{FCFCFF}0 \\
PAUQ~\cite{dataset-pauq} &
  \cellcolor[HTML]{F9FBFD}9876 &
  \cellcolor[HTML]{F2F8F6}5497 &
  \cellcolor[HTML]{FBFCFE}1.8 &
  \cellcolor[HTML]{FCFCFF}166 &
  \cellcolor[HTML]{FCFCFF}876 &
  \cellcolor[HTML]{F8FBFC}5.28 &
  \cellcolor[HTML]{CFEAD8}5.14 &
  \cellcolor[HTML]{FCFCFF}9693 &
  \cellcolor[HTML]{ECF6F1}1.82 &
  \cellcolor[HTML]{E7F4ED}1.17 &
  \cellcolor[HTML]{B9E1C5}0.53 &
  \cellcolor[HTML]{FCFCFF}0 &
  \cellcolor[HTML]{FCFCFF}0 \\
knowSQL~\cite{dou2023towards} &
  \cellcolor[HTML]{F4F9F8}28468 &
  - &
  - &
  \cellcolor[HTML]{FCFCFF}488 &
  - &
  - &
  - &
  - &
  - &
  - &
  - &
  - &
  - \\
Dr.Spider~\cite{dataset-dr-spider} &
  \cellcolor[HTML]{F8FBFB}15269 &
  \cellcolor[HTML]{F5FAF9}3847 &
  \cellcolor[HTML]{F8FBFC}4 &
  \cellcolor[HTML]{FCFCFF}549 &
  \cellcolor[HTML]{FCFCFF}2197 &
  \cellcolor[HTML]{F9FBFD}4 &
  \cellcolor[HTML]{CBE8D5}5.54 &
  \cellcolor[HTML]{FCFCFF}28460 &
  \cellcolor[HTML]{ECF6F1}1.81 &
  \cellcolor[HTML]{E4F3EA}1.19 &
  \cellcolor[HTML]{BBE2C7}0.52 &
  \cellcolor[HTML]{FCFCFF}0 &
  \cellcolor[HTML]{FCFCFF}0 \\
BIRD~\cite{dataset-bird} &
  \cellcolor[HTML]{F9FBFD}10962 &
  \cellcolor[HTML]{E8F4EE}10840 &
  \cellcolor[HTML]{FCFCFF}1 &
  \cellcolor[HTML]{FCFCFF}80 &
  \cellcolor[HTML]{FCFCFF}611 &
  \cellcolor[HTML]{F6FAFA}7.64 &
  \cellcolor[HTML]{B9E1C5}7.14 &
  \cellcolor[HTML]{F8FBFB}4585335 &
  \cellcolor[HTML]{E6F4EC}2.07 &
  \cellcolor[HTML]{F1F8F6}1.09 &
  \cellcolor[HTML]{A9DBB8}0.61 &
  \cellcolor[HTML]{BEE3CA}0.20 &
  \cellcolor[HTML]{9FD7AF}0.27 \\
AmbiQT~\cite{dataset-ambiqt} &
  \cellcolor[HTML]{FCFCFF}3046 &
  \cellcolor[HTML]{F7FAFA}3128 &
  \cellcolor[HTML]{FCFCFF}1 &
  \cellcolor[HTML]{FCFCFF}166 &
  \cellcolor[HTML]{FCFCFF}876 &
  \cellcolor[HTML]{F8FBFC}5.28 &
  \cellcolor[HTML]{CFEAD8}5.14 &
  \cellcolor[HTML]{FCFCFF}9665 &
  \cellcolor[HTML]{EBF5F0}1.85 &
  \cellcolor[HTML]{E7F4ED}1.17 &
  \cellcolor[HTML]{BDE3C8}0.51 &
  \cellcolor[HTML]{FCFCFF}0 &
  \cellcolor[HTML]{F9FBFD}0.01 \\
ScienceBenchmark~\cite{dataset-sciencebenchmark} &
  \cellcolor[HTML]{FBFCFE}5031 &
  \cellcolor[HTML]{F6FAFA}3654 &
  \cellcolor[HTML]{FCFCFF}1.4 &
  - &
  - &
  - &
  - &
  - &
  \cellcolor[HTML]{F3F9F7}1.45 &
  \cellcolor[HTML]{FCFCFF}1 &
  \cellcolor[HTML]{F1F8F5}0.24 &
  \cellcolor[HTML]{FCFCFF}0 &
  \cellcolor[HTML]{E7F4ED}0.07 \\
BULL~\cite{dataset-bull} &
  \cellcolor[HTML]{FAFCFD}7932 &
  \cellcolor[HTML]{F1F8F6}5864 &
  \cellcolor[HTML]{FCFCFF}1.4 &
  \cellcolor[HTML]{FCFCFF}3 &
  \cellcolor[HTML]{FCFCFF}78 &
  \cellcolor[HTML]{E3F2EA}26 &
  \cellcolor[HTML]{94D2A5}14.96 &
  \cellcolor[HTML]{FCFCFF}85631 &
  \cellcolor[HTML]{F8FBFC}1.22 &
  \cellcolor[HTML]{FCFCFF}1 &
  \cellcolor[HTML]{FCFCFF}0.18 &
  \cellcolor[HTML]{B8E1C4}0.42 &
  \cellcolor[HTML]{EDF6F2}0.05 \\
BookSQL~\cite{dataset-booksql} &
  \cellcolor[HTML]{E4F3EA}78433 &
  \cellcolor[HTML]{B1DEBF}39530 &
  \cellcolor[HTML]{FBFCFE}2 &
  \cellcolor[HTML]{FCFCFF}1 &
  \cellcolor[HTML]{FCFCFF}7 &
  \cellcolor[HTML]{F6FAFA}7 &
  \cellcolor[HTML]{A6DAB5}8.86 &
  \cellcolor[HTML]{FBFCFF}1012948 &
  \cellcolor[HTML]{F7FAFB}1.25 &
  \cellcolor[HTML]{EDF6F2}1.12 &
  \cellcolor[HTML]{88CD9B}0.78 &
  \cellcolor[HTML]{B8E1C4}0.39 &
  \cellcolor[HTML]{B8E1C4}0.22 \\ 
Archer~\cite{dataset-archer} &
  \cellcolor[HTML]{FCFCFF}518 &
  \cellcolor[HTML]{FCFCFF}260 &
  \cellcolor[HTML]{FBFCFE}2 &
  \cellcolor[HTML]{FCFCFF}10 &
  \cellcolor[HTML]{FCFCFF}68 &
  \cellcolor[HTML]{F7FAFA}6.8 &
  \cellcolor[HTML]{BDE3C9}6.81 &
  \cellcolor[HTML]{FCFCFF}31365.3 &
  \cellcolor[HTML]{C1E4CC}3.89 &
  \cellcolor[HTML]{97D3A8}3.07 &
  \cellcolor[HTML]{69C181}1.77 &
  \cellcolor[HTML]{DDF0E5}0.1 &
  \cellcolor[HTML]{63BE7B}3.55 \\
Spider2-Lite~\cite{dataset-spider2-lite} &
  \cellcolor[HTML]{FCFCFF}527 &
  \cellcolor[HTML]{FCFCFF}527 &
  \cellcolor[HTML]{FBFCFE}1 &
  \cellcolor[HTML]{FCFCFF}264 &
  \cellcolor[HTML]{FCFCFF}6259 &
  \cellcolor[HTML]{E5F3EC}23.71 &
  \cellcolor[HTML]{63BE7B}35.61 &
  \cellcolor[HTML]{FCFCFF}- &
  \cellcolor[HTML]{8ACE9D}6.53 &
  \cellcolor[HTML]{63BE7B}5.10 &
  \cellcolor[HTML]{63BE7B}3.57 &
  \cellcolor[HTML]{63BE7B}1.60 &
  \cellcolor[HTML]{7EC992}2.94 \\ \hline
\end{tabular}%
}
\end{table*}

\stitle{Single-Domain Text-to-SQL Datasets.}
Early \nlsql datasets focused on specific domains with relatively simple \sql queries, such as ATIS~\cite{dataset-atis} for flight information and GeoQuery~\cite{dataset-geoquery} for U.S. geographical facts. Recently, larger single-domain datasets~\cite{dataset-mimicsql, sen2020athena++, dataset-sede, dataset-sciencebenchmark, dataset-bull, dataset-booksql} have been introduced, featuring more complex databases and \sql queries tailored to specific scenarios. This shift reflects an increased emphasis on assessing \nlsql systems' performance and practical utility within particular domains.

\stitle{Cross-Domain Text-to-SQL Datasets.}
After the development of early single-domain datasets, the \nlsql field shifted toward cross-domain datasets to test systems' generalization across diverse \sql queries and databases.  WikiSQL~\cite{dataset-wikisql} was the first cross-domain dataset, drawing tables from Wikipedia across various domains. Subsequently, Spider~\cite{dataset-spider} was introduced, containing more complex relational databases with multiple tables. Recently, BIRD~\cite{dataset-bird} has further advanced complexity by including \sql functions and operations absent in Spider, providing a greater challenge for \nlsql.

\stitle{Multi-Turn Text-to-SQL Datasets.}
With advancements in \nlsql, multi-turn datasets have been developed to support interactive dialogues. SParC~\cite{dataset-sparc} is a cross-domain and multi-turn dataset with about $4.3K$ \nlq questions, totaling over $12K$ (\nlq, \sql) pairs, each \nlq question requiring contextual understanding across turns. CoSQL~\cite{dataset-cosql}, collected using a Wizard-of-Oz setup, includes over $30K$ turns and introduces additional challenges like unanswerable questions, further testing context comprehension. %

\stitle{Text-to-SQL Datasets with Robustness Testing.}
In real-world applications, \nlsql systems must handle diverse user groups and databases, emphasizing robustness.
Spider-Syn~\cite{spider-syn} simulates user unfamiliarity with schemas by using synonyms in \nlq questions, while Dr.Spider~\cite{dataset-dr-spider} applies 17 types of perturbations to databases, \nlq questions, and \sql queries for comprehensive robustness evaluation.

\stitle{Text-to-SQL Datasets with SQL Efficiency Testing.}
Real-world databases often hold vast amounts of data, and a single \nlq may correspond to multiple \sql queries with different execution efficiencies. BIRD~\cite{dataset-bird} introduces a metric for evaluating SQL execution efficiency called the Valid Efficiency Score (VES), which will be further discussed in Section~\ref{sec:evaluation}.

\stitle{Knowledge-Augmented Text-to-SQL Datasets.}
Domain-specific knowledge is essential for \nlsql systems to perform well in real-world applications.
KaggleDBQA~\cite{dataset-kaggledbqa} includes database documents, such as column and table descriptions. Similarly, Spider-DK~\cite{dataset-spider-dk} expands the Spider development set by adding five types of domain knowledge to \nlq questions, testing systems' ability to use this information.

\stitle{Text-to-SQL Datasets with Ambiguous Questions.}
In real-world \nlsql tasks, ambiguities often arise, such as semantic ambiguities in \nlq and overlapping database schemas, making ambiguity-focused evaluation increasingly important. AmbiQT~\cite{dataset-ambiqt} is the first dataset designed to assess ambiguity coverage, comprising four ambiguity types. Each \nlq question maps to two valid \sql queries, reflecting specific ambiguities.

\stitle{Synthetic Text-to-SQL Datasets.} 
MIMICSQL~\cite{dataset-mimicsql} employs a template-based approach to generate initial template questions and corresponding \sql queries, though manual refinement is required to make questions more natural. ScienceBenchmark~\cite{dataset-sciencebenchmark} also uses templates for initial \sql generation but leverages GPT-3 for \sql-to-\nlq translation.

\subsection{In-depth Analysis of Existing Text-to-SQL Datasets}
\label{subsec:dataset_stats}

To analyze and compare \nlsql datasets complexity, we use the NL2SQL360~\cite{nlsql360} system for statistical evaluation, as shown in Table~\ref{tab:datasets},
We measure the \textit{Redundancy},  including the number of \nlq questions, \sql queries, and their ratio.
DB Complexity covers the total databases, total tables, average tables per database, average columns per table, and average records per database.
Query Complexity measures the average number of tables, {\tt SELECT} keywords, aggregate functions, scalar functions, and mathematical computations in each \sql query.
For datasets without public dev/test splits, such as CHASE~\cite{dataset-chase}, only statistics for public splits are reported. For datasets without publicly available data, like knowSQL~\cite{dou2023towards}, values in Table~\ref{tab:datasets} are marked with ``--''.

From the Redundancy Measure perspective, we observe a trend from early datasets to recent ones where datasets have grown in size.
Specifically, MT-TEQL~\cite{dataset-mt-teql} stands out with the highest number of \nlq questions and the largest ratio of \nlq questions to \sql queries due to its automated transformation of \nlq questions, generating a large volume of variants.

In terms of Database Complexity, the number of databases and tables within each dataset aligns with its intended task. Single-domain datasets, such as BookSQL~\cite{dataset-booksql}, generally contain fewer databases, while those aimed at robustness evaluation, like Dr.Spider~\cite{dataset-dr-spider} and MT-TEQL~\cite{dataset-mt-teql}, include a larger number of databases.

Regarding Query Complexity, datasets like FIBEN~\cite{sen2020athena++} and SEDE~\cite{dataset-sede} feature \sql queries with multiple tables and aggregate functions, mirroring complexities in the real-world financial domains and Stack Exchange sites. Recent datasets also emphasize Scalar Functions and Mathematical Computations, adding structural challenges.

\etitle{Discussion.} Despite the increasing number of datasets proposed by the \nlsql community, a gap in \sql complexity remains compared to real-world scenarios. Current datasets typically feature fewer {\tt SELECT} keywords, indicating a lack of nested queries and complex set operations. Additionally, challenges involving Scalar Functions and Mathematical Computations require further focus. We encourage the community to propose datasets addressing these complexities.

\section{Evaluation and Error Analysis}
\label{sec:evaluation}

In this section, we introduce key evaluation metrics for \nlsql solutions (Section~\ref{subsec:metrics}), review toolkits for low-cost and comprehensive evaluation (Section~\ref{subsec:eval_tool}),  
and provide an error taxonomy for analyzing \sql errors in the \nlsql process (Section~\ref{subsec:errors}).

\subsection{Evaluation Metrics}
\label{subsec:metrics}

Evaluation metrics are crucial for measuring \nlsql performance.
We define $N$ as the dataset size, $Q_i$ as the \nlq question of the $i$-th example, $V_i$ as the execution result set of the ground-truth \sql query $Y_i$ and $\hat{V}_i$ as the execution result set of the \sql query $\hat{Y}_i$ generated by the \nlsql solution.

\etitle{Execution Accuracy (EX)~\cite{dataset-spider}.} %
This metric evaluates the performance of the \nlsql system by comparing whether the execution result sets of the ground-truth \sql queries and the predicted \sql queries are identical. It can be computed by:
%
    $EX = \frac{ \sum_{i=1}^{N} \mathbbm{1} ( V_i = \hat{V}_i ) }{N},$
where $\mathbbm{1}(\cdot)$ is an indicator function that equals 1 if the condition inside is satisfied, and 0 otherwise.  Note that false negatives could occur because different \sql queries corresponding to semantically different \nlq queries may produce identical execution result sets. 

\etitle{String-Match Accuracy (SM)~\cite{dataset-wikisql}.} %
This metric, also called Logical Form Accuracy, simply compares whether the ground-truth \sql query and the predicted \sql query are identical as strings. It may penalize \sql queries that produce the correct execution result sets but do not have the exact string match with the ground-truth \sql queries. It can be computed as follows:
    $SM = \frac{ (\sum_{i=1}^{N} \mathbbm{1} ( Y_i = \hat{Y}_i )) }{N}.$

\etitle{Component-Match Accuracy (CM)~\cite{dataset-spider}.} %
This metric evaluates the detailed performance of the \nlsql system by measuring the exact matching of different \sql components such as {\tt SELECT}, {\tt WHERE} and others between the ground-truth \sql query and the predicted \sql query. For a specific \sql component $C$. The computation can be formalized as follows:
\begin{equation*}
    CM^C = \frac{ \sum_{i=1}^{N} \mathbbm{1} ( Y_i^C = \hat{Y}_i^C ) }{N},
\end{equation*}
where $Y_i^C$ is the component of \sql query $Y_i$. To correctly determine if an \sql component matches, some \sql components ({\eg {\tt WHERE}}) do not consider order constraints.

\etitle{Exact-Match Accuracy (EM)~\cite{dataset-spider}.} %
This metric is based on the Component-Match Accuracy (CM) and measures whether all \sql components $\mathbb{C} = \{C_k\}$ of the predicted \sql query match the ground-truth \sql query. It can be computed as follows:
\begin{equation*}
    EM = \frac{ \sum_{i=1}^{N} \mathbbm{1} ( \bigwedge_{C_k \in \mathbb{C}} Y_i^{C_k} = \hat{Y}_i^{C_k} ) }{N}.
\end{equation*}

\etitle{Valid Efficiency Score (VES)~\cite{dataset-bird}.} %
This metric measures the execution efficiency of valid \sql queries. It considers both the accuracy and efficiency of \sql execution, which can be computed as follows:
\begin{equation*} \small
    VES = \frac{ \sum_{i=1}^{N} \mathbbm{1} ( V_i = \hat{V}_i ) \cdot \mathbf{R} (Y_i, \hat{Y_i}) }{N},~
    \mathbf{R}(Y_i, \hat{Y_i}) = \sqrt{\frac{\mathbf{E}(Y_i)}{\mathbf{E}(\hat{Y}_i)}},
\end{equation*}
where $\mathbf{R}(\cdot)$ measures the relative execution efficiency of the predicted \sql query compared to the ground-truth \sql query, eliminating uncertainties due to machine status. $\mathbf{E}(\cdot)$ measures the efficiency of specific \sql query, which can refer to execution time, memory usage and more.

\etitle{Query Variance Testing (QVT)~\cite{nlsql360}.} %
This metric measures the robustness of a \nlsql system in handling variations in \nlq queries.  For a given \sql query $Y_i$, there are often multiple corresponding \nlq queries, represented as pairs  \{($Q_1$, $Y_i$), ($Q_2$, $Y_i$), \ldots, ($Q_m$, $Y_i$)\}. The QVT metric is calculated as:
\begin{equation*}
QVT = \frac{1}{N} \sum_{i=1}^{N} \left( \frac{\sum_{j=1}^{m_i} \mathbbm{1} \left( \mathcal{F}(Q_{ij}) = Y_i \right)}{m_i} \right),
\end{equation*}
where $m_i$ is the number of different \nlq variations for the \sql query $Y_i$, and $\mathcal{F}(Q_{ij})$ is the predicted \sql query for the $j$-th \nlq variation of $Y_i$.

\subsection{Text-to-SQL Evaluation Toolkits}
\label{subsec:eval_tool}

Recent \nlsql solutions have achieved remarkable performance on various \nlsql benchmarks. However, in real-world applications, variations in \nlq query styles, database schemas, and \sql query characteristics across domains make it difficult to fully assess system robustness using standard benchmark metrics alone. To address this, recent toolkits~\cite{dataset-mt-teql, nlsql360} have been developed to provide a more comprehensive evaluation of \nlsql systems in practical scenarios.

MT-TEQL~\cite{dataset-mt-teql} is a unified framework for evaluating the performance of \nlsql systems in handling real-world variations in \nlq queries and database schemas. It is based on a metamorphic testing approach, implementing semantic-preserving transformations of \nlq queries and database schemas to generate their variants without manual efforts automatically. 
It includes four types of transformations for \nlq queries (e.g. {\em Prefix Insertion}) and eight types of transformations for database schemas: (e.g. {\em Table Shuffle}).

NL2SQL360~\cite{nlsql360} is a multi-angle evaluation framework offering fine-grained assessments of \nlsql systems across diverse scenarios (Figure~\ref{fig:overview}(c)). Unlike MT-TEQL, it emphasizes varied \sql query characteristics in different applications, such as aggregate functions, nested queries, or top-$k$ queries typical of the Business Intelligence scenario. Comprising six core components,  {\em Dataset}, {\em Model Zoo}, {\em Metrics}, {\em Dataset Filter}, {\em Evaluator}, and {\em Analysis}, NL2SQL360 provides a unified, model-agnostic interface for systematic evaluations. Users can apply both public and private datasets, customize metrics for specific scenarios, and analyze performance on subsets with scenario-specific \sql characteristics, offering valuable insights into \nlsql system effectiveness across applications.

\subsection{A Taxonomy for Text-to-SQL Errors Analysis}
\label{subsec:errors}

Error analysis involves examining model errors to identify limitations and guide corrective actions for improved performance.
In this section, we first review the existing \nlsql error taxonomies. We then propose design principles and introduce a two-level \nlsql errors taxonomy.

\stitle{Existing Taxonomies for Text-to-SQL Errors Analysis.}
Recent \nlsql research~\cite{liu2025nl2sql,narechania2021diy, ning2023empirical, talaei2024chess, pourreza2024din, sun2023sql} has increasingly incorporated error analysis, proposing various error taxonomies. 
Ning et al.~\cite{ning2023empirical} introduced a detailed error taxonomy based on two dimensions:
(1)~\textit{Syntactic dimension} identifies specific \sql parts where errors occur, organized by keywords such as \texttt{WHERE} and \texttt{JOIN}.
(2)~\textit{Semantic dimension} indicates misinterpretations of the natural language description, such as errors in understanding table names. SQL-PaLM~\cite{sun2023sql} categorizes errors into five types: 
(1)~\textit{Schema Linking}, irrelevant or missing table/column selection; 
(2)~\textit{Database Content}, misinterpreting data values; 
(3)~\textit{Knowledge Evidence}, failing to utilize external hints; 
(4)~\textit{Reasoning}, lacking intermediate logical steps; and 
(5)~\textit{Syntax}, invalid SQL format.
 NL2SQL-BUGs~\cite{liu2025nl2sql} focuses on the analysis of semantic errors, organizing them into 9 main categories and 31 subcategories. It further proposes a new benchmark for evaluating models’ error detection capabilities, advancing automated error analysis in \nlsql.

\begin{figure*}[t!]
	\centering
	\includegraphics[width=\linewidth]{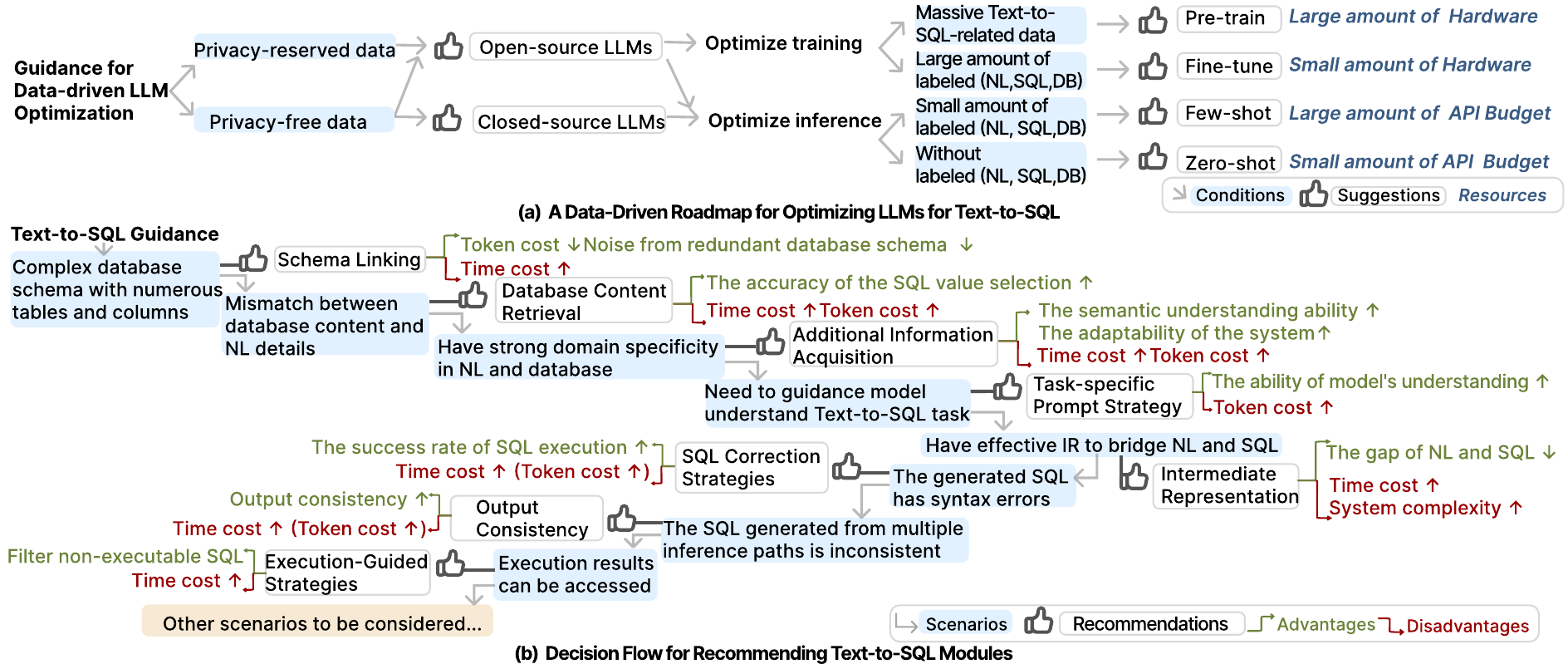}
	\caption{A Data-Driven Roadmap and a Decision Flow for Recommending Text-to-SQL modules.}
	\label{fig:NL2SQL_Guidance}
\end{figure*}

\marginpar{AE C2}
\stitle{\addd{Taxonomy Principles for Text-to-SQL Errors Analysis.}}
\addd{Current error taxonomies in \nlsql are often specific to particular datasets, limiting their general applicability.  To address these issues, a standardized and effective taxonomy is essential. We propose the following principles~\cite{taipalus2023framework} to guide the development of a \nlsql error taxonomy:
}
\bi
    \item \addd{\textit{Comprehensiveness}: The taxonomy should cover all possible error types in the \nlsql translation process.}
    \item \addd{\textit{Mutual Exclusivity}: Each error type should be clearly distinct to avoid classification ambiguity.}
    \item \addd{\textit{Extensibility}: The taxonomy should be adaptable to include emerging error types as \nlsql evolves.}
    \item \addd{\textit{Practicality}: It should be practical, enabling users to diagnose and address errors in real-world scenarios.}
\ei

\stitle{\addd{Our Taxonomy for Text-to-SQL Errors Analysis.}}
\addd{
Following these principles, we developed a two-level \nlsql error analysis taxonomy:}
\bi
    \item \addd{\textit{Error Localization}: The first level identifies specific \sql components where errors occur, such as the {\tt SELECT} or {\tt WHERE} clause. Pinpointing error locations enables targeted adjustments and enhances correction efficiency.}
    \item \addd{\textit{Cause of Error}: The second level focuses on the underlying reasons for the error. For instance, errors in the {\tt WHERE} clause values may indicate the model's limitations in database content retrieval or interpretation. }
\ei

\etitle{\addd{Discuss the Application of the Two-level Error Taxonomy.}}
\addd{We collected and classified errors from DIN-SQL~\cite{pourreza2024din} on the Spider~\cite{dataset-spider} using our proposed taxonomy. As shown in Figure~\ref{fig:overview}(d), only \textit{1.8\%} of the errors fall into the Others category, suggesting our taxonomy is practical and effective.}

\addd{Nonetheless, we recognize that developing a complete and universally applicable Text-to-SQL error taxonomy is inherently iterative. We encourage continued efforts from the community to refine and expand this taxonomy over time.}

\section{Practical Guidance for Text-to-SQL}
\label{sec:roadmap}
In this section, we provide practical guidance for developing \nlsql solutions, considering key factors and scenarios.

\subsection{Data-Driven Roadmap for Text-to-SQL}
\label{subsec: NLP_Guidance}
In Figure~\ref{fig:NL2SQL_Guidance}(a), we outline a strategic roadmap designed to optimize LLMs for \nlsql task, based on data privacy and data volume. Data privacy affects the choice of open-source and closed-source LLMs, while data volume affects the strategies for optimization for training and inference.

\etitle{Condition 1: Data Privacy.} 
For privacy-sensitive data, open-source LLMs are preferable, as closed-source models typically use external APIs, potentially exposing data to external servers. Open-source models allow full control over local training and inference, providing stronger data privacy protection.

\etitle{Condition 2: Data Volume.}
For open-source LLMs, optimization is possible in both training and inference phases, while closed-source LLMs allow only inference-stage optimization due to limited access. With extensive \nlsql data, pre-training enhances performance; fine-tuning is suitable for datasets with hundreds to thousands of (\nlq, \sql) pairs. In low-data scenarios, few-shot learning is recommended, while zero-shot methods are essential when labeled data is unavailable. Hardware resources and API costs are also important considerations in selecting the best optimization strategy.

\subsection{Decision Flow of Selecting Text-to-SQL Modules}
\label{subsec: NL2SQL_Guidance}
In Figure~\ref{fig:NL2SQL_Guidance}(b), we present recommendations for choosing \nlsql modules based on specific scenarios, highlighting both benefits and trade-offs. Below, we outline two examples.

\etitle{Scenario 1:} {\em Complex Database Schema with Numerous Tables and Columns.} 
In this case, using Schema Linking strategies is advisable. This reduces token costs and minimizes noise from irrelevant schema elements, enhancing efficiency. However, it also incurs additional time costs.

\etitle{Scenario 2:} {\em  Execution Results Can be Accessed.} 
Here, Execution-Guided Strategies are recommended, as they improve system performance by filtering out non-executable \sql queries. The downside is the increased time required for query execution, which can be substantial with large databases.

In summary, while each module offers unique advantages for specific \nlsql scenarios, it is essential to balance these benefits with the potential drawbacks in system design.

\section{\addd{Limitations and Open Problems}}
\label{sec:openproblem}
\addd{We analyze the limitations of LLM-based methods and propose corresponding open problems, highlighting unresolved challenges and suggesting directions for future research.
}

\marginpar{\text{R3 C2}}
\stitle{\addd{Limitations of Current LLM-based Solutions.}}
\addd{
Although recent LLM-based Text-to-SQL methods have made significant progress, they still face several challenges when dealing with complex queries in real-world scenarios. First, existing methods are typically trained and executed on a single, fixed database, which limits their ability to handle open environments that require cross-database queries and multi-source data aggregation. Second, although LLMs have strong natural language understanding capabilities, they incur high token consumption during inference, leading to high costs and low efficiency. In addition, most Text-to-SQL methods lack interpretability and debugging mechanisms, making it difficult for users to understand how the model generates SQL or to detect and fix potential semantic errors. Finally, current methods show limited adaptability to new domains and rely heavily on high-quality training data; how to automatically generate targeted training samples based on model feedback remains an open problem. These limitations reveal shortcomings in cross-database scalability, inference efficiency, system reliability, and data adaptability, highlighting the need for more efficient, trustworthy, and scalable Text-to-SQL solutions.
}

\stitle{Open-Domain Text-to-SQL Problem.}
In real-world scenarios like government open data platforms, citizens may ask questions that require querying multiple databases and aggregating results.
For example, answering ``What is the average processing time for tax returns in the last five years?'' requires retrieving tables from multiple databases (e.g., \textit{tax records}, \textit{processing logs}, and \textit{statistical reports}) and generating multiple \sql queries over them. Unlike traditional  \nlsql,  where a single target database is specified by the user, Open \nlsql may need to generate multiple \sql queries that access different databases for a single \nlq.

Thus, the Open \nlsql problem introduces unique challenges, including: 
(1) \textit{database retrieval}: accurately identifying and retrieving relevant databases from a vast array of data sources; 
(2) \textit{handling heterogeneous schemas}: integrating data with varied structures and terminologies, requiring advanced schema matching and linking techniques;
(3) \textit{answer aggregation}: inferring final answers from multiple \sql queries across databases, which demands methods to plan query order, resolve conflicts, and ensure consistency;
(4) \textit{domain adaptation}: generalizing models across domains to address differences in terminology and structure;
(5) \textit{scalability and efficiency}: managing large data volumes while maintaining performance; and 
(6) \textit{evaluating and benchmarking}: developing metrics and datasets that accurately reflect real-world complexity for Open \nlsql solutions.

\stitle{\addd{Develop Cost-effective Text-to-SQL Methods.}}  
\marginpar{AE C3}
\marginpar{R1 C2}
\addd{
LLM-based \nlsql methods show great potential but are limited by high token consumption, leading to increased costs and slower inference times. In contrast, PLM-based \nlsql methods excel at handling complex \sql queries and accurately interpreting database schemas.
A promising approach is to combine the strengths of both, developing modular \nlsql solutions or using a multi-agent framework to integrate LLMs and PLMs for the \nlsql task (as shown in Table~\ref{tab:spider_resource_usage}). 
In parallel, efforts have aimed to improve LLM-based efficiency. EllieSQL~\cite{zhu2025elliesqlcostefficienttexttosqlcomplexityaware} employs complexity-aware routing to enhance cost-efficiency by assigning queries to suitable LLM-based generators. 
}

\begin{table}[]
\centering
\caption{\addd{Resource Consumption Statistics on the Spider.}}
\label{tab:spider_resource_usage}
\resizebox{\linewidth}{!}{%
\begin{tabular}{|c|c|c|c|}
\hline
\textbf{Method} &
  \textbf{Base Model} &
  \multicolumn{1}{c|}{\textbf{API Tokens/SQL}} &
  \multicolumn{1}{c|}{\textbf{Latency/SQL (s)}} \\ \hline
\textbf{RESDSQL~\cite{li2023resdsql}}       & PLM     & -              & 1.91          \\ \hline
\textbf{RESDSQL+NatSQL~\cite{li2023resdsql}} & PLM     & -              & 1.97          \\ \hline
\textbf{ZeroNet~\cite{gu2023interleaving}}       & PLM+LLM & 377            & 3.72          \\ \hline
\textbf{DIN-SQL~\cite{pourreza2024din}}       & LLM     & \textbf{3579} & \textbf{10.34} \\ \hline
\end{tabular}%
}
\end{table}

\stitle{Make Text-to-SQL Solutions Trustworthy.} 
Ensuring \nlsql solutions are trustworthy is essential for generating accurate and reliable \sql, mitigating risk, and reducing the need for manual intervention. Topics include the following:

{\em \underline{Interpreting Text-to-SQL Solutions.}}
Understanding the reasoning behind a \nlsql model's performance enhances confidence in its reliability. Explainable AI techniques~\cite{DBLP:journals/inffus/AliAEMACGSRH23, DBLP:journals/tist/ZhaoCYLDCWYD24}, such as surrogate models~\cite{DBLP:journals/corr/abs-2307-08678}
and saliency maps~\cite{DBLP:conf/icml/ShrikumarGK17}, aim to reveal model decisions. However, their effectiveness in \nlsql contexts, especially with combined LLMs and PLMs, remains an open question.
In addition, multi-agent LLM frameworks~\cite{zhang2024aflowautomatingagenticworkflow} improve reliability by splitting \nlsql into specialized sub-tasks. Although this approach improves robustness, coordinating agents to ensure consistent and optimized performance remains a major challenge.

{\em \underline{Text-to-SQL Debugging Tools.}}
Inspired by compiler design, a debugger for \nlsql could improve accuracy and reliability by measuring semantic and syntactic errors in generated \sql queries. Such tools would detect potential errors, enable users to examine the \sql generation process and identify mismatches~\cite{liu2025nl2sql,yang2025automated}. However, achieving this goal presents significant challenges. Traditional code compilers primarily capture syntactic errors, while \nlsql debugging must also address semantic errors, \ie ensuring that the generated \sql query accurately reflects the intent of the \nlq query.

{\em \underline{Interactive Text-to-SQL Tools.}} These tools are essential for empowering professional users (\eg DBAs) to create complex \sql queries that span multiple databases, often exceeding 50 lines of code. A key feature is the model’s ability to decompose complex queries into manageable sub-queries, reducing cognitive load and enabling DBAs to focus on each part before reassembling them. 
Supporting both bottom-up and top-down workflows, such tools enable users to iteratively refine outputs, align \sql generation with intent, and integrate model assistance with domain expertise.

\stitle{Adaptive Training Data Synthesis.}
Learning-based \nlsql models often fail to generalize to unseen domains, partly due to limited training data coverage, quality, and diversity.
Therefore, an interesting research problem is to automatically and incrementally generate (\nlq,~\sql) pairs based on the model performance. 
Specifically, by incorporating insights from evaluation metrics and evaluation results, we can identify specific weaknesses of the model. Using this information, we can synthesize training data that continually evolves with the help of LLMs to cover a broader range of domains. 

\section{Conclusion}
\label{sec:conclusion}

In this paper, we comprehensively review \nlsql techniques from a lifecycle perspective in the LLM era. We formally define the \nlsql task, discuss key challenges, and propose a taxonomy based on underlying language models. We summarize key modules of language model-driven methods, including pre-processing, translation, and post-processing strategies. Furthermore, we analyze benchmarks and evaluation metrics, highlighting their characteristics and common errors. We also offer a practical roadmap for adapting LLMs to \nlsql tasks and maintain an online handbook with the latest advancements, discussing ongoing challenges and open problems.


\section*{Acknowledgments}
This paper was supported by the National Key R\&D Program of China (2023YFB4503600); the NSF of China (62402409, 62525202, 62232009, 62436010, and 62441230); the Guangdong Basic and Applied Basic Research Foundation (2023A1515110545); the Guangzhou Basic and Applied Basic Research Foundation (2025A04J3935); the Guangzhou-HKUST(GZ) Joint Funding Program (2025A03J3714); and the Guangdong provincial project (2023CX10X008).

\bibliographystyle{IEEEtran}
\bibliography{reference} 

\begin{thebibliography}{100}
\providecommand{\url}[1]{#1}
\csname url@samestyle\endcsname
\providecommand{\newblock}{\relax}
\providecommand{\bibinfo}[2]{#2}
\providecommand{\BIBentrySTDinterwordspacing}{\spaceskip=0pt\relax}
\providecommand{\BIBentryALTinterwordstretchfactor}{4}
\providecommand{\BIBentryALTinterwordspacing}{\spaceskip=\fontdimen2\font plus
\BIBentryALTinterwordstretchfactor\fontdimen3\font minus \fontdimen4\font\relax}
\providecommand{\BIBforeignlanguage}[2]{{%
\expandafter\ifx\csname l@#1\endcsname\relax
\typeout{** WARNING: IEEEtran.bst: No hyphenation pattern has been}%
\typeout{** loaded for the language `#1'. Using the pattern for}%
\typeout{** the default language instead.}%
\else
\language=\csname l@#1\endcsname
\fi
#2}}
\providecommand{\BIBdecl}{\relax}
\BIBdecl

\bibitem{gu2023few}
Z.~Gu, J.~Fan, N.~Tang, and et~al., ``Few-shot text-to-sql translation using structure and content prompt learning,'' \emph{SIGMOD}, 2023.

\bibitem{DBLP:conf/acl/ChenCWMPSS023}
Z.~Chen, S.~Chen, M.~White, R.~J. Mooney, and et~al., ``Text-to-sql error correction with language models of code,'' in \emph{{ACL}}, 2023.

\bibitem{DBLP:conf/kdd/WangQHLYWLSHSL22}
L.~Wang and et~al., ``Proton: Probing schema linking information from pre-trained language models for text-to-sql parsing,'' in \emph{{KDD}}, 2022.

\bibitem{DBLP:conf/kdd/LiuH0W22}
A.~Liu, X.~Hu, L.~Lin, and L.~Wen, ``Semantic enhanced text-to-sql parsing via iteratively learning schema linking graph,'' in \emph{{KDD}}, 2022.

\bibitem{pourreza2024din}
M.~Pourreza and D.~Rafiei, ``Din-sql: Decomposed in-context learning of text-to-sql with self-correction,'' \emph{NeurIPS}, 2024.

\bibitem{gao2023text}
D.~Gao, H.~Wang, Y.~Li, and et~al., ``Text-to-sql empowered by large language models: A benchmark evaluation,'' \emph{Proc. {VLDB} Endow.}, 2024.

\bibitem{li2023resdsql}
H.~Li, J.~Zhang, C.~Li, and H.~Chen, ``Resdsql: Decoupling schema linking and skeleton parsing for text-to-sql,'' in \emph{AAAI}, 2023.

\bibitem{DBLP:conf/cidr/0001YF0LH24}
N.~Tang, C.~Yang, J.~Fan, and et~al., ``Verifai: Verified generative {AI},'' in \emph{{CIDR}}.\hskip 1em plus 0.5em minus 0.4em\relax www.cidrdb.org, 2024.

\bibitem{DBLP:journals/corr/abs-2406-07815}
Y.~Zhu, S.~Du, B.~Li, and et~al., ``Are large language models good statisticians?'' \emph{CoRR}, 2024.

\bibitem{DBLP:journals/corr/abs-2406-11033}
Y.~Xie, Y.~Luo, G.~Li, and N.~Tang, ``Haichart: Human and {AI} paired visualization system,'' \emph{Proc. {VLDB} Endow.}, 2023.

\bibitem{DBLP:journals/tvcg/ShenSLYHZTW23}
L.~Shen, E.~Shen, Y.~Luo, and et~al., ``Towards natural language interfaces for data visualization: {A} survey,'' \emph{{IEEE} TVCG.}, 2023.

\bibitem{DBLP:journals/tkde/LuoQCTLL22}
Y.~Luo, X.~Qin, C.~Chai, and et~al., ``Steerable self-driving data visualization,'' \emph{{IEEE} Trans. Knowl. Data Eng.}, 2022.

\bibitem{DBLP:journals/tvcg/LuoTLTCQ22}
Y.~Luo, N.~Tang, G.~Li, and et~al., ``Natural language to visualization by neural machine translation,'' \emph{{IEEE} Trans. Vis. Comput. Graph.}, 2022.

\bibitem{DBLP:conf/sigmod/TangLOLC22}
J.~Tang, Y.~Luo, M.~Ouzzani, and et~al., ``Sevi: Speech-to-visualization through neural machine translation,'' in \emph{SIGMOD}, 2022.

\bibitem{DBLP:conf/sigmod/Luo00CLQ21}
Y.~Luo, N.~Tang, G.~Li, and et~al., ``Synthesizing natural language to visualization {(NL2VIS)} benchmarks from {NL2SQL} benchmarks,'' in \emph{SIGMOD}, 2021.

\bibitem{DBLP:journals/vldb/QinLTL20}
X.~Qin, Y.~Luo, N.~Tang, and et~al., ``Making data visualization more efficient and effective: a survey,'' \emph{{VLDB} J.}, 2020.

\bibitem{DBLP:conf/icde/LuoQ0018}
Y.~Luo, X.~Qin, N.~Tang, and et~al., ``Deepeye: Towards automatic data visualization,'' in \emph{{ICDE}}.\hskip 1em plus 0.5em minus 0.4em\relax {IEEE} Computer Society, 2018.

\bibitem{DBLP:conf/sigmod/LuoQ00W18}
Y.~Luo, X.~Qin, N.~Tang, G.~Li, and X.~Wang, ``Deepeye: Creating good data visualizations by keyword search,'' in \emph{SIGMOD}.\hskip 1em plus 0.5em minus 0.4em\relax {ACM}, 2018.

\bibitem{DBLP:journals/dase/ZhouSL24}
X.~Zhou, Z.~Sun, and G.~Li, ``{DB-GPT:} large language model meets database,'' \emph{Data Sci. Eng.}, 2024.

\bibitem{DBLP:journals/sigmod/AmerYahiaBCLSXY23}
S.~Amer{-}Yahia and et~al., ``From large language models to databases and back: {A} discussion on research and education,'' \emph{{SIGMOD}}, 2023.

\bibitem{zhang2024natural}
W.~Zhang, Y.~Wang, Y.~Song, and et~al., ``Natural language interfaces for tabular data querying and visualization: A survey,'' \emph{TKDE}, 2024.

\bibitem{katsogiannis2023survey}
G.~Katsogiannis-Meimarakis and G.~Koutrika, ``A survey on deep learning approaches for text-to-sql,'' \emph{The VLDB Journal}, 2023.

\bibitem{deng-etal-2022-recent}
N.~Deng, Y.~Chen, and Y.~Zhang, ``Recent advances in text-to-sql: A survey of what we have and what we expect,'' in \emph{COLING}, 2022.

\bibitem{kim2020natural}
H.~Kim, B.-H. So, W.-S. Han, and et~al., ``Natural language to sql: Where are we today?'' \emph{Proc. {VLDB} Endow.}, 2020.

\bibitem{shi2024surveyemployinglargelanguage}
L.~Shi, Z.~Tang, N.~Zhang, X.~Zhang, and Z.~Yang, ``A survey on employing large language models for text-to-sql tasks,'' 2024.

\bibitem{mohammadjafari2024naturallanguagesqlreview}
A.~Mohammadjafari, A.~S. Maida, and R.~Gottumukkala, ``From natural language to sql: Review of llm-based text-to-sql systems,'' 2024.

\bibitem{zhu2024largelanguagemodelenhanced}
X.~Zhu, Q.~Li, L.~Cui, and Y.~Liu, ``Large language model enhanced text-to-sql generation: A survey,'' 2024.

\bibitem{hong2024next}
Z.~Hong, Z.~Yuan, Q.~Zhang, and et~al., ``Next-generation database interfaces: A survey of llm-based text-to-sql,'' \emph{arXiv:2406.08426}, 2024.

\bibitem{Hozcan2020state}
F.~{\H{O}}zcan, A.~Quamar, J.~Sen, and et~al., ``State of the art and open challenges in natural language interfaces to data,'' in \emph{SIGMOD}, 2020.

\bibitem{DBLP:conf/sigmod/LiR17}
Y.~Li and D.~Rafiei, ``Natural language data management and interfaces: Recent development and open challenges,'' in \emph{SIGMOD}, 2017.

\bibitem{DBLP:journals/pvldb/Katsogiannis-Meimarakis23}
G.~Katsogiannis{-}Meimarakis and et~al., ``Natural language interfaces for databases with deep learning,'' \emph{Proc. {VLDB} Endow.}, 2023.

\bibitem{DBLP:conf/sigmod/Katsogiannis-Meimarakis21}
G.~Katsogiannis-Meimarakis and G.~Koutrika, ``A deep dive into deep learning approaches for text-to-sql systems,'' in \emph{SIGMOD}, 2021.

\bibitem{liu2025nl2sql}
X.~Liu, S.~Shen, B.~Li, and et~al., ``Nl2sql-bugs: A benchmark for detecting semantic errors in nl2sql translation,'' \emph{arXiv}, 2025.

\bibitem{yang2025automated}
Y.~Yang, Z.~Wang, Y.~Xia, Wei, and et~al., ``Automated validation and fixing of text-to-sql translation with execution consistency,'' 2025.

\bibitem{DBLP:journals/corr/abs-2204-00498}
N.~Rajkumar and et~al., ``Evaluating the text-to-sql capabilities of large language models,'' \emph{CoRR}, 2022.

\bibitem{10.1145/2588555.2594519}
F.~Li and H.~V. Jagadish, ``Nalir: an interactive natural language interface for querying relational databases,'' in \emph{ACM SIGMOD}, 2014.

\bibitem{yu2021grappa}
T.~Yu, C.-S. Wu, X.~V. Lin, and et~al., ``Grappa: Grammar-augmented pre-training for table semantic parsing,'' in \emph{ICLR}, 2021.

\bibitem{xiao2016sequence}
C.~Xiao, M.~Dymetman, and C.~Gardent, ``Sequence-based structured prediction for semantic parsing,'' in \emph{ACL}, 2016.

\bibitem{lin2019grammar}
K.~Lin, B.~Bogin, M.~Neumann, and et~al., ``Grammar-based neural text-to-sql generation,'' \emph{arXiv:1905.13326}, 2019.

\bibitem{bogin2019representing}
B.~Bogin, M.~Gardner, and et~al., ``Representing schema structure with graph neural networks for text-to-sql parsing,'' \emph{arXiv:1905.06241}, 2019.

\bibitem{devlin-etal-2019-bert}
J.~Devlin and et~al., ``Bert: Pre-training of deep bidirectional transformers for language understanding,'' in \emph{NAACL}, 2019.

\bibitem{raffel2020exploring}
C.~Raffel, N.~Shazeer, A.~Roberts, and et~al., ``Exploring the limits of transfer learning with a unified text-to-text transformer,'' \emph{JMLR}, 2020.

\bibitem{li2023graphix}
J.~Li and et~al., ``Graphix-t5: Mixing pre-trained transformers with graph-aware layers for text-to-sql parsing,'' \emph{arXiv:2301.07507}, 2023.

\bibitem{gu2023interleaving}
Z.~Gu, J.~Fan, and et~al., ``Interleaving pre-trained language models and large language models for zero-shot nl2sql generation.'' \emph{arXiv}, 2023.

\bibitem{dataset-spider}
T.~Yu, R.~Zhang, K.~Yang, and et~al., ``Spider: A large-scale human-labeled dataset for complex and cross-domain semantic parsing and text-to-sql task,'' in \emph{EMNLP}, 2018.

\bibitem{nlsql360}
B.~Li, Y.~Luo, C.~Chai, and et~al., ``The dawn of natural language to sql: Are we fully ready?'' \emph{Proc. {VLDB} Endow.}, 2024.

\bibitem{zhao2023survey}
W.~X. Zhao, K.~Zhou, J.~Li, and et~al., ``A survey of large language models,'' \emph{arXiv:2303.18223}, 2023.

\bibitem{minaee2024largelanguagemodelssurvey}
S.~Minaee, T.~Mikolov, N.~Nikzad, and et~al., ``Large language models: A survey,'' \emph{arXiv:2402.06196}, 2024.

\bibitem{li2024codes}
H.~Li, J.~Zhang, H.~Liu, and et~al., ``Codes: Towards building open-source language models for text-to-sql,'' \emph{SIGMOD}, 2024.

\bibitem{dataset-bull}
C.~Zhang and et~al., ``Finsql: Model-agnostic llms-based text-to-sql framework for financial analysis,'' in \emph{SIGMOD}, 2024.

\bibitem{li2023starcoder}
R.~Li, L.~B. Allal, Y.~Zi, and et~al., ``Starcoder: may the source be with you!'' \emph{arXiv:2305.06161}, 2023.

\bibitem{dataset-bird}
J.~Li and et~al., ``Can {LLM} already serve as {A} database interface? {A} big bench for large-scale database grounded text-to-sqls,'' in \emph{NIPS}, 2023.

\bibitem{lei2020re}
W.~Lei, W.~Wang, Z.~Ma, and et~al., ``Re-examining the role of schema linking in text-to-sql,'' in \emph{EMNLP}, 2020.

\bibitem{wang2023mac}
B.~Wang, C.~Ren, J.~Yang, and et~al., ``Mac-sql: A multi-agent collaborative framework for text-to-sql.'' \emph{CoRR}, 2023.

\bibitem{pourreza2024chase}
M.~Pourreza, H.~Li, R.~Sun, and et~al., ``Chase-sql: Multi-path reasoning and preference optimized candidate selection in text-to-sql,'' 2024.

\bibitem{li2025alpha}
B.~Li, J.~Zhang, J.~Fan, Y.~Xu, C.~Chen, N.~Tang, and Y.~Luo, ``Alpha-sql: Zero-shot text-to-sql using monte carlo tree search,'' in \emph{Forty-Second International Conference on Machine Learning, {ICML} 2025, Vancouver, Canada, July 13-19, 2025}, 2025.

\bibitem{xie2025opensearch}
X.~Xie, G.~Xu, L.~Zhao, and R.~Guo, ``Opensearch-sql: Enhancing text-to-sql with dynamic few-shot and consistency alignment,'' 2025.

\bibitem{qin2024route}
Y.~Qin, C.~Chen, Z.~Fu, and et~al., ``Route: Robust multitask tuning and collaboration for text-to-sql,'' \emph{arXiv preprint arXiv:2412.10138}, 2024.

\bibitem{talaei2024chess}
S.~Talaei, M.~Pourreza, Y.-C. Chang, and et~al., ``Chess: Contextual harnessing for efficient sql synthesis,'' \emph{arXiv:2405.16755}, 2024.

\bibitem{pourreza2024dts}
M.~Pourreza and D.~Rafiei, ``Dts-sql: Decomposed text-to-sql with small large language models,'' \emph{arXiv:2402.01117}, 2024.

\bibitem{qu2024before}
G.~Qu, J.~Li, B.~Li, and et~al., ``Before generation, align it! a novel and effective strategy for mitigating hallucinations in text-to-sql generation,'' \emph{arXiv}, 2024.

\bibitem{li2024pet}
Z.~Li and et~al., ``Pet-sql: A prompt-enhanced two-stage text-to-sql framework with cross-consistency,'' \emph{arXiv:2403.09732}, 2024.

\bibitem{zhang2024coe}
H.~Zhang, R.~Cao, H.~Xu, and et~al., ``Coe-sql: In-context learning for multi-turn text-to-sql with chain-of-editions,'' \emph{arXiv:2405.02712}, 2024.

\bibitem{ren2024purple}
T.~Ren, Y.~Fan, Z.~He, , and et~al., ``Purple: Making a large language model a better sql writer,'' \emph{arXiv:2403.20014}, 2024.

\bibitem{fan2024metasql}
Y.~Fan, Z.~He, T.~Ren, and et~al., ``Metasql: A generate-then-rank framework for natural language to sql translation,'' 2024.

\bibitem{xie2024decomposition}
Y.~Xie and et~al., ``Decomposition for enhancing attention: Improving llm-based text-to-sql through workflow paradigm,'' \emph{arXiv}, 2024.

\bibitem{dong2023c3}
X.~Dong, C.~Zhang, Y.~Ge, and et~al., ``C3: Zero-shot text-to-sql with chatgpt,'' \emph{arXiv:2307.07306}, 2023.

\bibitem{rai-etal-2023-improving}
D.~Rai, B.~Wang, Y.~Zhou, and et~al., ``Improving generalization in language model-based text-to-sql semantic parsing: Two simple semantic boundary-based techniques,'' in \emph{ACL}, 2023.

\bibitem{zhang2023act}
H.~Zhang and et~al., ``Act-sql: In-context learning for text-to-sql with automatically-generated chain-of-thought,'' in \emph{EMNLP(Findings)}, 2023.

\bibitem{chang2023selective}
S.~Chang and E.~Fosler-Lussier, ``Selective demonstrations for cross-domain text-to-sql,'' in \emph{Findings of EMNLP}, 2023.

\bibitem{fu2023catsql}
H.~Fu, C.~Liu, B.~Wu, and et~al., ``Catsql: Towards real world natural language to sql applications,'' \emph{Proc. {VLDB} Endow.}, 2023.

\bibitem{bazaga2023sqlformer}
A.~Bazaga, P.~Li{\`o}, and et~al., ``Sqlformer: Deep auto-regressive query graph generation for text-to-sql translation,'' \emph{arXiv:2310.18376}, 2023.

\bibitem{xiang2023g3r}
Y.~Xiang, Q.-W. Zhang, X.~Zhang, and et~al., ``G3r: A graph-guided generate-and-rerank framework for complex and cross-domain text-to-sql generation,'' in \emph{Findings of ACL}, 2023.

\bibitem{zhao2022importance}
Y.~Hu, Y.~Zhao, J.~Jiang, and et~al., ``Importance of synthesizing high-quality data for text-to-sql parsing,'' in \emph{Findings of {ACL}}, 2023.

\bibitem{zeng2023n}
L.~Zeng, S.~H.~K. Parthasarathi, and D.~Hakkani-Tur, ``N-best hypotheses reranking for text-to-sql systems,'' in \emph{SLT}, 2023.

\bibitem{qi2022rasat}
J.~Qi, J.~Tang, Z.~He, and et~al., ``Rasat: Integrating relational structures into pretrained seq2seq model for text-to-sql,'' \emph{arXiv:2205.06983}, 2022.

\bibitem{scholak2021picard}
T.~Scholak and et~al., ``Picard: Parsing incrementally for constrained auto-regressive decoding from language models,'' 2021.

\bibitem{gao2022towards}
C.~Gao, B.~Li, W.~Zhang, and et~al., ``Towards generalizable and robust text-to-sql parsing,'' \emph{arXiv:2210.12674}, 2022.

\bibitem{hui2022s}
B.~Hui and et~al., ``S2sql: Injecting syntax to question-schema interaction graph encoder for text-to-sql parsers,'' in \emph{Findings of ACL}, 2022.

\bibitem{DBLP:conf/acl/WangSLPR20}
B.~Wang, R.~Shin, X.~Liu, and et~al., ``{RAT-SQL:} relation-aware schema encoding and linking for text-to-sql parsers,'' in \emph{{ACL}}, 2020.

\bibitem{rubin2020smbop}
O.~Rubin and J.~Berant, ``Smbop: Semi-autoregressive bottom-up semantic parsing,'' \emph{arXiv:2010.12412}, 2020.

\bibitem{huang2021relation}
J.~Huang, Y.~Wang, Y.~Wang, and et~al., ``Relation aware semi-autoregressive semantic parsing for nl2sql,'' \emph{arXiv:2108.00804}, 2021.

\bibitem{lin2020bridging}
X.~V. Lin and et~al., ``Bridging textual and tabular data for cross-domain text-to-sql semantic parsing,'' \emph{arXiv:2012.12627}, 2020.

\bibitem{guo2019towards}
J.~Guo, Z.~Zhan, Y.~Gao, and et~al., ``Towards complex text-to-sql in cross-domain database with intermediate representation,'' in \emph{ACL}, 2019.

\bibitem{yu2018typesql}
T.~Yu, Z.~Li, Z.~Zhang, and et~al., ``Typesql: Knowledge-based type-aware neural text-to-sql generation,'' \emph{arXiv:1804.09769}, 2018.

\bibitem{brunner2021valuenet}
U.~Brunner and K.~Stockinger, ``Valuenet: A natural language-to-sql system that learns from database information,'' in \emph{ICDE}.\hskip 1em plus 0.5em minus 0.4em\relax IEEE, 2021.

\bibitem{damerau1964technique}
F.~J. Damerau, ``A technique for computer detection and correction of spelling errors,'' \emph{Communications of the ACM}, 1964.

\bibitem{dong2019data}
Z.~Dong, S.~Sun, H.~Liu, and et~al., ``Data-anonymous encoding for text-to-sql generation,'' in \emph{EMNLP-IJCNLP}, 2019.

\bibitem{brown2020language}
T.~Brown, B.~Mann, N.~Ryder, and et~al., ``Language models are few-shot learners,'' \emph{NeurIPS}, 2020.

\bibitem{lee2024mcs}
D.~Lee, C.~Park, J.~Kim, and et~al., ``Mcs-sql: Leveraging multiple prompts and multiple-choice selection for text-to-sql generation,'' 2024.

\bibitem{yin2020tabert}
P.~Yin, G.~Neubig, W.-t. Yih, and et~al., ``Tabert: Pretraining for joint understanding of textual and tabular data,'' \emph{arXiv:2005.08314}, 2020.

\bibitem{speer2012representing}
R.~Speer, C.~Havasi \emph{et~al.}, ``Representing general relational knowledge in conceptnet 5.'' in \emph{LREC}, 2012.

\bibitem{indyk1998approximate}
P.~Indyk and R.~Motwani, ``Approximate nearest neighbors: towards removing the curse of dimensionality,'' in \emph{STOC}, 1998.

\bibitem{10.1561/1500000019}
S.~Robertson, H.~Zaragoza \emph{et~al.}, ``The probabilistic relevance framework: Bm25 and beyond,'' \emph{FOUND TRENDS INF RET}, 2009.

\bibitem{aho1975efficient}
A.~V. Aho and M.~J. Corasick, ``Efficient string matching: an aid to bibliographic search,'' \emph{Communications of the ACM}, 1975.

\bibitem{li2014schema}
F.~Li and et~al., ``Schema-free sql,'' in \emph{SIGMOD}, 2014.

\bibitem{yu2018syntaxsqlnet}
T.~Yu and et~al., ``Syntaxsqlnet: Syntax tree networks for complex and cross-domaintext-to-sql task,'' \emph{arXiv:1810.05237}, 2018.

\bibitem{lee1999semql}
J.-O. Lee and D.-K. Baik, ``Semql: a semantic query language for multidatabase systems,'' in \emph{CIKM}, 1999.

\bibitem{zhang2019editing}
R.~Zhang and et~al., ``Editing-based sql query generation for cross-domain context-dependent questions,'' \emph{arXiv:1909.00786}, 2019.

\bibitem{NatSQL}
Y.~Gan and et~al., ``Natural {SQL}: Making {SQL} easier to infer from natural language specifications,'' in \emph{Findings of EMNLP}, 2021.

\bibitem{wolfson2020break}
T.~Wolfson, M.~Geva, A.~Gupta, and et~al., ``Break it down: A question understanding benchmark,'' \emph{TACL}, 2020.

\bibitem{dou2023towards}
L.~Dou, Y.~Gao, X.~Liu, and et~al., ``Towards knowledge-intensive text-to-sql semantic parsing with formulaic knowledge,'' in \emph{EMNLP}, 2022.

\bibitem{karpukhin2020dense}
V.~Karpukhin, B.~O{\u{g}}uz, S.~Min, and et~al., ``Dense passage retrieval for open-domain question answering,'' \emph{arXiv:2004.04906}, 2020.

\bibitem{DBLP:conf/acl/LiuYZGZL21}
Q.~Liu, D.~Yang, J.~Zhang, J.~Guo, B.~Zhou, and J.~Lou, ``Awakening latent grounding from pretrained language models for semantic parsing,'' in \emph{{ACL/IJCNLP} (Findings)}, 2021.

\bibitem{sui2023reboost}
G.~Sui, Z.~Li, Z.~Li, and et~al., ``Reboost large language model-based text-to-sql, text-to-python, and text-to-function--with real applications in traffic domain,'' \emph{arXiv:2310.18752}, 2023.

\bibitem{clark2020electra}
K.~Clark, ``Electra: Pre-training text encoders as discriminators rather than generators,'' \emph{arXiv:2003.10555}, 2020.

\bibitem{cao2021lgesql}
R.~Cao and et~al., ``Lgesql: line graph enhanced text-to-sql model with mixed local and non-local relations,'' \emph{arXiv:2106.01093}, 2021.

\bibitem{velivckovic2017graph}
P.~Veli{\v{c}}kovi{\'c}, G.~Cucurull, A.~Casanova, A.~Romero, P.~Lio, and Y.~Bengio, ``Graph attention networks,'' \emph{arXiv:1710.10903}, 2017.

\bibitem{xu2017sqlnet}
X.~Xu and et~al., ``Sqlnet: Generating structured queries from natural language without reinforcement learning,'' \emph{arXiv:1711.04436}, 2017.

\bibitem{dataset-wikisql}
V.~Zhong and et~al., ``Seq2sql: Generating structured queries from natural language using reinforcement learning,'' \emph{CoRR}, 2017.

\bibitem{bi2024deepseek}
X.~Bi, D.~Chen, G.~Chen, and et~al., ``Deepseek llm: Scaling open-source language models with longtermism,'' \emph{arXiv:2401.02954}, 2024.

\bibitem{wei2022chain}
J.~Wei, X.~Wang, D.~Schuurmans, and et~al., ``Chain-of-thought prompting elicits reasoning in large language models,'' \emph{NeurIPS}, 2022.

\bibitem{tai2023exploring}
C.-Y. Tai, Z.~Chen, T.~Zhang, and et~al., ``Exploring chain-of-thought style prompting for text-to-sql,'' \emph{arXiv:2305.14215}, 2023.

\bibitem{dataset-academic}
F.~Li and H.~V. Jagadish, ``Constructing an interactive natural language interface for relational databases,'' \emph{Proc. {VLDB} Endow.}, 2014.

\bibitem{eyal2023semantic}
B.~Eyal, A.~Bachar, O.~Haroche, and et~al., ``Semantic decomposition of question and sql for text-to-sql parsing,'' \emph{arXiv:2310.13575}, 2023.

\bibitem{lee2019clause}
D.~Lee, ``Clause-wise and recursive decoding for complex and cross-domain text-to-sql generation,'' \emph{arXiv:1904.08835}, 2019.

\bibitem{wang2022self}
X.~Wang and et~al., ``Self-consistency improves chain of thought reasoning in language models,'' \emph{arXiv:2203.11171}, 2022.

\bibitem{renze2024effect}
M.~Renze and E.~Guven, ``The effect of sampling temperature on problem solving in large language models,'' \emph{arXiv:2402.05201}, 2024.

\bibitem{kelkar2020bertrand}
A.~Kelkar, R.~Relan, V.~Bhardwaj, and et~al., ``Bertrand-dr: Improving text-to-sql using a discriminative re-ranker,'' \emph{arXiv:2002.00557}, 2020.

\bibitem{DBLP:journals/pacmmod/LuoZ00CS23}
Y.~Luo, Y.~Zhou, N.~Tang, and et~al., ``Learned data-aware image representations of line charts for similarity search,'' \emph{SIGMOD}, 2023.

\bibitem{zhang2023refsql}
K.~Zhang and et~al., ``Refsql: A retrieval-augmentation framework for text-to-sql generation,'' in \emph{Findings of EMNLP}, 2023.

\bibitem{dataset-atis}
D.~A. Dahl, M.~Bates, M.~K. Brown, and et~al., ``Expanding the scope of the atis task: The atis-3 corpus,'' in \emph{Workshop on HLT}, 1994.

\bibitem{dataset-geoquery}
J.~M. Zelle and R.~J. Mooney, ``Learning to parse database queries using inductive logic programming,'' in \emph{AAAI}, 1996.

\bibitem{dataset-restaurant}
L.~R. Tang and R.~J. Mooney, ``Automated construction of database interfaces: Intergrating statistical and relational learning for semantic parsing,'' in \emph{EMNLP}, 2000.

\bibitem{dataset-yelp-imdb}
N.~Yaghmazadeh, Y.~Wang, I.~Dillig, and T.~Dillig, ``Sqlizer: query synthesis from natural language,'' \emph{Proc. {ACM} Program. Lang.}, 2017.

\bibitem{dataset-scholar}
S.~Iyer, I.~Konstas, A.~Cheung, J.~Krishnamurthy, and L.~Zettlemoyer, ``Learning a neural semantic parser from user feedback,'' in \emph{ACL}, 2017.

\bibitem{dataset-advising}
C.~Finegan{-}Dollak, J.~K. Kummerfeld, L.~Zhang, and et~al., ``Improving text-to-sql evaluation methodology,'' in \emph{ACL}, 2018.

\bibitem{dataset-sparc}
T.~Yu, R.~Zhang, M.~Yasunaga, and et~al., ``Sparc: Cross-domain semantic parsing in context,'' in \emph{ACL}, 2019.

\bibitem{dataset-cosql}
T.~Yu, R.~Zhang, H.~Er, and et~al., ``Cosql: {A} conversational text-to-sql challenge towards cross-domain natural language interfaces to databases,'' in \emph{EMNLP-IJCNLP}, 2019.

\bibitem{dataset-cspider}
Q.~Min, Y.~Shi, and Y.~Zhang, ``A pilot study for chinese {SQL} semantic parsing,'' in \emph{EMNLP-IJCNLP}, 2019.

\bibitem{dataset-mimicsql}
P.~Wang, T.~Shi, and C.~K. Reddy, ``Text-to-sql generation for question answering on electronic medical records,'' in \emph{WWW}, 2020.

\bibitem{dataset-squall}
T.~Shi and et~al., ``On the potential of lexico-logical alignments for semantic parsing to {SQL} queries,'' in \emph{Findings of EMNLP}, 2020.

\bibitem{sen2020athena++}
J.~Sen, C.~Lei, A.~Quamar, and et~al., ``Athena++ natural language querying for complex nested sql queries,'' \emph{Proc. {VLDB} Endow.}, 2020.

\bibitem{dataset-vitext2sql}
A.~T. Nguyen, M.~H. Dao, and D.~Q. Nguyen, ``A pilot study of text-to-sql semantic parsing for vietnamese,'' in \emph{Findings of EMNLP}, 2020.

\bibitem{dataset-dusql}
L.~Wang, A.~Zhang, K.~Wu, and et~al., ``Dusql: A large-scale and pragmatic chinese text-to-sql dataset,'' in \emph{EMNLP}, 2020.

\bibitem{dataset-portuguese-spider}
M.~A. Jos{\'e} and F.~G. Cozman, ``mrat-sql+ gap: a portuguese text-to-sql transformer,'' in \emph{BRACIS}, 2021.

\bibitem{dataset-chase}
J.~Guo, Z.~Si, Y.~Wang, and et~al., ``Chase: {A} large-scale and pragmatic chinese dataset for cross-database context-dependent text-to-sql,'' in \emph{ACL/IJCNLP}, 2021.

\bibitem{spider-syn}
Y.~Gan, X.~Chen, Q.~Huang, and et~al., ``Towards robustness of text-to-sql models against synonym substitution,'' in \emph{ACL/IJCNLP}, 2021.

\bibitem{dataset-spider-dk}
Y.~Gan, X.~Chen, and M.~Purver, ``Exploring underexplored limitations of cross-domain text-to-sql generalization,'' in \emph{EMNLP}, 2021.

\bibitem{dataset-spider-realistic}
X.~Deng, A.~H. Awadallah, C.~Meek, and et~al., ``Structure-grounded pretraining for text-to-sql,'' in \emph{NAACL-HLT}, 2021.

\bibitem{dataset-kaggledbqa}
C.~Lee, O.~Polozov, and M.~Richardson, ``Kaggledbqa: Realistic evaluation of text-to-sql parsers,'' in \emph{ACL/IJCNLP}, 2021.

\bibitem{dataset-sede}
M.~Hazoom and et~al., ``Text-to-sql in the wild: A naturally-occurring dataset based on stack exchange data,'' \emph{arXiv:2106.05006}, 2021.

\bibitem{dataset-mt-teql}
P.~Ma and S.~Wang, ``Mt-teql: Evaluating and augmenting neural {NLIDB} on real-world linguistic and schema variations,'' \emph{VLDB}, 2021.

\bibitem{dataset-pauq}
D.~Bakshandaeva, O.~Somov, E.~Dmitrieva, V.~Davydova, and E.~Tutubalina, ``{PAUQ:} text-to-sql in russian,'' in \emph{Findings of {EMNLP}}, 2022.

\bibitem{dataset-dr-spider}
S.~Chang, J.~Wang, M.~Dong, and et~al., ``Dr.spider: {A} diagnostic evaluation benchmark towards text-to-sql robustness,'' in \emph{ICLR}, 2023.

\bibitem{dataset-ambiqt}
A.~Bhaskar, T.~Tomar, A.~Sathe, and S.~Sarawagi, ``Benchmarking and improving text-to-sql generation under ambiguity,'' in \emph{EMNLP}, 2023.

\bibitem{dataset-sciencebenchmark}
Y.~Zhang, J.~Deriu, G.~Katsogiannis{-}Meimarakis, and et~al., ``Sciencebenchmark: {A} complex real-world benchmark for evaluating natural language to {SQL} systems,'' \emph{Proc. {VLDB} Endow.}, 2023.

\bibitem{dataset-booksql}
R.~Kumar, A.~R. Dibbu, S.~Harsola, and et~al., ``Booksql: A large scale text-to-sql dataset for accounting domain,'' \emph{arXiv:2406.07860}, 2024.

\bibitem{dataset-archer}
D.~Zheng and et~al., ``Archer: {A} human-labeled text-to-sql dataset with arithmetic, commonsense and hypothetical reasoning,'' in \emph{EACL}, 2024.

\bibitem{dataset-spider2-lite}
F.~Lei, J.~Chen, Y.~Ye, and et~al., ``Spider 2.0: Evaluating language models on real-world enterprise text-to-sql workflows,'' 2024.

\bibitem{narechania2021diy}
A.~Narechania, A.~Fourney, B.~Lee, and et~al., ``Diy: Assessing the correctness of natural language to sql systems,'' in \emph{IUI}, 2021.

\bibitem{ning2023empirical}
Z.~Ning, Z.~Zhang, T.~Sun, and et~al., ``An empiricmacsqlal study of model errors and user error discovery and repair strategies in natural language database queries,'' in \emph{IUI}, 2023.

\bibitem{sun2023sql}
R.~Sun, S.~O. Arik, H.~Nakhost, and et~al., ``Sql-palm: Improved large language modeladaptation for text-to-sql,'' \emph{arXiv:2306.00739}, 2023.

\bibitem{taipalus2023framework}
T.~Taipalus and H.~Grahn, ``Framework for sql error message design: A data-driven approach,'' \emph{TOSEM}, 2023.

\bibitem{zhu2025elliesqlcostefficienttexttosqlcomplexityaware}
Y.~Zhu, R.~Jiang, B.~Li, , and et~al., ``Elliesql: Cost-efficient text-to-sql with complexity-aware routing,'' 2025.

\bibitem{DBLP:journals/inffus/AliAEMACGSRH23}
S.~Ali, T.~Abuhmed, S.~H.~A. El{-}Sappagh, and et~al., ``Explainable artificial intelligence {(XAI):} what we know and what is left to attain trustworthy artificial intelligence,'' \emph{Inf. Fusion}, 2023.

\bibitem{DBLP:journals/tist/ZhaoCYLDCWYD24}
H.~Zhao, H.~Chen, F.~Yang, and et~al., ``Explainability for large language models: {A} survey,'' \emph{{ACM} Trans. Intell. Syst. Technol.}, 2024.

\bibitem{DBLP:journals/corr/abs-2307-08678}
Y.~Chen, R.~Zhong, N.~Ri, and et~al., ``Do models explain themselves? counterfactual simulatability of natural language explanations,'' 2023.

\bibitem{DBLP:conf/icml/ShrikumarGK17}
A.~Shrikumar, P.~Greenside, and A.~Kundaje, ``Learning important features through propagating activation differences,'' in \emph{{ICML}}, 2017.

\bibitem{zhang2024aflowautomatingagenticworkflow}
J.~Zhang, J.~Xiang, Z.~Yu, , and et~al., ``{AF}low: Automating agentic workflow generation,'' in \emph{ICLR}, 2025.

\end{thebibliography}

\end{document}